\documentclass[12pt]{iopart}
\usepackage{graphicx}
\usepackage{color}
\begin{document}
\review[Strong coupling between surface plasmon polaritons and emitters]{Strong coupling between surface plasmon polaritons and emitters}

\author{P T\"orm\"a$^1$ and W L Barnes$^{2,3}$}

\address{$^1$ COMP Centre of Excellence, Department of Applied Physics, Aalto University, FI-00076 Aalto, Finland}
\address{$^2$ School of Physics and Astronomy, University of Exeter, Stocker Road, Exeter, Devon, EX4 4QL, United Kingdom}
\address{$^3$ Complex Photonic Systems (COPS), MESA+ Institute for Nanotechnology, University of Twente, 7500 AE Enschede, The Netherlands}
\eads{\mailto{paivi.torma@aalto.fi}, \mailto{w.l.barnes@exeter.ac.uk}}
\begin{abstract}
In this review we look at the concepts and state-of-the-art concerning the strong coupling of surface plasmon-polariton modes to states associated with quantum emitters such as excitons in J-aggregates, dye molecules and quantum dots. We explore the phenomenon of strong coupling with reference to a number of examples involving electromagnetic fields and matter. We then provide a concise description of the relevant background physics of surface plasmon polaritons. An extensive overview of the historical background and a detailed discussion of more recent relevant experimental advances concerning strong coupling between surface plasmon polaritons and quantum emitters is then presented. Three conceptual frameworks are then discussed and compared in depth: classical, semi-classical and fully quantum mechanical; these theoretical frameworks will have relevance to strong coupling beyond that involving surface plasmon polaritons. We conclude our review with a perspective on the future of this rapidly emerging field, one we are sure will grow to encompass more intriguing physics and will develop in scope to be of relevance to other areas of science.
\end{abstract}
\submitto{\RPP}
\maketitle

\section{Introduction}

Many sources of light involve transitions between the electronic energy levels of a well-defined quantum system, for example dye molecules and quantum dots. It is now widely known that the rate of the emission process can be modified by placing the emitter in a structured environment, e.g. in front of a mirror \cite{Drexhage_ProgOpt_1974_12_163}. Usually the interaction between the emitter and its local optical environment is such that only the spontaneous emission rate is modified, the emission frequency remaining unaltered. However, if the interaction is strong enough then the energy levels responsible for the emission are also altered, they become inextricably linked with the levels (modes) of the local optical environment. If this happens the energy levels of this hybrid system may be very different from those of the emitter or the optical system individually. This situation is known as {\it strong coupling} and is the subject of this review. Not only do these new hybrid systems offer an exciting arena in which to explore light-matter interactions, they also offer the prospect of exploiting nano-fabrication techniques to design quantum optical systems. 

The paradigm model of strong coupling is that of two coupled harmonic oscillators, they may become coupled if there is some way for them to exchange energy. The dynamics of the coupled system is influenced not only by the original frequencies of the oscillators but also by the exchange process involved in the coupling. The energy spectrum of the system is modified: indeed, one obtains new modes whose frequencies are different from those of the original oscillator modes, the difference between the original and the new frequencies depends on the strength of the coupling. If the coupling is very small compared to other relevant energy scales, the modification of the original energies due to the coupling is negligible: this is the weak coupling regime. On the other hand if the coupling is large compared to other energies, the coupling modifies the energy spectrum of the total system qualitatively: this is the strong coupling regime. The new energies in the strong coupling regime correspond to modes that are hybrids of the original modes of the two oscillators \cite{Haroche_1992_CQED}. 

Strong coupling phenomena are observed in, and are of importance to several fields of physics and technology. A large variety of strong coupling phenomena are observed in case of the light-matter interaction: the light field of a certain frequency is one of the oscillators and a material (atom, molecule, semiconductor, etc.) with a well-defined optical transition provides the other oscillator. The phenomenon has been realized also for systems where the field frequency is not in the optical domain. 

One of the attractions of the combination of having excitons as one of the oscillators and plasmon modes as the other is the very extensive control we have over the plasmon modes supported by metallic nanostructures. This control derives from a combination of impressive nano-fabrication techniques \cite{Gates_ChemRev_2005_105_1171} and a very good understanding of the relationship between the details of the nanostructure and the nature of the associated plasmon modes \cite{Kern_ACSNano_2012_6_9828}.

We will now discuss the basics of the strong coupling phenomenology with the aid of a simple example, that of two coupled classical harmonic oscillators. More elaborate descriptions, in the specific context of this review article, are given in sections \ref{classical} and \ref{quantum}.

\subsection{Strong coupling basics: two coupled harmonic oscillators}

\begin{figure}
\includegraphics[width=1.0\textwidth]{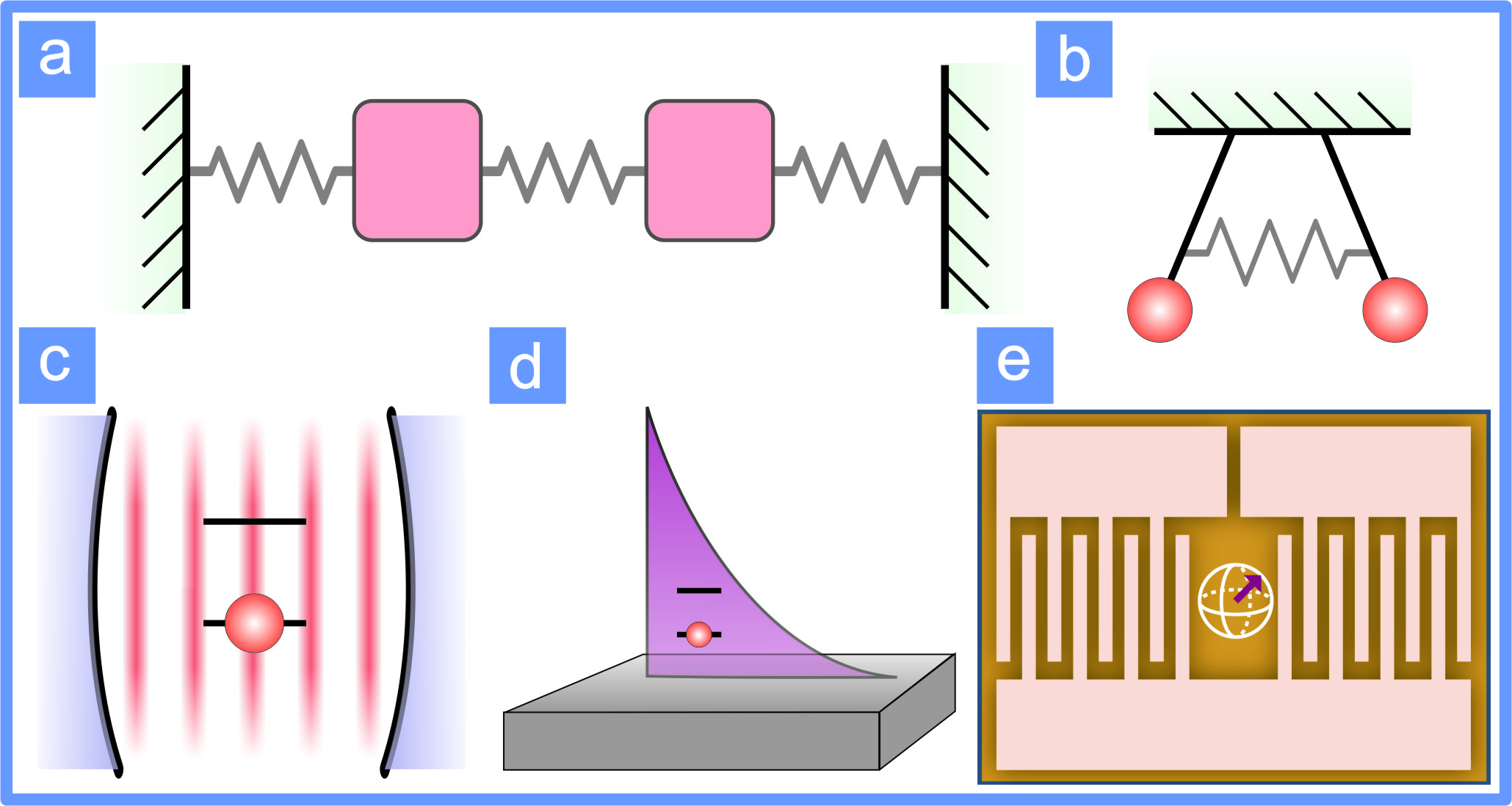}
\caption{Two coupled harmonic oscillators. Two coupled strings (a) and coupled pendulums (b) are given as examples from the macroscopic world. In the microscopic and nanoscale, for instance atoms/molecules interacting with cavity (c) or surface plasmon fields (d), or superconducting circuits interacting with microwave resonators (e) can be approximated as coupled oscillators.}
\label{oscillatorfigure}
\end{figure}

Let us consider two harmonic oscillators that are coupled. Examples of such systems include: two coupled pendula (oscillating at small frequencies), an optical field coupled to a dipolar two-level transition of an atom/molecule, and a microwave field coupled to a resonating circuit. Here, for simplicity, we set the resonance condition, that is, the frequencies of the oscillators, to be the same. The treatment can be easily generalized to the case of the individual oscillators having different frequencies. Using the theory of simple harmonic motion the dynamics of the coupled system is described by (see also figure \ref{oscillatorfigure}) \cite{Novotny_AJP_2010_78_1199}, 
\begin{eqnarray}
\frac{d^2 x_1}{dt^2} + \omega^2 x_1 + \Omega^2 (x_1 - x_2) &=& 0 , \\
\frac{d^2 x_2}{dt^2} + \omega^2 x_2 + \Omega^2 (x_2 - x_1) &=& 0 . 
\end{eqnarray}
Here $\omega$ is the angular frequency of the oscillators and $\Omega$ gives the strength of the coupling between the two oscillators, which we refer to by the labels 1 and 2. These differential equations can be solved to give the time evolution for the positions of the oscillators:
\begin{equation}
x_1(t) = A \sin (\omega_+ t + C) + B \sin (\omega_- t + D) ,  \label{HOsolutions1}
\end{equation}
\begin{equation}
x_2(t) = - A \sin (\omega_+ t + C) + B \sin (\omega_- t + D) . \label{HOsolutions2} 
\end{equation}
The constants $A$, $B$, $C$ and $D$ may be determined from the initial conditions. The new frequencies that appear here are, 
\begin{eqnarray}
\omega_+^2 &=& \omega_c^2 + \Omega^2 \\
\omega_-^2 &=& \omega_c^2 - \Omega^2 \\
\omega_c^2 &=& \omega^2 + \Omega^2  .
\end{eqnarray}
Here $\omega_c^2$ is the frequency that one of the oscillators would have if the other one was held fixed. The new frequencies $\omega_+$ and $\omega_-$ are the {\it normal modes} of the coupled oscillator system. Let us now assume that we want to find the frequencies $\omega_+$ and $\omega_-$ by observing the dynamics of the system. From (\ref{HOsolutions1}-\ref{HOsolutions2}) one obtains
\begin{eqnarray}
\sin (\omega_+ t + C) &=& \frac{1}{2A} (x_1(t) - x_2(t)) , \\
\sin (\omega_- t + C) &=& \frac{1}{2B} (x_1(t) + x_2(t)) .
\end{eqnarray} 
We see that the normal modes $\omega_+$ and $\omega_-$ are not related to the motion (position) of either of the single oscillators alone, instead, to find the normal modes one needs to examine the time evolution of the motion of both oscillators. In other words, the normal modes are hybrid modes of the two original oscillators. As a result it is no longer adequate to describe the system in terms of the original oscillators, rather one should use the normal, hybrid modes. Note that the energy (frequency) separation of the normal modes depends on the coupling $\Omega$. 

\begin{figure}
\includegraphics[width=1.0\textwidth]{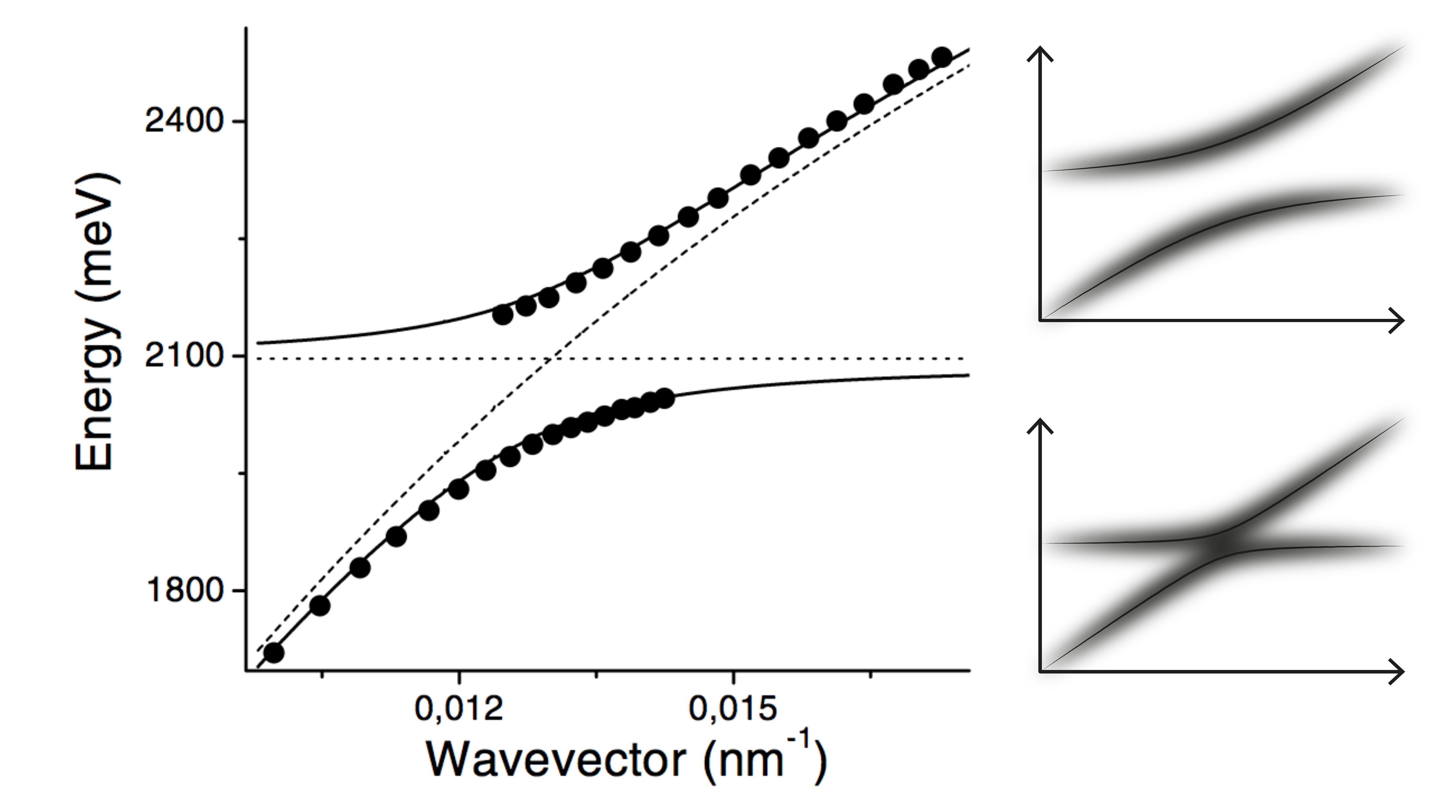}
\caption{Left figure: Strong coupling between a surface plasmon-polariton mode (diagonal dotted line) and an excitonic mode (horizontal dashed line). The energies of these two modes are shown as a function of in-plane wavevector. The system is a metal film (that supports the SPPmode) overcoated with a film of polymer containing aggregated dye molecules. The solid lines and the data (circles) show how these modes interact to produce an avoided crossing, the dashed lines show the dispersion expected in the absence of strong coupling. Figure reproduced with permission from \cite{Bellessa_PRL_2004_93_036404}. Figures on the right: strong coupling regime can be defined as the splitting being large enough compared to the linewidths of the coupled states so that it is actually experimentally visible as in the upper picture; in the lower one, the splitting is hidden under the linewidths.}
\label{Bellessa_figure}
\end{figure}

One can show, as will be done in sections \ref{classical} and \ref{quantum}, that the effect of the coupling becomes smaller the further away one is from the resonance condition. Let us say the oscillators have different frequencies, $\omega_1$ and $\omega_2$; their difference is defined as $\delta = \omega_1 - \omega_2$ and is called the {\it detuning}. Assume now that one of the original oscillator frequencies is not tuneable: for instance in the case of an atom interacting with a light field it would be natural to have the energy of the electronic transition fixed, the other frequency can be varied. In the atom+light example, it would be possible to tune the frequency of the light field. Naturally, when one of the frequencies changes, the detuning $\delta$ changes. The behaviour of the energies (frequencies) of the coupled system is now shown in figure \ref{Bellessa_figure}. The data show the avoided crossing of an excitonic transition associated with J-aggregated dye molecules, and a surface plasmon-polariton mode taken from the work of Bellessa {\it et al.} and discussed in section \ref{J-aggregates} below.  The behaviour expected if there were no strong coupling is shown by the dashed lines. Far away from the resonance, the energies (frequencies) of the original oscillators are practically unchanged from the un-coupled case. Near the crossing point (of the non-coupled case) the new normal modes appear as a result of the coupling: a so called avoided crossing is observed. The energy separation between the normal modes at the avoided crossing is called {\it the normal mode splitting} or, equivalently, the Rabi split or {\it the Rabi splitting} (the origin of the name will become clearer in section \ref{quantum}).      

Although the new normal modes always appear for the coupled system, whether the phenomenon is significant depends on the strength of the coupling, compared to other relevant energies. Accordingly two regimes may be defined: {\it the weak coupling} and {\it the strong coupling regimes}. How the regimes are defined, however, depends to some extent on the context. For mechanical oscillators, like strings, strong coupling is sometimes defined as $\Omega > \omega$: the coupling modifies the oscillation frequency considerably when it is of the order of the frequency itself. However, in some other contexts the frequencies of the oscillators can be really high, as is the case for instance with light fields, $~10^{15}Hz$, and a lot of interesting physics can be observed already for couplings that are much, much smaller than this (actually, in many contexts $\Omega > \omega$ is the condition for the so-called {\it ultrastrong coupling} regime). Perhaps a more useful comparison energy scale is the transition linewidths, the pictures on the right side of figure \ref{Bellessa_figure} illustrate this. The width of the energy line schematically represents the linewidth. The Rabi split becomes significant, i.e.\ experimentally observable, only when the coupling is large enough compared to the linewidth. Thus the strong coupling regime is sometimes defined as the range where the coupling exceeds the linewidths of the two coupled systems. This issue is however a subtle one and will be discussed in section \ref{DyePT}. The definition of strong coupling is thus somewhat context sensitive, depending among other things on the conventions of the particular field of physics involved. For the purposes of this article, we define the term strong coupling in a pragmatic way: the system is in the strong coupling regime whenever the Rabi split is experimentally observable.     

Strong coupling physics is intimately linked with the concept of coherence. In the harmonic oscillator example above we ignored damping, if we now consider the oscillators to be damped then it is easy to show that if the damping is much stronger than the coupling between the oscillators, then the normal modes approach the frequencies of the original oscillators i.e.\ the effect of the coupling on the energies is negligible. The difference in the normal mode frequencies, i.e the splitting, was given by the strength of the coupling, the associated timescale is related to the inverse of the coupling strength. Now, to resolve such an effect by Fourier analysis into a notable feature in the energy spectrum one would need oscillatory motion for a timescale longer than the inverse of the coupling. For damping stronger than the coupling this does not happen. In the context of light interacting with matter (and also for many other systems in the microscopic world) the undamped oscillatory behaviour is referred to by the word {\it coherence}. One can have either classical coherence, such as coherent light with a well-defined phase, 
and/or quantum coherence such as quantum superpositions between different quantum states (e.g. the ground and excited electronic states of an atom). To observe strong coupling the system in question has to have coherence times that are larger than the inverse of the coupling. 

\subsection{Relevance and applications of strong coupling}

Whenever one observes strong coupling, it is a signature of entering a regime where coherence phenomena play a role. Strong coupling is thus intimately connected to important phenomena such as stimulated emission, gain and lasing. In the strong coupling regime, it may be possible to realize, for example, tresholdless lasing \cite{Rice1994}.  Apart from coherence, an interesting issue is the hybrid nature of the normal modes. The normal mode possesses properties of both of the oscillators, which in the case of, for instance light and matter, can be very different. This leads to interesting phenomenology and possibilities for applications. An example of this was the strong coupling of excitons and light into hybrid modes called polaritons in semiconductor systems. By making clever use of both the excitonic (that is, simply electrons of the semiconductor) and the light part of the polariton, it was possible to realize Bose Einstein condensates of polaritons \cite{Kasprzak2006}. In the context of quantum information technology, achieving the strong coupling regime or, in other words, quantum coherent oscillations between the coupled systems, is a prerequisite for quantum information processing. Furthermore, the ability of strong coupling to modify the electromagnetic environment of an emitter has been used to modifying chemical landscapes and to control chemical reaction rates \cite{Hutchison2012}. In the context of superconducting qubits (two-level systems) interacting with microwave radiation, the coupling between a qubit and the electromagnetic modes of an infinite quasi-one-dimensional transmission line results in interference effects between the incoming wave and the wave emitted by the qubit. This effect gets stronger as the coupling to the line increases, thus rendering the decay into the line as the dominant decay channel of the qubit. This type of one-dimensional fluorescence effect has been observed experimentally \cite{Astafiev2010} and it could have applications for single-photon switches \cite{Li2012}, routers \cite{Hoi2011}, photon detectors \cite{Romero2009}, and interferometers with single-atom 
mirrors \cite{Paraoanu2010}.

The focus so far has been on strong coupling between a quantum emitter and a field mode, what will happen if we look at how the exchange of energy between two emitters (resonant energy transfer) is modified when the two emitter system is strongly coupled to a surface plasmon polariton (SPP) mode. The use of SPP modes associated with planar metal films to extend the range over which energy transfer between two emitters may take place, a distance which is typically only of order a few nm, was reported by Andrew and Barnes in 2004 \cite{Andrew_Science_2004_306_1002}. Martin-Cano {\it et al.} made use of finite element modelling to show that energy transfer between emitters could also be achieved using plasmon waveguides, such as channel and wedge plasmon modes\cite{Martin-Cano_NL_2010_10_3129}.  To-date there appear to be no reports of how resonant energy transfer is modified when both donor and acceptor are strongly coupled to an SPP mode. It will be interesting to see how this topic develops, the role of the local optical density of states in resonant energy transfer is still a controversial one \cite{Andrew_Science_2000_290_785,Blum_2012_PRL_109_203601}.

It is useful at this stage to remind ourselves about strong coupling in microscopic systems, e.g. atoms in cavities. It was in such systems that strong coupling was first observed and it still provides the foundation against which to compare other systems, such as the quantum emitter + SPP mode considered here.

\subsection{Observations of strong coupling in microscopic systems}

We do not concern ourselves here with macroscopic mechanical systems showing strong coupling but focus instead on the strong coupling phenomena in the microscopic world. There, the two systems that are coupled can be described quantum mechanically, although often a purely classical or semiclassical description is sufficient for a quantitative account of experimental observations.

Concerning light and matter, the coupling between them is usually rather weak. This is fortunate when we consider applications such as spectroscopy: if the aim is to measure the optical transitions of the material, one does not want the coupling with the probe light to modify them! However, to utilize coherence and other strong coupling benefits, it has been a long-term goal to reach the strong coupling regime of light-matter interactions. This was achieved at first at microwave frequencies for a single Rydberg atom \cite{Rempe1987} and for a few atoms \cite{Brune1987} in a superconducting cavity in 1987, following earlier many-atom strong coupling \cite{Kaluzny1983} and one-atom maser \cite{Meschede1985} studies. 
Strong coupling was realized in the optical regime first with many atoms \cite{Raizen1989,Zhu1990,Rempe1991}, and finally in 1992 with a single atom \cite{Thompson1992}
inside an optical high-finesse cavity. Later laser cooling and other techniques allowed the creation of ultracold gases, even forming Bose Einstein condensates \cite{Anderson1995,Davis1995,Andrews1997} and there one can achieve coherent behaviour of a macroscopic cloud of atoms described by a single wavefunction. Atoms have been coupled strongly with light in various shapes of cavities
such as waveguide structures and whispering gallery mode microresonators, leading to e.g.\ strong coupling with non-transversal fields as in \cite{Junge2013}, see also references therein.

These early reports on strong coupling involved atoms in vacuo. In the 1990's, Rabi splittings of the order of 1-10 meV were observed for light interacting with emitters in solid-state systems, specifically inorganic semiconductors where integrated microcavities were employed at cryogenic temperatures \cite{Weisbuch1992,Houdre1994,Norris1994,Houdre1995} (strong coupling features, even at room temperature were observed in \cite{Houdre19942}) for a review see e.g.\ \cite{Skolnick1998}. There it was referred as {\it multi-atom vacuum Rabi splitting}, since the semiconductor material, not single atoms or molecules, was interacting with the light field. The difference between the single-emitter and multi-emitter case will be discussed in sections \ref{classical} and \ref{quantum}. Solid state single-emitter strong coupling was later achieved as well: in 2004, a single quantum dot interacting with a photonic crystal cavity mode was shown to display a vacuum Rabi split \cite{Reithmaier_Nature_2004_432_197,Yoshle_Nature_2004_432_200}, for a review see \cite{Khitrova2006}. In later experiments, the quantum nature of the single quantum dot / microcavity strong-coupling was demonstrated to the extent that these systems would be feasible for quantum information processing \cite{Hennessy2007}. 

In 1998 reports concerning organic semiconducting materials for realizing the strong coupling regime emerged: large Rabi splits of the order of 100 meV were reported and, importantly, using organic materials it was possible to obtain this strong coupling at {\it room temperature} \cite{Lidzey1998,Lidzey2000}. In 2002 it was shown that a system that included emitters based on organic semiconductors placed within a metal microcavity may lead to a 300 meV Rabi split at room temperature \cite{Hobson2002}. This splitting was greater than that achieved for cavities based on dielectric mirrors owing to the larger optical fields arising from the tighter confinement of the field in all-metal microcavities. Strong coupling in microcavities was observed also for small molecules \cite{Holmes2004} (for a review see \cite{Holmes2007}). These developments suggested there were advantages in using organic materials in the search for robust, room temperature strong coupling phenomena \cite{Lidzey_Nature_1999_395_53,Lidzey_PRL_1999_82_3316,Lidzey_Science_2000_288_1620}. This work paved the way for the development of strong coupling involving surface plasmon polaritons, the topic of this review article.

We note here that strong coupling has also been explored in a number of systems that do not involve emitters. One area that is very topical is that of nanomechanical systems. Strong coupling of an optical cavity to a mechanical resonator was reported by Gr\"oblacher {\it et al.} \cite{Groblacher_Nature_2009_460_724} in 2009, and microwave strong coupling with nanomechanical systems was observed in 2010 \cite{OConnell2010}. Recently microwave amplification \cite{Massel2011}, hybrid circuit-cavity-quantum-electrodynamics/micromechanical-resonator systems \cite{Pirkkalainen2013}, and a room temperature optoelectromechanical transducer \cite{Bagci2014} have been realized in such systems. Another interesting area is that of strong coupling using superconducting components. Attaining the strong coupling regime is made easier there because dissipation is low in superconducting systems. It is possible in well designed circuit elements made of superconducting components, such as coplanar waveguide resonators and LC oscillators, that the coupling, realized either inductively or capacitively, is larger than the decay rate of the oscillators.  In this strong coupling regime, it is possible to observe the coherent transfer of quanta between the two systems.  The first demonstrations were reported in \cite{Chiorescu2004,Wallraff}.  In the first case, a flux qubit (a superconducting loop interrupted by two Josephson junctions) was coupled to an oscillator (a SQUID structure). In the second case, a microwave resonator (a cavity) was fabricated as a segment of a coplanar waveguide and a charge qubit (a Cooper-pair box) was embedded in the  gap between the signal line and the ground. For this design, the strong coupling regime was reached due to the combination of the large electrical dipole of the Cooper pair box and the large electrical field strength of the quasi-one dimensional cavity. Since these experiments, the strong coupling regime and vacuum Rabi oscillations have been routinely reproduced in various labs with all the combinations of qubits and oscillators. These systems offer one way to create quantum gates that are needed for quantum processing tasks. Furthermore, the availability of strong coupling and the possibility of designing it by microwave engineering allows the realization of complex quantum circuits that could be used in principle as simulators for quantum many-body and quantum field theory systems, for a review see e.g.\ \cite{Paraoanu2014}.  

The degree of freedom that couples strongly with the electromagnetic field may also be the {\it electron spin}. Recently, nitrogen vacancies (NV centers) in diamond have emerged as very promising spin systems where the relevant degree of freedom is the electron spin of the NV center. Electron spin may be coupled to a microwave field. Strong coupling of a spin ensemble to a superconducting resonator has been observed since 2010 \cite{Kubo2010,Amsuss2011}. It has been achieved also for ensembles of electron spins in other solid state materials such as ruby \cite{Schuster2010} and rare-earth materials \cite{Probst2013}. Although it goes beyond our focus on field-matter strong coupling, it is worth mentioning in this context that coherent coupling of a superconducting flux qubit to an electron spin ensemble in diamond was reported in 2011 \cite{Zhu2011}. Reaching the single spin strong coupling regime in all these systems is challenging.

Strong coupling is not an easy regime to achieve. For instance with trapped ions, which are an extremely promising system for quantum information processing \cite{Cirac1995,Monroe1995,Schmidt-Kaler2003,Leibfried2003}, thus far only many-ion strong coupling with light has been realized in experiments \cite{Herskind2009}; it is difficult to make a cavity around a trapped ion that is small enough to enable single-ion strong coupling to be achieved. The quantum information processing applications in these systems do not require light-matter strong coupling, however, strong coupling would certainly bring a new and interesting degree of freedom to the system. The challenge is to create small enough optical cavities around the trapped ions. A step towards this has been taken in \cite{Steiner2013} where the ion trap was built inside an optical fiber resonator with a small mode volume. 
 
In the following section we look at what happens when one of the oscillators is a plasmon mode. More specifically, we consider the case of emitters coupled to a plasmon mode (rather than, for example, a cavity mode).

\subsection{Strong coupling between surface plasmon polaritons and emitters} 
\label{sppstrongcouplingbriefly}

Recently, strong coupling of {\it surface plasmon polaritons} with emitters has become an active topic of research. Surface plasmon polaritons (SPP) are hybrid modes involving electron oscillations in a metal in conjunction with an oscillating light field on the metal surface (more details are given in section \ref{basics}). The energies can be in the optical domain. Due to their near-field character, the light field component of a SPP may be confined to dimensions smaller than the free-space wavelength of light at the same frequency. This allows one to enter the world of nano-optics where light can be confined to dimensions similar to other nano-objects \cite{Gramotnev_NatPhot_2010_4_83,Gramotnev_NatPhot_2014_8_13}. This field confinement also leads to a field enhancement, something that is used to great effect in, for instance surface-enhanced Raman spectroscopy (SERS) \cite{LRandE}. The sensitivity of SPP resonances to the  refractive index of the medium adjacent to the metal has enabled commercial biosensing applications \cite{Rich_JMolRec_2011_24_892}. Attaining the strong coupling regime using SPPs was reported for various types of emitter: J-aggregates \cite{Bellessa2004,Dintinger2005,Dintinger2006,Sugawara2006,Wurtz2007,Chovan2008,Fofang2008,Bellessa2009}, dye molecules \cite{Pockrand1978,Hakala2009,Vakevainen2013} and quantum dots \cite{Gomez2010a,Gomez2010b}. A more comprehensive list of references is given in section \ref{experiments}. Strong coupling of SPP modes with systems other than emitters, namely photonic modes, has also been realized \cite{Ameling2010}. 

Let us now compare the strong coupling in SPP systems with the other systems described above. The strength of the coupling between matter and light can be increased in two ways: by increasing the dipole moment (oscillator strength) of the matter part, or by confining the light into smaller volume. To reach the strong coupling regime, the coupling has to be large compared to the energy widths of the individual resonances, that is, the linewidth of the optical transition and/or the linewidth of the optical field. Thus in addition to, or instead of, increasing the coupling, one may aim to decrease the linewidths. In the atom + optical cavity systems, nothing could be done about the dipole moment of the atom which is given by nature, but the cavity helped to decrease the mode volume and the light linewidth. Due to the small value of the dipole moment of single atoms, very high finesse cavities are required to realize strong coupling, something that is technically demanding. In the strong coupling studies where semiconductor microcavities are employed, many dipoles contributed to the effect and consequently the effective dipole moment was larger. Naturally, the microcavities help by decreasing the mode volume. Yet the coupling in inorganic semiconductor materials is so weak that one has to cool the samples to liquid nitrogen temperatures to decrease the line widths so that Rabi splittings become visible. For inorganic semiconductor materials, room temperature strong coupling is achievable, but microcavities are still needed.

In the case of SPPs, the mode volumes are extremely small since light is confined in the nanoscale rather than micronscale \cite{Gramotnev_NatPhot_2010_4_83}. The SPP field intensity is enhanced due to the resonant character of the SPP excitation (see section \ref{basics}). Moreover, relatively large effective dipole moments can be achieved by using high concentrations of the optically active materials (molecules, quantum dots). Indeed, Rabi splits of 450 meV were observed in 2009 \cite{Bellessa2009}, and in 2011 even 650 meV which is already in the ultrastrong coupling regime \cite{Schwartz2011}. For these reasons it is possible to observe the vacuum Rabi splitting for SPP and emitters {\it at room temperature} AND {\it without the need for a closed cavity}. This is a major technical advantage that may be very important for potential applications. Another important difference is that the SPP + emitter system allows one to probe strong coupling phenomena in the nano-world in the sense that both the light and the matter part of the strongly coupled hybrid can be confined to the nanoscale. Due to the dissipative nature of SPP modes, the lifetimes of the hybrid modes created by strong coupling are much shorter than is the case for optical or microcavity systems. In other words fewer Rabi oscillations will take place before coherence is lost. The coherence times of SPP are typically of the order of $10-100$ femtoseconds \cite{van_Exter_PRL_1988_60_49,Sonnichsen_2002_PRL_88_077402}. An interesting possibility to increase the coherence times, that is, the Q-values of plasmonic modes is to use collective resonances (so-called surface lattice resonances) possible in metal nanoparticle arrays \cite{Zou2004,Auguie2008,GarciadeAbajo2007}. 

\subsubsection{Status of the research}

The first reports of strong coupling for SPP systems concerned observations of splittings in the energy spectrum, and used organic molecules called J-aggregates for the emitters \cite{Bellessa2004,Dintinger2005,Dintinger2006,Sugawara2006,Wurtz2007,Chovan2008,Fofang2008,Bellessa2009}. The special feature of J-aggregates is that they have a relatively narrow absorption line. Interestingly, strong coupling is possible also for molecules with a broader absorption line, such as Rhodamine 6G \cite{Hakala2009,Cade2009,Rodriguez2013,Vakevainen2013,Shi2014}. Strong coupling has been demonstrated for other dye molecules \cite{Valmorra2011,Baieva2012} as well as for photochromic molecules \cite{Schwartz2011}. Strong coupling between quantum dots and SPP was achieved in \cite{Gomez2010a}. The dynamics of the phenomenon have also been explored \cite{Hakala2009,Vasa_ACSNano_2010_4_7559,Vasa2013}. A more complete list of references on these topics will be given in section \ref{experiments}. Recently, the coherence over large distances that is typical for strong coupling was demonstrated \cite{AberraGuebrou2012} in the strong coupling regime. The transition from weak-to-strong coupling was studied in \cite{Shi2014}, showing how the wave-function (mode-function), and the corresponding spatial coherence properties, of the strongly coupled hybrid evolves when going from weak to strong coupling. It has become of interest to explore the quantum origin of the strong coupling phenomenon in plasmonics, and indeed the theoretical quantum descriptions of strong coupling between SPPs and many emitters have been developed \cite{Gonzalez-Tudela2013}. The existing experiments, at the level of the present experimental accuracy, are qualitatively consistent with classical, semiclassical and fully quantum descriptions. While it is natural to use quantum theory as the most accurate microscopic description of the physics of atoms, molecules and photons, it is also of interest to try to identify phenomena where the classical theory completely fails. One of the goals of this review is to present the classical, semiclassical and quantum theories of strong coupling in a compact way that allows easy comparision of the similarities and differences between these descriptions: the details will be discussed in sections \ref{classical} and \ref{quantum}, and a summary is made in the concluding remarks in section \ref{conclusions}.

In this review article, we will present the theoretical background necessary for understanding and quantitatively describing the SPP strong coupling phenomena, we will review the main experimental and theoretical developments in the field, and outline future directions and challenges. The article is organized as follows. Section \ref{basics} presents the basis of surface plasmon polariton physics. In sections \ref{classical} and \ref{quantum} the classical and quantum descriptions of SPP strong coupling phenomena are presented, respectively. Section \ref{quantum} covers both the semiclassical description, and the one where the SPP field is also quantized. The experiments on SPP strong coupling are reviewed in section \ref{experiments}. Finally, we outline our views for the future of this fascinating research area in section \ref{conclusions}.

In the following section we provide an overview of surface plasmon polaritons with a focus on those aspects that are important in the context of strong coupling. A number of excellent more extensive introductions to surface plasmon polaritons are available in the literature 
\cite{Zayats_PhysRep_2005_408_131,LRandE,Pitarke_RepProgPhys_2007_70_1,Maier}.

\section{Surface Plasmon Polaritons --- overview} \label{basics}

In much the same way that light can be guided by an optical fibre, it may also be guided by the interface between a metal and a dielectric. More specifically, under appropriate conditions, light may interact with the free electrons (the plasma) in the surface of a metal to yield a combined electron/light oscillation mode known as a surface plasmon-polariton. Surface plasmon polaritons are one of a wider class of surface modes where the interaction between light and matter leads to the possibility of a bound surface mode, other possibilities include surface phonon polaritons 
\cite{Huber_APL_2005_87_081103} and surface exciton polaritons \cite{Gentile_NL_2014}. In addition to these propagating surface plasmon polaritons, there is also an interesting class of surface plasmon polaritons that are spatially confined. This confinement is a result of being associated with metallic nanostructures such as nano-spheres, nano-discs etc., such modes are called localised surface plasmon polaritons; both propagating and localised modes will be of interest to us here. Before looking at these modes in detail, let us identify the attributes that SPPs possess, attributes that make them interesting in the strong-coupling context.

\begin{figure}
\includegraphics[width=1.0\textwidth]{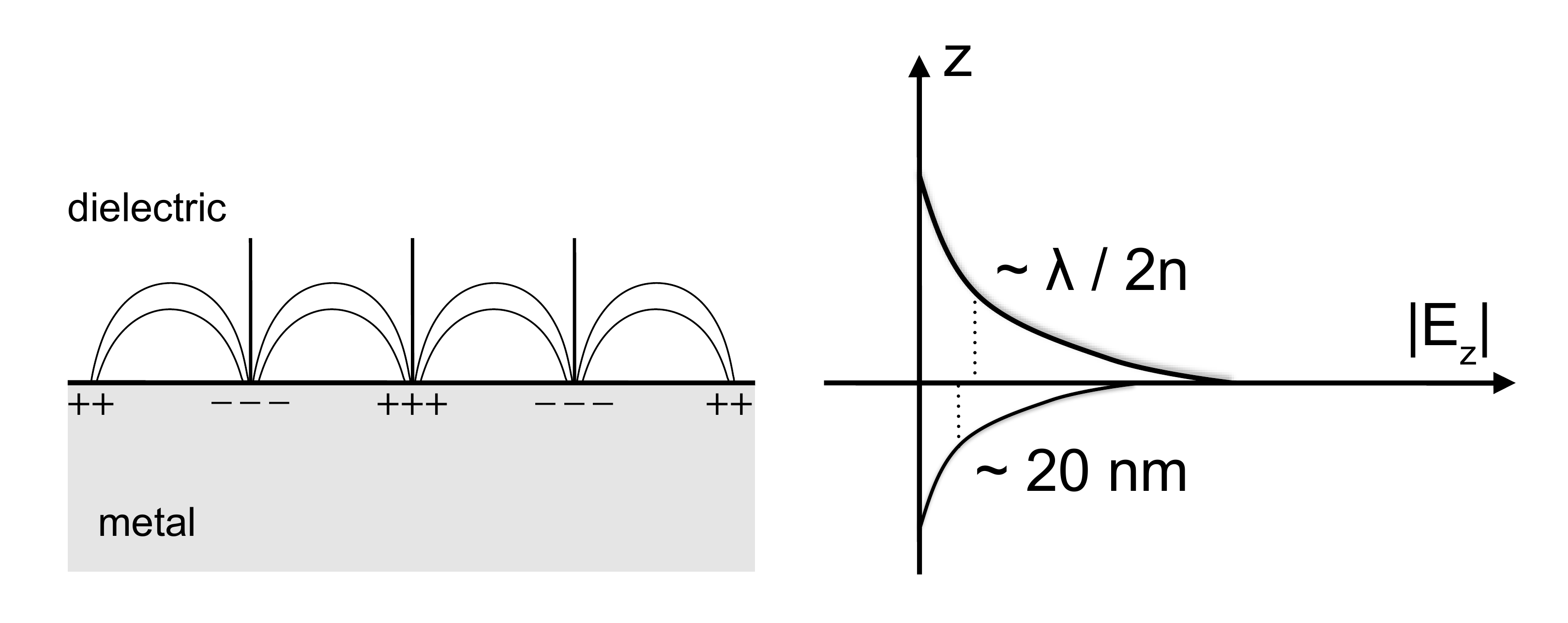}
\caption{Schematic of electric field and charge ditribution associated with the surface plasmon polartion mode on a planar surface (left). The strength of the field associated with the SPP mode decays exponentially with distance away from the surface. In the metal this is the skin depth (right).} 
\label{fig_SPP1}
\end{figure}

There are three important attributes of SPPs that we consider here: optical field confinement, optical field enhancement, and near-field character; all three are related to the fact that SPP modes are bound to interfaces. The spatial distribution of the electric field (it is primarily the electric field rather than the magnetic field that is of interest in strong coupling phenomena) is shown in figure \ref{fig_SPP1}. The strength of the field decays exponentially into the surrounding media. The decay length into the dielectric is $\sim\lambda/2n$ where $n$ is the refractive index of the dielectric. The decay length into the metal is $\sim 20\ nm$ (for most metals of interest in the visible part of the spectrum) \cite{Barnes_JOptA_2006_8_S87}. The confinement of the optical field to the proximity of the surface is evident from the evanescent nature of these fields. Associated with this confinement is an enhancement in the strength of the electric field adjacent to the surface \cite{LRandE}. Figure \ref{fig_SPP1} also shows the charge density and associated electric field distribution. The near-field character is important for both propagating and localised SPPs since, as we will see, it facilitates coupling between quantum emitters and SPP modes. Let us now look at these modes in more detail.

\subsection{Propagating Surface Plasmon Polaritons} \label{propagationSPP}

\begin{figure}
\includegraphics[width=0.8\textwidth]{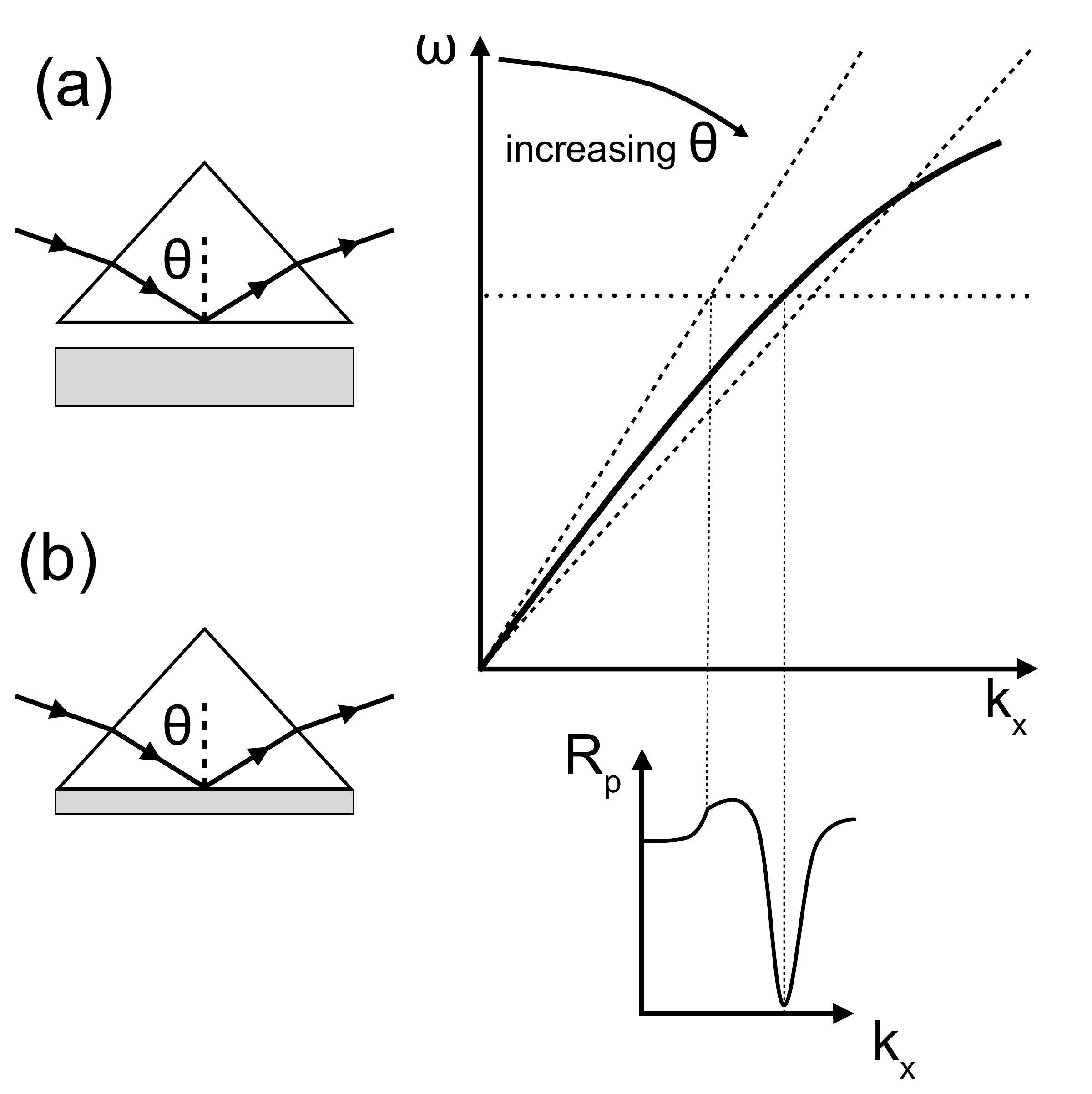}
\caption{(c) Surface plasmon-polariton dispersion curve, i.e. frequency of the SPP mode as a function of in-plane wavevetor, i.e. wavevector along the surface (here chosen as the x-direction). The light line ($\omega=ck_0$) represents light that propagates along the surface. Notice that the SPP mode is always to the high in-plane wavector side of the light line. Coupling can be achieved between an incident plane wave and the SPP mode through prism coupling, either in the Otto geometry (a) or the Kretschmann-Raether geometry (b). 
By measuring the reflectivity of p-polarised light evidence for coupling to the SPP mode can be seen as a dip in the reflectance as the in-plane wavevector (angle of incidence) is scanned.} 
\label{fig_SPP2}
\end{figure}

We are concerned here with surface plasmon polaritons that propagate along the interface between a metal and a dielectric. The field distribution associated with these modes, together with their dispersion relation, are calculated by looking for solutions to the Maxwell equations in a source-free region of space comprising two semi-infinite media, the metal and the dielectric. Some of the fields are shown schematically in figure \ref{fig_SPP1}. The dispersion relation, i.e. how the in-plane wavevector of the SPP mode varies with frequency of the mode, is obtained by solving the Maxwell equations under appropriate boundary conditions and looking for a solution that takes the form of a surface wave, the result is \cite{Raether},
\\
\begin{equation}
k_{SPP} = \frac{\omega}{c}\sqrt{\frac{\epsilon_1\epsilon_2}{\epsilon_1+\epsilon_2}}, \label{k_SPP}
\end{equation}

\noindent where $\epsilon_1$ and $\epsilon_2$ are the frequency dependent relative permittivity of the two media. If we take the simplest model for the permittivity of a metallic plasma, 
i.e. the Drude model,
\\
\begin{equation}
\epsilon=1-\omega_P^2/\omega^2,
\label{Drude}
\end{equation}

\noindent then the dispersion relation above, equation (\ref{k_SPP}) takes the form shown in figure \ref{fig_SPP2}c below. A key feature to note from figure \ref{fig_SPP2}c is that the in-plane wavevector (and hence momentum of the SPP mode) is always greater than that of light propagating in the same plane. A direct consequence is that freely propagating light in the dielectric can not couple to the SPP mode. For coupling to occur some kind of momentum-matching scheme is required. Several such schemes are available and include: prism coupling, grating coupling, near-field coupling and non-linear coupling \cite{Renger_PRL_2009_103_266802}. We will look at the first three of these schemes below, before doing so though we want first to discuss how the character of the SPP mode varies with the region of the associated part of the dispersion curve, and introduce the way the presence of emitters may alter the dispersion.

\subsubsection{The character of propagating SPP modes and the dispersion diagram}  \label{CharacterOfSPPDispersionPaivi}

Let us first consider in more detail the SPP phenomenon for propagating SPP modes, especially as it relates to the dispersion diagram. The dispersion relation was given above for SPPs, equation (\ref{k_SPP}), and we will make use of the Drude model for the permittivity of the metal, equation (\ref{Drude}), and take the dielectric half space to be vacuum (air). The dispersion relation can then be solved analytically to give (here $k\equiv k_{SPP}$),

\begin{equation}
\omega^2 = c^2 k^2 + \frac{\omega_P^2}{2} \pm \sqrt{ c^4 k^4 + \frac{\omega_P^4}{4}} . \label{SPPdispPaivi}
\end{equation}

\begin{figure}
\includegraphics[width=0.7\textwidth]{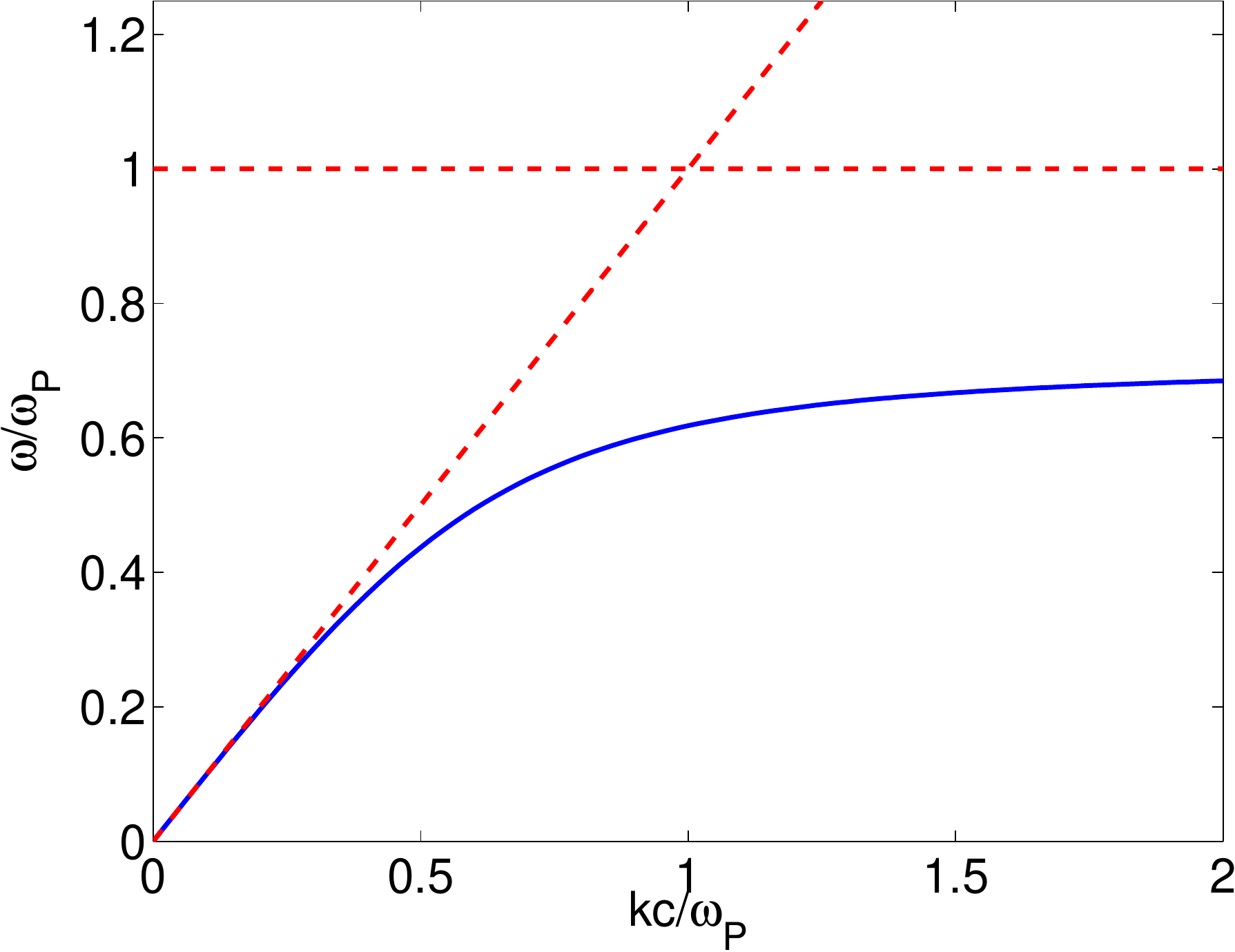}
\caption{The SPP dispersion (blue line) based on the Drude model for the metal and vacuum for the dielectric, equation (\ref{SPPdispPaivi}). The SPP dispersion corresponds to a mode 
at a metal-dielectric interface. It approaches the light line (diagonal dashed line) for small wavevectors. When the dispersion approaches a resonance in the system, here the surface plasma frequency $\omega_P/{\sqrt{2}}$, the dispersion curve bends.} 
\label{SPPdispersionPT}
\end{figure}

The solution has two branches, corresponding to the $\pm$ signs. For small values of the in-plane wavevector, $k$, the lower branch -- the solution obtained by taking the minus sign in equation (\ref{SPPdispPaivi}) -- behaves like light, i.e. the dispersion of the mode lies close to the light-line. For large values of $k$, the frequency $\omega$ for this solution approaches $\omega_P/\sqrt{2}$. 
In this regime the SPP mode is very different from light, see figure \ref{SPPdispersionPT}. The upper branch, corresponding to the + sign in equation (\ref{SPPdispPaivi}) is unphysical; one should not confuse the upper branch of equation (\ref{SPPdispPaivi}) with what is usually called "upper branch" in the context of SPPs, namely the transverse wave propagating in bulk metal
which has a dispersion $\omega^2= c^2 k^2 + \omega_p^2$ (see e.g.\ Chapter 6., especially 6.2 in \cite{Gaponenko2010}).  

In general, strong absorption at a particular frequency is likely to lead to changes in the dispersion curve, avoided crossings and/or energy bands with no purely real solutions. 
These changes in dispersion correspond to changes in the character of the mode, it becomes more clearly hybrid in nature, containing more of the character of the excitation associated with the strong absorption (a resonance). SPPs are hybrid modes of light and collective charge (plasmon) oscillations. In this sense the usual SPP dispersion is related to strong coupling phenomena: the bending of the dispersion is due to a resonance in the system, and indeed the SPP mode is a hybrid of the extreme ends ($k=0$ and $k= \infty$) of the dispersion: light and plasma oscillations. Let us return now to the three means that we wanted to discuss by which light and SPP modes may be coupled. 

\subsubsection{Coupling light and propagating SPP modes}

First, prism coupling makes use of attenuated total reflection, a phenomenon well known in optics. Turbadur \cite{Turbadar_ProcPhysSoc_1959_73_40} appears to have been the first to employ prism coupling to excite surface plasmon polaritons (although it seems he did not know that surface plasmon polaritons were responsible for the phenomenon he observed), but the technique became wide-spread following the work of Otto \cite{Otto_ZPhys_1968_216_398} and of Kretschmann and Raether \cite{Kretschmann_ZNatur_1968_23_2135} in 1968. Both the Otto and the Kretschmann and Raether schemes are shown schematically in figure \ref{fig_SPP2}. In the Otto configuration, figure \ref{fig_SPP2}a, light is incident (from within) on the base of a glass prism. In glass of refractive index $n$ light has a wavevector (momentum) that is enhanced over its free-space value ($k_0$) to $nk_0$. When the angle of incidence on the base of the prism is greater than the critical angle, total internal reflection occurs. The optical field does not fall immediately to zero at the interface, rather an evanescent field is produced that decays in strength exponentially with distance from the prism base. If now a metal surface is brought up to within a wavelength or so of the prism base then the evanescent field that extends beyond the base of the prism (when total internal reflection occurs) may couple to the surface plasmon mode. By adjusting the angle of incidence the in-plane wavevector of the evanescent field may be adjusted to match that of the SPP mode (figure \ref{fig_SPP2}c). Light coupled into the SPP mode can couple back out by the same process, but the phase of this re-radiated light is out of phase with the specular reflection \cite{Herminghaus_OL_1994_19_293}. Power is instead eventually lost to heat in the metal film and, if the coupling is adjusted properly -- through a careful choice of the size of the air gap between the prism and the metal surface -- a sharp dip in the reflectivity is observed figure \ref{fig_SPP2}d. In the Kretschmann-Raether scheme, figure \ref{fig_SPP2}b, a metal film is deposited directly onto the base of the prism and the evanescent field that is generated upon total internal reflection extends through the metal film to couple to the SPP mode on the metal surface away from the prism. In practice this is generally a more convenient approach than the Otto configuration, a metal of the required thickness (of order the skin depth in the metal) is easily made using vacuum deposition techniques and avoids the need for making a wavelength-scale air-gap.

\begin{figure}
\includegraphics[width=1.0\textwidth]{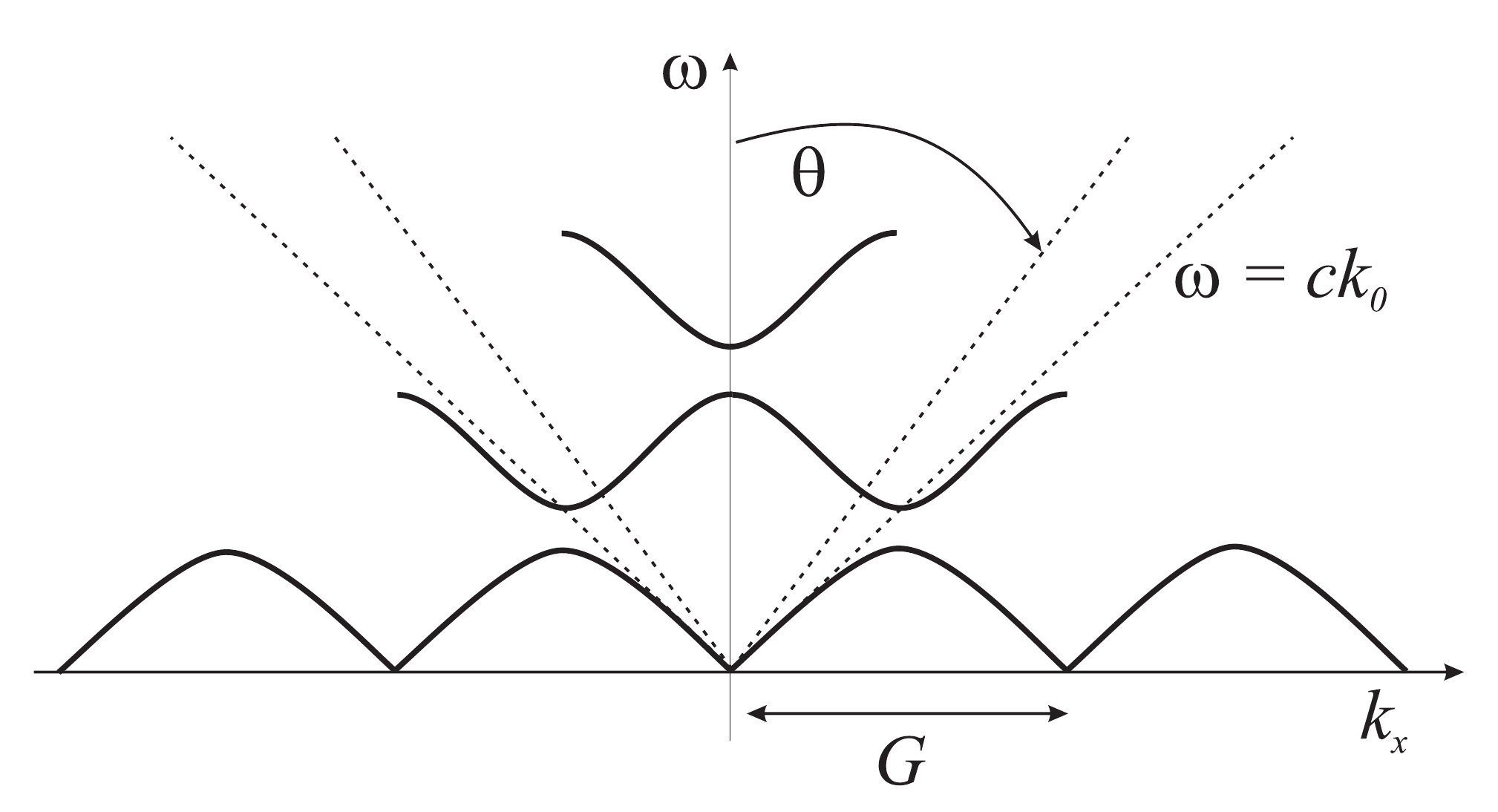}
\caption{Surface plasmon-polariton dispersion curve for SPP propagation on the surface of a metallic diffraction grating. The SPP dispersion curve is replicated at $\pm G$, where $G=2\pi/a$, $a$ being the grating period.} 
\label{fig_SPP3}
\end{figure}

Second, in grating coupling the surface of the metal film is modified to take the form of a diffraction grating. For a suitable period of grating, diffraction can produce an evanescent diffracted order that is able to momentum-match to the SPP mode. As with prism coupling, when momentum matching happens the reflectivity falls, enabling the effect of coupling to the SPP mode to be monitored. Wood unwittingly observed such reflectivity dips more than a century ago \cite{Wood_PhilMag_1902_4_396}. The effect of a grating on the SPP dispersion curve is shown in figure \ref{fig_SPP3}. There are two things to notice. First, the SPP dispersion curve has now been replicated inside the light line, incident light can couple via diffraction to the SPP mode. Second, a gap opens up where the different scattered SPP modes cross, i.e. there is an anti-crossing. The gap is strong where counter propagating SPP modes can be linked by a first-order ($\pm G$) scattering process, region (a) in the figure; weaker where a second-order scattering process is required ($\pm 2G$) (b), where $G=2\pi/a$, $a$ being the grating period. One could view the anti-crossing behaviour as a form of strong coupling, 
as we alluded to above, section \ref{CharacterOfSPPDispersionPaivi}. The profile and amplitude of the grating determines the strength of the scattering/coupling \cite{Barnes_PRB_1996_54_6227}. Thus, in addition to enabling coupling of SPP modes and propagating light, periodic structures may also act to introduce band gaps in the dispersion (propagation) of SPPs \cite{Kitson_PRB_1995_52_11441}. In the context of strong coupling this is an interesting phenomenon since at the band edge the local density of optical states is high, enhancing light-matter interactions \cite{Turnbull_PRB_2001_64_125122}. 

The metallic hole array is an extreme form of grating and is a key ingredient in accounting for the extraordinary transmission shown by some metallic hole arrays \cite{Ebbesen_Nature_1998_391_667}. The diffractive nature of metallic hole arrays plays a key role in coupling light to the surface plasmon polaritons supported by such structures. Incident light couples to the SPP on the input side of the structure via grating coupling. As the incident light is scattered by the periodic structure it gains/looses momentum in the plane of the surface associated with the grating, i.e. it gains/looses a wavevector $G$. If the scattered light has an in-plane wavevector that matches $k_{SPP}$ (at some wavelength) then incident light may couple to the SPP. The evanescent fields associated with the SPP span across the metal film, and can be scattered by the periodic structure on the output side, thereby enabling the SPP to be scattered 
into transmitted light \cite{Garcia-Vidal_RMP_2010_82_729}.

The complementary structure of the hole array is an array of particles, for example a periodic array of metallic nanodiscs. When the distance between particles is of the same order as the wavelength associated with the localised plasmon resonance, coherent effects are possible, notably lattice surface resonances \cite{Zou_JChemPhys_2004_121_12606} \cite{Auguie_PRL_2008_101_143902}. Such coherent effects are also interesting, offering rich opportunities for controlling light matter interactions.

Third, near-field coupling to SPPs is particularly important in the context of strong coupling between quantum emitters and optical modes. Consider an excited dye molecule placed close to a metal surface. In the near-field of such an emitter the field-distribution has contributions that cover a wide range of wave-vectors, some of which will match that of the SPP mode. The near-field thus provides a pathway by which the excited molecule may couple to the SPP mode \cite{Andrew_PRB_2001_64_125405}, and this coupling can be very efficient, depending on the distance between the emitter and the surface \cite{Barnes_JMO_1998_45_661}. One might have assumed that closer is better, i.e. getting an emitter as close as possible to a metallic surface will maximise the effectiveness of coupling to an SPP mode. However, this is not the case. If the emitter is too close then its energy will predominantly be lost more directly as heat in the metal \cite{Barnes_JMO_1998_45_661}. The optimum distance for coupling to SPPs is $\sim 10 - 20\ $ nm for a planar surface \cite {Pockrand_CPL_1980_69_499}, and more generally depends on the surface morphology \cite{Johansson_PRB_2005_72_035427}. The near-field provides the means that allows plasmon modes and emitters to couple, and the strength of the coupling can be sufficient to allow strong-coupling to occur.

\subsection{Localised Surface Plasmon Polaritons}

Plasmon modes can also be sustained by metallic nanostructures, a useful example is that of a nano-sphere. Light incident on a metallic nano-sphere will act to drive the mobile conduction electrons into oscillation. When electrons are displaced relative to the positive charge of the static cores, opposite sides of the sphere will take on opposite charges. 
The Coulomb interaction between these opposite charges provides a restoring force that, in common with all restoring forces, leads to a natural frequency of oscillation. When the incident light is of the same frequency, energy is coupled into the plasmon mode. Unlike the propagating SPP associated with the planar surfaces discussed above, there is no momentum mismatch to overcome in this situation since the sphere breaks the translational invariance associated with the planar surface. As with the planar surface though, the fields associated with the plasmon mode are confined to the vicinity of the nano-sphere, typically on a length scale comparable to the radius of curvature of the sphere, making them very useful in confining light to sub-wavelength volumes. Gold nano-spheres formed the basis of the gold colloids studied by Faraday in the nineteenth century \cite{Faraday_PhilTrans_1857_147_145}, the colours he observed arose from the plasmon modes supported by  gold particles. Many other structures can support localised surface plasmon modes including rods, discs, holes and voids. These structures may support higher-order plasmon modes as well, especially as their size increases.

We have focused here on providing some basic background concerning surface plasmon polaritons, and have concentrated on SPPs associated with planar surfaces and small metallic nanoparticles. Before we move on to look at the combination of SPPs and strong coupling we should perhaps note that the field of SPPs is still a fast moving and diverse field. SPPs are being used to guide energy, for example using metallic nanowires \cite{Charbonneau_OL_2000_25_844}, to enhance the absorption of light \cite{Polyakov_PRL_2013_110_076802}, and to improve the efficiency of light-emitting diodes 
\cite{Smith_AdvFuncMat_2005_15_1839}.

Armed with this rudimentary knowledge of surface plasmon polaritons we are now better placed to look at the strong coupling between SPPs and matter.

\section{Classical description of the strong coupling between SPPs and matter} \label{classical}

\begin{figure}
\includegraphics[width=0.7\textwidth]{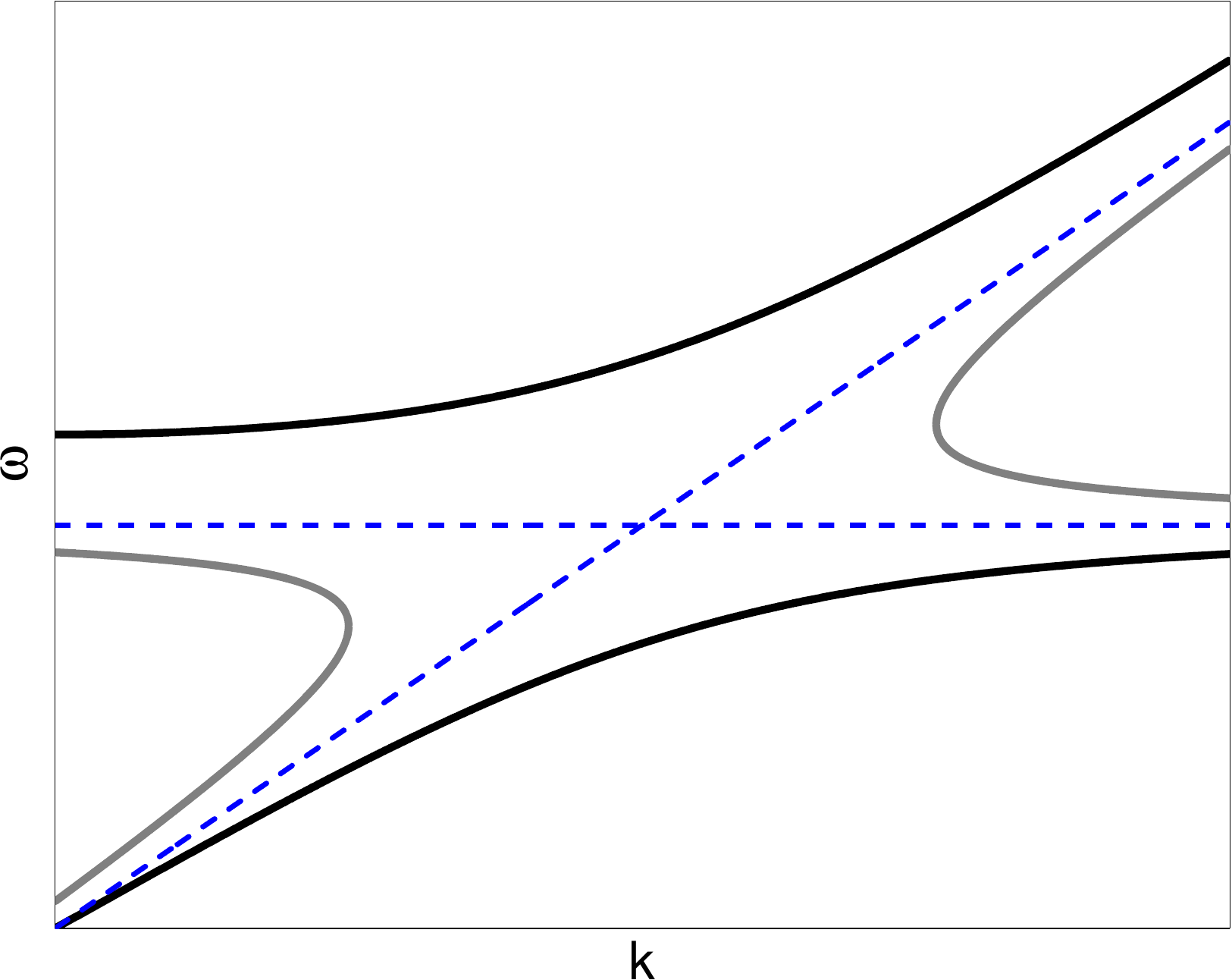}
\caption{The material wherein the propagating light mode resides absorbs at frequency $\omega_0$ (horizontal dashed line). Since light cannot propagate if it gets totally absorbed, the dispersion $d\omega/dk$ that gives the group velocity of light has to approach zero, i.e.\  a horizontal line in this graph. Far from $\omega_0$, the dispersion follows the light line (diagonal dashed line) of the material. The type of bendings for which $d\omega/dk$ always stays below the speed of light are marked with black lines: these are the physically possible solutions. The grey lines correspond to unphysical cases where the group velocity would become larger than speed of light. From these general considerations, we can anticipate what kind of behaviour the existence of emitters, with a clear maximum absorption frequency in the vicinity of the SPP modes, will lead to.} 
\label{basicBendingPT}
\end{figure}

In this section, we describe the strong coupling between SPPs and emitters when the dielectric in the vicinity of the metal (c.f. section \ref{basics}) contains emitters with a well defined absorption/emission spectrum. Whenever the frequency of light (SPPs) is close to an absorption frequency ($\omega_0$) the absorption will hinder the propagation of SPPs. Associated with this absorption there will be a slowing of the SPP (we assume we are on the low-frequency side of the absorption frequency) and corresponding decrease of the group velocity $d\omega/dk$ towards zero, causing bending of the dispersion. Approaching $\omega_0$ from above, the dispersion has to bend as well. However, in both cases the bending has to be such that, at all stages, $d\omega/dk$ 
is less than or equal to the speed of light: other solutions are unphysical. Figure \ref{basicBendingPT} illustrates this.

Next we consider the case when the absorption is due to emitters which are described as classical Lorentzian oscillators in the medium. This simple description is adequate to describe the basic physics of SPPs interacting with, e.g. molecules, atoms, quantum dots etc.\ located near the metal surface. A more microscopic, quantum mechanical description will be given later in section \ref{quantum}.

\subsection{Strong coupling of SPPs and Lorentzian emitters}  \label{SPP+Lorentzian}

Let us consider the situation where we have emitters on top of the metal. As a practical example we could consider a polymer film containing molecules with a suitable optical absorption/emission spectrum spin-coated on top of the metal. One could also think about individual emitters like quantum dots positioned or dispersed in the vicinity of the metal where the SPPs reside. We now consider this situation with a simple description which only takes into account classical electrodynamics. This description assumes that the molecule or other emitter can be described as a classical Lorentzian oscillator, i.e.\ that it can be described by the dynamics of an electron (upon which the polarizability of the molecule is based).

Let us consider an electron (of the molecule/emitter) of charge $e$ and mass $m$ as a harmonically bound, damped oscillator that is driven by an electromagnetic (EM) field $E(r,t)$: the problem of movement of a charge in an EM field. The equation of motion is

\begin{equation}
m (\ddot{r} + \gamma \dot{r} + \omega_0^2 r) = - e E(r,t)  .
\end{equation}

\noindent Here $\omega_0$ is the frequency of the harmonic oscillator and $\gamma$ describes damping. We consider for simplicity a one-dimensional system. Within the usual dipole approximation, let us assume that the EM field is constant in $r$ since the electron movement is small compared to the wavelength of the EM field. Furthermore, we assume the EM field is harmonic, i.e.\ $E(r,t) = E_0 e^{-i \omega t}$. Then the steady-state solution becomes (this is easy to verify by taking the time-derivatives),

\begin{equation}
r = -\frac{e}{m} \frac{1}{ \omega_0^2 - \omega^2 - i\gamma \omega} E_0 e^{-i \omega t} .
\end{equation}

\noindent The dipole moment of the electron motion is given by the product of its charge and position, thus we have,

\begin{equation}
p = - e r = \frac{e^2}{m} \frac{1}{ \omega_0^2 - \omega^2 - i\gamma \omega} E.
\end{equation} 

For a medium comprising many ($N$) dipole moments the macroscopic polarization density (or simply polarization) $P$ is defined as the average dipole moment per unit volume ($V$) and now
becomes,

\begin{equation}
P = \frac{N e^2}{Vm} \frac{1}{ \omega_0^2 - \omega^2 - i\gamma \omega} E.
\label{Pwlb}
\end{equation} 

\noindent Here $N/V$ is the number density of dipole moments. Now, the macroscopic polarization is defined as,
 
\begin{equation}
P = \epsilon_0 \chi E,
\end{equation}

\noindent where $\chi$ is the susceptibility. Thus the macroscopic electric susceptibility is,

\begin{equation}
\chi (\omega) = \frac{N e^2}{V\epsilon_0 m} \frac{1}{ \omega_0^2 - \omega^2 - i\gamma \omega} .  \label{chiPT}
\end{equation}

The imaginary part of the susceptibility describes dissipation and gives the absorption coefficient of the material. In the limit $\omega \gg \gamma$ 
and close to resonance we have,

\begin{equation}
\chi'' (\omega) = Im \chi (\omega) \simeq \frac{N e^2 \gamma}{4 V \epsilon_0 m\omega_0} \frac{1}{ (\omega - \omega_0)^2 + \frac{\gamma^2}{4}}.
\end{equation}

\noindent Similarly, the real part becomes,
 
\begin{equation}
\chi' (\omega) = Re \chi (\omega) \simeq -\frac{N e^2 }{2 V \epsilon_0 m\omega_0} \frac{\omega-\omega_0}{ (\omega - \omega_0)^2 + \frac{\gamma^2}{4}}.
\end{equation}

\noindent The permittivity $\epsilon(\omega)$ is related to the susceptibility through,

\begin{equation}
\epsilon(\omega) = 1 + \chi (\omega) ,  \label{epsilon_PT}
\end{equation}

\noindent and the refractive index consequently will have a real as well as an imaginary part, often denoted as $n$ and $\kappa$, respectively. The relations of these to the real and imaginary parts $\epsilon'$ and $\epsilon''$ of the permittivity are given by,

\begin{eqnarray}
\epsilon' = n^2 - \kappa^2, \label{epsilon_real_PT} \\
\epsilon'' = 2 n \kappa. \label{epsilon_imag_PT}
\end{eqnarray}

Let us now look at the SPP dispersion as given in section \ref{basics}, i.e.

\begin{equation}
k = \frac{\omega}{c} \sqrt{\frac{\epsilon_1 \epsilon_2}{\epsilon_1 + \epsilon_2}}.  \label{SPPdispersionAgainPaivi}
\end{equation}

\noindent We aim first to obtain a set of analytical results to understand the basic phenomenon; later, numerical treatment is considered. Numerically, one could just substitute $\epsilon_1$ and $\epsilon_2$ into Eq.\ref{SPPdispersionAgainPaivi} and obtain the dispersion, but there is quite a lot that one can do analytically, provided reasonable assumptions are made. First, let us consider the case where the metal dielectric function $\epsilon_1$ is assumed constant. This is true far away from the plasma frequency and provided we consider only a relatively small frequency range around the central frequency of the oscillator, $\omega_0$. Second, we use the fact that typically $\epsilon_1$ for metals is negative and rather large in absolute value: even when $\epsilon_2$ has a Lorentzian contribution, i.e. can be reasonably large and positive, we assume that $\epsilon_1 + \epsilon_2$ always stays negative, and furthermore, that the functional dependence of $\epsilon_2$ on frequency has negligible significance for $\epsilon_1 + \epsilon_2$, whereas in the numerator where it appears as $\epsilon_1 \epsilon_2$,  $\epsilon_2$ influences the dispersion much more strongly. Based on these considerations, we may write the dispersion as,
 
\begin{equation}
k^2 = \frac{\omega^2}{c^2} \frac{|\epsilon_1|}{|\epsilon_1 + \epsilon_2|} (1 + \chi(\omega)).
\end{equation}

\noindent We then scale the momentum to $\kappa^2 = k^2 \frac{|\epsilon_1 + \epsilon_2|c^2}{|\epsilon_1|}$ to obtain,

\begin{equation}
\kappa^2 =\omega^2(1 + \chi(\omega)) =\omega^2 \left( 1 + \frac{A}{ \omega_0^2 - \omega^2 - i\gamma \omega} \right) ,
\label{cleanDispersionPT}
\end{equation}

\noindent where we have written $A = \frac{N e^2}{V \epsilon_0 m}$. Note that $4\pi A^2 = \omega_p^2$ where $\omega_p$ is the plasma frequency of the free electron gas. The physics we consider here is different from plasma oscillations. However, this connection arises because both in the present case and in the free electron gas, the parameters entering the model are $e$, $m$, $\epsilon_0$, 
and the density $N/V$, and there is only one frequency that can be constructed from these on dimensional grounds.   

Now consider first the dissipationless case $\gamma = 0$. Then clearly $1 + \frac{A}{ \omega_0^2 - \omega^2}$ must be positive for a real solution to exist. We can use this to define areas of no real solutions just as for the case of the SPP dispersion in section \ref{CharacterOfSPPDispersionPaivi}. The function $1 + \frac{A}{ \omega_0^2 - \omega^2}$ is positive at plus and minus infinity and changes sign (in this section, we always consider positive $\omega$ and do not discuss $-\omega$ explicitly) at two places: first at $\omega=\omega_0$ via infinity, and then at $1 + \frac{A}{ \omega_0^2 - \omega^2}=0$ via zero. The latter equation gives $\omega = 
\sqrt{A + \omega_0^2}$. Between these values the function is negative and no real solutions exist. There is thus something akin to a stop-band whose width is $\Delta \omega = \sqrt{A + \omega_0^2} - \omega_0 \simeq A/(2\omega_0)$. This is not a genuine stop-band since for the realistic $\gamma\neq0$ case there are solutions in this area, however, they are suppressed in amplitude. Note that if $\omega_0\sim0$ (rather unrealistic in optics), then $\Delta \omega \sim \sqrt{A}$. The width of the stop band thus depends both on $\omega_0$ and $A$, extending between $\sqrt{A}$ and $A/(2\omega_0)$. We can thus anticipate that, just as for the SPP-plasma frequency case, the existence of an absorption maximum will produce an area without purely real solutions.

\begin{figure}
\includegraphics[width=0.7\textwidth]{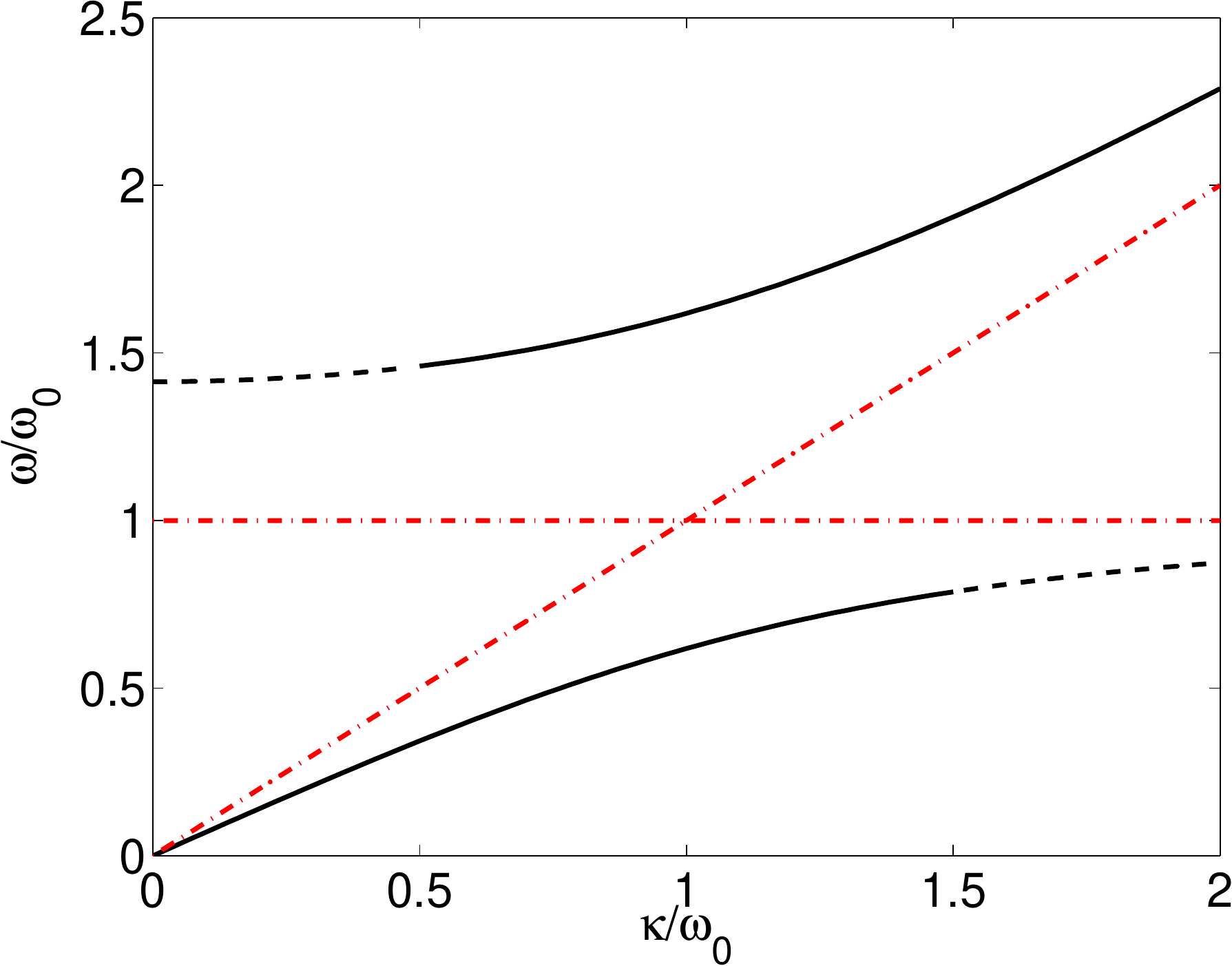}
\caption{The dispersion of an SPP - emitter system as given by equation (\ref{cleanDispersionPT}) is depicted by the black lines. The red lines are the emitter energy and the SPP dispersion, that is, equation (\ref{cleanDispersionPT}) for $A=0$. The black lines turning from solid to dashed reminds that, in case of finite damping, the nearly flat modes cease to be well defined further away from the crossing point. The parameter $A$ was taken as $\omega_0^2$ here.} 
\label{SPPavoidedScetch}
\end{figure}

We see from the dispersion relation, equation (\ref{cleanDispersionPT}), that an infinity in the dielectric function means infinity of $\kappa$, and the zero point means $\kappa = 0$. Far away from $\omega_0$, one should recover the linear dispersion. The dispersion relation, equation (\ref{cleanDispersionPT}), is shown in figure \ref{SPPavoidedScetch} for the case of no damping, $\gamma=0$. Note that the size of the area with no real solutions is not usually used for characterizing the behaviour of the dispersion, since it refers basically to what happens for zero and infinite momenta. Namely, when $k$ approaches infinity or zero in the nearly flat modes around the splitting, these modes cease to be well defined, due to damping, \cite{Agranovich2003}. This is illustrated in figure \ref{SPPavoidedScetch} by making the lines dashed. Instead, one usually characterizes the avoided crossing by the splitting at resonance, i.e.\ at the point $\kappa = \omega_0$, c.f.\ figure \ref{SPPavoidedScetch}. We can calculate the size of this splitting from the dispersion. 

For the sake of clarity, we will first present a simple approximative derivation, and later an exact one (exact in the $\gamma=0$ case). Let us write the dispersion relation, equation (\ref{cleanDispersionPT}), in the following form,

\begin{equation}
\frac{(\kappa+\omega)(\kappa-\omega)}{\omega^2} = \frac{A}{(\omega_0+\omega)(\omega_0-\omega)}.
\end{equation}

\noindent Then we assume that $\omega$ (and thus $\kappa$) is quite close to $\omega_0$, so that one can approximate $(\kappa+\omega) \sim 2\omega_0$, $(\omega_0+\omega)\sim 2\omega_0$ and $\omega^2 \sim \omega_0^2$. The equation then becomes,

\begin{equation}
(\kappa-\omega)(\omega_0-\omega)= \frac{A}{4}. \label{apprNodampingPT}
\end{equation}

\noindent This equation produces two solutions, corresponding to two normal modes, of the form,

\begin{equation}
\omega_\pm = \frac{\kappa}{2} + \frac{\omega_0}{2} \pm \frac{1}{2}\sqrt{A+(\kappa-\omega_0)^2}  . \label{apprModesPT}
\end{equation}

For very large and very small $\kappa$ one can approximate the square root and see that the two solutions approach the light line of the SPP. But the larger the value of $A$ and the closer one is to the resonance $\kappa = \omega_0$, the greater the distortion of the dispersion from the light line. The difference in the energies of $\omega_+$ and $\omega_-$ at the resonance point $\kappa = \omega_0$ gives the so-called normal-mode splitting, denoted by $\Omega$. It turns out that this is similar to the vacuum Rabi splitting derived from fully quantum theory, as will be discussed below in section \ref{quantum}. At resonance ($\kappa=\omega_0$), one has $\omega = \omega_0 \pm \sqrt{A}/2$ which means that the normal-mode splitting (corresponding to the vacuum Rabi splitting) becomes,

\begin{equation}
\Omega = \sqrt{A} = \sqrt{\frac{N}{V}} \frac{e}{\sqrt{\epsilon_0 m}}.  \label{OmegaClassicalPT2}
\end{equation}

The splitting is proportional to $\frac{e}{\sqrt{\epsilon_0 m}}$ and to the square root of the number density (concentration) of the emissive species. The quantum theory of strong coupling (\cite{Agarwal1984} and section \ref{quantum}) gives exactly this dependence on the concentration of oscillators (emitters). This means that the size of the splitting as a function of the density of oscillators does not allow us to distinguish between having quantized or classical fields, or between quantum (two-level system) or classical (Lorentzian oscillator) emitters.

Note that the other term in $\Omega$, namely $\frac{e}{\sqrt{\epsilon_0 m}}$, is not specific to any atom or molecule but only depends on electron charge and mass. This is because our derivation was based on a simple bound electron picture. In general, in both the semiclassical and quantum cases, the dipole moment specific to the atom/molecule as well as $\hbar$ will appear in the expression for $\Omega$. (It can be shown (see e.g.\ section 3.3 in \cite{Lucarini2005}) that by calculating the high-frequency asymptotic of the linear susceptibility from quantum theory of the atomic/molecular dipole moment (the field still being classical) one obtains a result where the plasma frequency $\omega_p^2 = 4 \pi A = 4 \pi Ne^2/(V \epsilon_0 m)$ is the only physical parameter appearing, that is, $\hbar$ is missing and the system responds, at high frequencies and in the linear regime, as a classical free electron gas. But this is only the high-frequency asymptote far away from any resonance; in general, the quantum mechanical transition dipole moment specific to an atom/molecule as well as $\hbar$ appear in the result in the semiclassical and quantum case.) In summary the $\sqrt{N/V}$ dependence shown here for the classical case will also appear the semiclassical and quantum treatments. However, the other factor in $\Omega$ will differ from $\frac{e}{\sqrt{\epsilon_0 m}}$. At this point we would like to note that sometimes when the classical susceptibility (\ref{chiPT}) is used in the literature, the $\frac{N}{V}\frac{e^2}{\epsilon_0 m}$ term is replaced by $\frac{N}{V}\frac{f_0 e^2}{\epsilon_0 m}$ where $f_0$ is an oscillator strength. Some works further replace $f_0$ by the oscillator strength derived from a quantum mechanical two level system (e.g.\ using the relation $f_0 = 2m\omega_0 d^2/(3\hbar e^2)$ 
\cite{Liboff2003} where $d$ is the dipole moment of the two-level system) which leads to the splitting $\sqrt{A}$ to be proportional to $\sqrt{\frac{N\omega_0}{V\epsilon_0 \hbar}}d$. Such an approach can be considered semiclassical since it uses the quantum mechanically calculated oscillator strength (dipole moment). Indeed we will obtain this dependence for the splitting from the semiclassical treatment in section \ref{semiclassical}.    

Now let us see what happens if we relax the approximations made above. Equation (\ref{cleanDispersionPT}) can be solved exactly since it will have terms of fourth, second and zeroth order in $\omega$. The result is (remember we consider only positive frequencies),

\begin{equation}
\omega_\pm = \sqrt{\frac{1}{2} (\omega_0^2 + \kappa^2 + A)\pm \frac{1}{2} \sqrt{(\kappa^2-\omega_0^2)^2 + A^2 + 2 A (\kappa^2 
+ \omega_0^2)}}. \label{exactModesPT}
\end{equation} 
  
\noindent At resonance we obtain (to prove the penultimate equality, take squares of both sides),
 
\begin{eqnarray}
\Omega &=& \omega_+(\kappa=\omega_0) - \omega_-(\kappa=\omega_0) \\
&=& \sqrt{\omega_0^2 + \frac{A}{2} + \frac{1}{2} \sqrt{A^2 + 4 A \omega_0^2}}
- \sqrt{\omega_0^2 + \frac{A}{2} - \frac{1}{2} \sqrt{A^2 + 4 A \omega_0^2}} \\
&=& \sqrt{A} = \sqrt{\frac{N}{V}} \frac{e}{\sqrt{\epsilon_0 m}} . 
\end{eqnarray}

\begin{figure}
\includegraphics[width=0.7\textwidth]{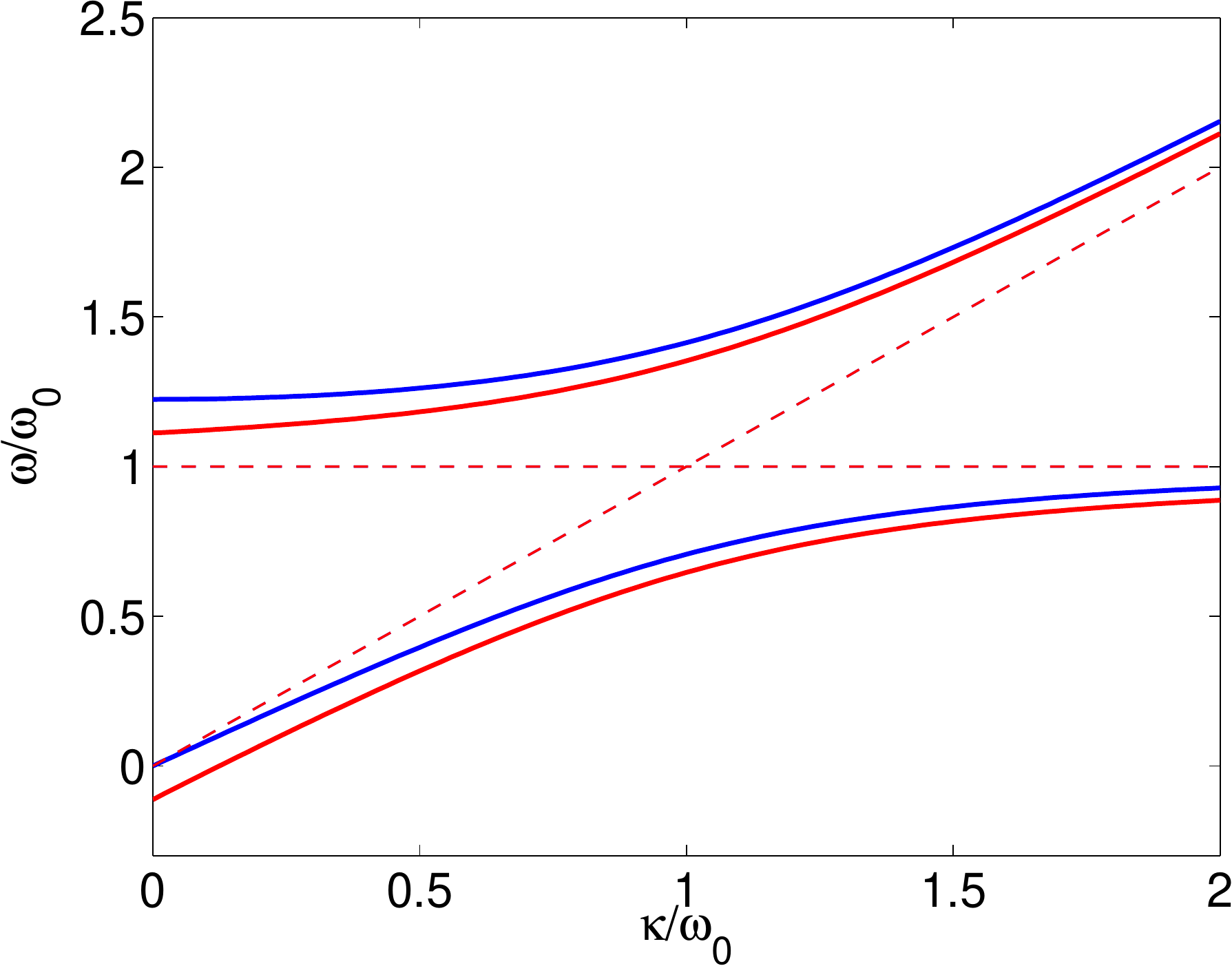}
\caption{Exact (blue) and approximate (red) normal modes as given by equations (\ref{exactModesPT}) and (\ref{apprModesPT}), respectively. The distance between the two branches at the resonance point $\kappa=\omega_0$ is exactly the same in both cases, as can be shown analytically.} 
\label{analyticalPlotPT}
\end{figure}

\noindent The exact result thus gives precisely the same result for the extent of the Rabi splitting at resonance as the approximate one. This can be seen in figure \ref{analyticalPlotPT} where we show the normal modes calculated using the exact and approximate treatments, equations (\ref{exactModesPT}) and (\ref{apprModesPT}), respectively. 

The calculations presented in this section are similar to those in \cite{Agranovich1974} where the possibility of surface plasmon polariton strong coupling was proposed for the first time although the calculations differ in some details, for instance the two solutions in Eq.(4) of \cite{Agranovich1974} have an implicit frequency dependence although the general form of the equation is similar to our result. 

It is important to note that the existence of a splitting in frequency implies the existence of dynamics in the time domain at frequencies that correspond to the splitting: in this case the coherent exchange of energy between the SPP field and the oscillators (emitters), see section \ref{Dynamics}.

\subsection{Influence of damping on SPP-emitter strong coupling}  \label{classicwithdampingPT}

Let us now consider the case of dispersion with damping, i.e.\ equation (\ref{cleanDispersionPT}).

\begin{equation}
\kappa^2 =\omega^2 \left( 1 + \frac{A}{ \omega_0^2 - \omega^2 - i\gamma \omega} \right) .
\label{cleanDispersionTwoPT}
\end{equation}

\noindent The $\omega$ in the damping term $i \gamma \omega$ now causes this equation to have a third order term, thus it is difficult to solve in closed form, unlike the case $\gamma = 0$. However we can again apply the approximations $(\kappa+\omega) \sim 2\omega_0$, $(\omega_0+\omega)\sim 2\omega_0$ and $\omega \sim \omega_0$ as above and obtain,
 
\begin{equation}
(\kappa-\omega)(\omega_0-\omega - i \gamma/2)= \frac{A}{4}.
\label{apprDampingPT}
\end{equation}

\noindent The solutions are,

\begin{equation}
\omega_\pm = \frac{\kappa}{2} + \frac{\omega_0}{2} - i \frac{\gamma}{4} \pm \frac{1}{2}\sqrt{A+(\kappa-\omega_0+i\gamma/2)^2} ,
\end{equation}

\noindent and at resonance,

\begin{equation}
\omega_\pm = \omega_0 - i \frac{\gamma}{4} \pm \frac{1}{2}\sqrt{A-\frac{\gamma^2}{4}}.
\end{equation}

\noindent The presence of damping thus diminishes the size of the Rabi splitting.

So far, we have, for simplicity, treated the SPP mode as being loss-less. This, of course, does not correspond to reality since the SPP modes are rather lossy, with linewidths corresponding to lifetimes of the order $10-100 fs$ \cite{van_Exter_PRL_1988_60_49,Sonnichsen_2002_PRL_88_077402}. We can incorporate SPP losses in the analytical model considered here by replacing $\kappa$ with $\kappa - i\gamma_{SPP}/2$. 
This gives the energies of the normal modes at resonance as,

\begin{equation}
\omega_\pm = \omega_0 - i \frac{\gamma}{4} - i \frac{\gamma_{SPP}}{4} \pm \frac{1}{2}\sqrt{A-\left(\frac{\gamma}{2}
-\frac{\gamma_{SPP}}{2}\right)^2}.  \label{ModesBothGamma}
\end{equation}

\noindent This gives a strict condition $A-\left(\frac{\gamma}{2}-\frac{\gamma_{SPP}}{2}\right)^2>0$
to keep the term in the square root positive. But obviously, if $\gamma \sim \gamma_{SPP}$, this is always valid. At first, this may look somewhat puzzling: can the decay constants indeed cancel each other inside the square root, and thus perhaps produce a bigger Rabi splitting? This is, however, not quite the case. Namely, since the energies $\omega_\pm$ are complex, we have to understand them as damped modes with linewidths characterized by $\frac{\gamma}{2} + \frac{\gamma_{SPP}}{2}$; the new normal modes inherit the damping from both the SPP and the oscillator modes. Now think about two Lorentzian (or Gaussian) distributions that have maxima and certain widths, and are so close that they essentially overlap: the double peak maxima will not be at the maxima of the individual distributions, but at positions shifted towards the middle of the overlap region. Similarly here, the actual Rabi splitting will be clearly visible only if the difference in the real part of the energies, given by the square root term, is bigger than the widths of the new modes $\frac{\gamma}{2} + \frac{\gamma_{SPP}}{2}$, that is $\sqrt{A-\left(\frac{\gamma}{2}-\frac{\gamma_{SPP}}{2}\right)^2}>\frac{\gamma}{2} + \frac{\gamma_{SPP}}{2}$ which gives $A>\frac{\gamma^2}{2} + \frac{\gamma^2_{SPP}}{2}$. The strong coupling condition is sometimes given also as $\sqrt{A}>\frac{\gamma}{2} + \frac{\gamma_{SPP}}{2}$; these two are obviously the same if $\gamma \sim \gamma_{SPP}$. Thus we may write the strong coupling condition as
\begin{equation}
\frac{Ne^2}{V\epsilon_0 m} > \frac{\gamma^2}{2}+\frac{\gamma_{SPP}^2}{2}.  \label{Strongcouplingcondition}
\end{equation}
Note that this 
should be understood more as an order of magnitude condition: for two Lorentzian distributions separated by some distance (for instance \cite{Agarwal1984} shows how strong coupling leads to double Lorentzian form for the susceptibility), a double peak structure is visible even when the widths of individual distributions are slightly larger than the separation. Note that this is the case for Lorentzian distributions but no so much for Gaussians. One should always make a careful connection from the complex normal modes (\ref{ModesBothGamma}) to the measured splitting, especially when aiming for precision measurements, broadening
may change the measured value of the splitting. The main point to keep in mind, in general,
is that although the condition required for strong coupling is often worded as 
"the splitting has to be larger than the widths of the modes", this is a rule of thumb and the actual 
measured splitting can be slightly smaller than the average width. Figure 2 of \cite{Khitrova2006} nicely 
illustrates this.     
 
The derivation presented here is similar to the cases of strong coupling in optical cavities and in semiconductor microcavity systems. We discuss here briefly the connection to some key literature in that context. Agranovich {\it et al.}\ \cite{Agranovich2003} present a quantum and a classical theory for organic semiconductors in microcavities. The derivation in the classical formalism starts from the dispersion of the transverse wave in the cavity, equation (2) of \cite{Agranovich2003}. To within a number of constants, there is one-to-one mapping to the approximate derivation we have given (equations (\ref{apprNodampingPT}) and (\ref{apprDampingPT}))  if we equate the cavity mode dispersion, their Eq.(1), with $\kappa$ (i.e.\ the momentum scaled to include the SPP dispersion) in our case. The style of the derivation is the same: to search for the new normal modes explicitly. Sometimes in the context of cavities a slightly different approach is applied, see e.g.\ \cite{Zhu1990}: The total transmission or reflection of the cavity is calculated, and it is shown to be dependent on the phase shift the light accumulates over a round-trip period in the cavity. In a Fabry-Perot type cavity, for instance, this phase has to have certain values (multiples of $\pi$) in order to obtain the constructive interference that defines the modes. When the emitter material is present in the cavity, the phase shift is in essence given by the corresponding refractive index, which can be modelled by a Lorentzian oscillator, as we have done here. This leads to non-trivial behaviour of the phase shift and eventually to an energy splitting visible in the transmission (reflectance) of the cavity. The basic physics of strong coupling is exactly the same, although the style of formulating the problem is different. The particularities of the cavity can be thought to have the same role as the specifics of the SPP dispersion here. 

Note that the polarization is proportional to the electric field $E$ through,

\begin{equation}
P = \frac{Ne^2}{V m} \frac{1}{\omega_0^2 - \omega^2 - i \gamma \omega}E.
\end{equation}

\noindent In the presence of a resonant mode, such as an SPP or micro-cavity mode, the appropriate field in the above equation is the enhanced field. The combination of strong confinement (defining $V$) of the electric field and the high concentration $n=N/V$ of molecular dipole materials (such as those to be described below) makes strong coupling between SPP modes and many quantum emitters easy to observe even with open cavities such as simple flat metal films, as we will see later. One of the attractive features of plasmonics is that deeply sub-wavelength effective volumes can be achieved. 
The question of whether the volume can be made small enough to see strong coupling with a single emitter 
naturally arises. We should note that for the highly concentrated emitter materials typically used, reducing the volume 
will keep $n=N/V$ constant since as the volume is reduced so is the number of emitters it contains. Only when a reduction in volume is such that the number of emitters does not decrease in proportion will $n=N/V$ rise. This will require very small mode volumes, achieving the single emitter limit, $N=1$, may be possible by using specially tailored SPP modes.  

\subsection{Abandoning the simplifications --- numerical treatment of SPP strong coupling}

\begin{figure}
\includegraphics[width=1.0\textwidth]{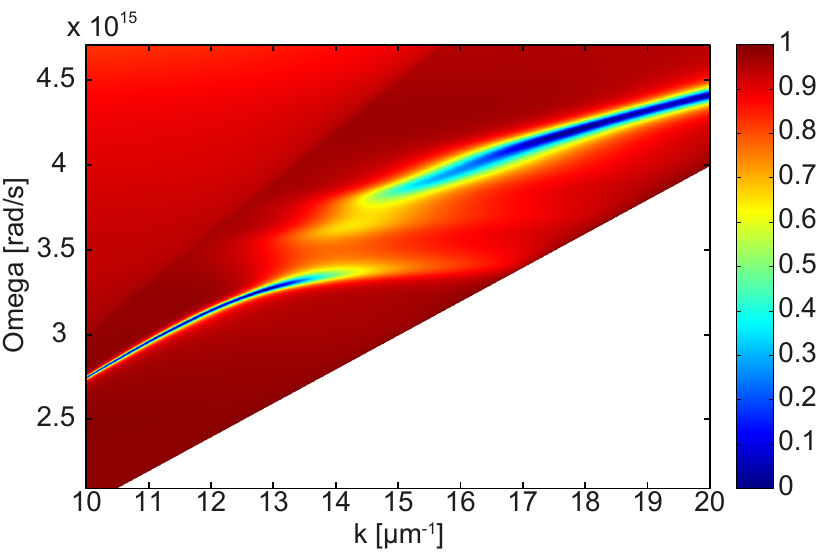}
\caption{Numerically calculated reflectance of a metal film (thickness 50 nm) and an absorbing film 
on top (thickness 30 nm). Reflectance is calculated in a Kretschmann-Raether configuration. The permittivity of the absorbing film is obtained from measured values for 200 mM concentration of Rhodamine 6 G (R6G) in PMMA and the data for silver permittivity is from \cite{Johnson1972}.} 
\label{numericalPlotPT}
\end{figure}

We used several simplifications in the above treatment. We assumed a simple Lorentzian oscillator, although the 
line-shapes of various emitters are often more complicated. Furthermore, we assumed that only the term $\epsilon_1\epsilon_2$ in equation (\ref{SPPdispersionAgainPaivi}) gives an interesting frequency dependence, and took $\epsilon_1+\epsilon_2$ to be constant. We also assumed the dielectric constant of the metal to be independent of frequency. This might to some extent be valid near the resonance $\omega_0$ if the resonance is sufficiently narrow. However, we made these and other approximations simply to obtain analytical results to guide understanding. To describe experiments, one usually has to use a numerical treatment. One can take the SPP dispersion,
 
\begin{equation}
k =  \frac{\omega}{c} \sqrt{\frac{\epsilon_1(\omega) \epsilon_2 (\omega)}{\epsilon_1 (\omega) + \epsilon_2 (\omega)}},
\end{equation}

\noindent and use the measured, tabulated values for $\epsilon_1(\omega)$, e.g.\ from \cite{Johnson1972,Palik1991,Li2013}, and the $\epsilon_2(\omega)$ describing the dielectric emitter material. Both dielectric functions will be complex, so damping will be taken into account. Often the emitter material is characterized by its absorption spectrum: one may find the absorption spectrum from the literature or measure it directly. In order to obtain the real part of the electric susceptibility from the absorption/extinction spectrum, one can use the Kramers-Kronig relations. The absorption spectrum $\alpha (\omega)$ is proportional to the imaginary part of the refractive index, $\kappa$. The real part, $n$, can be calculated using the Kramers-Kronig relations:

\begin{equation}
n(\omega) = 1 + \frac{2}{\pi} P \int_0^\infty \frac{\omega' \kappa(\omega')}{\omega'^2-\omega^2} d\omega' ,
\end{equation}

\noindent where $P$ denotes the principal value. The real and imaginary parts of the susceptibility are then given by equations (\ref{epsilon_PT}), (\ref{epsilon_real_PT}) and (\ref{epsilon_imag_PT}). The appendix of the book \cite{Lucarini2005} provides a nice Matlab code for using the Kramers-Kronig relations. Note also that the SPP dispersion given above is just the simple formula for the interface of two infinite half-spaces. For more complicated structures, e.g.\ several layers, one can use a Fresnel-type calculation for the reflectivity spectrum, using the susceptibility for the material layer that contains the oscillators, and obtain strong coupling phenomena. In figure \ref{numericalPlotPT}, we present an example of a numerical simulation of the reflectance of a silver film covered with an absorbing film, using experimentally realistic values. 

As a final note on the classical theory of strong coupling we would like to point out the following. It has long been known (at least since \cite{Zhu1990}) that a large number of field-matter interaction experiments displaying strong coupling, i.e.\ splittings in the energy spectrum, can be equally well described by a classical or a quantum theory. Note, however, that the classical and quantum treatments and their respective results are not exactly the same. One difference originates from the fact that a first order time derivative appears in the Schr\"odinger equation and a second order derivative typically in classical equations of motion. Therefore the dielectric susceptibility of a classical oscillator is of the form $1/(\omega_0^2-\omega^2-i\omega \gamma)$ whilst that of a corresponding quantum oscillator 
(a two-level system) would have a susceptibility of the form $1/(\omega_0-\omega-i\gamma)$. (Also saturation effects make 
a difference, as will be discussed in section \ref{quantum}.) Above, with some dirty tricks, which are justifiable close 
to $\omega=\omega_0$, we could make the classical susceptibility to look like this (equations (\ref{apprNodampingPT}) 
and (\ref{apprDampingPT})), and then same phenomenology as from the quantum case follows; a motivation for doing such 
approximations is indeed to relate the quantum and classical cases. Note, however, that in certain issues the quantum 
and classical theories give exactly the same result, a notable example is the size of the splitting at the resonance 
and its $\sqrt{N/V}$ dependence. Exactly at resonance, i.e. when $\omega=\omega_0$, both $\omega - \omega_0$ (quantum) 
and $\omega^2 - \omega_0^2$ (classical) vanish, so that the difference between the first and the second order derivative 
does not play a role. However, the stronger the coupling, the larger are the deviations of the exact classical normal 
modes from the approximate ones (which are the same as those given by quantum theory) 
(c.f.\ figure \ref{analyticalPlotPT}). One can ask whether at some point the predictions of the exact classical theory 
for the dispersion curves start to considerably deviate from the approximate ones (which are the same as given by quantum 
theory). Another issue is the size of the splitting. Here the factor multiplying $\sqrt{N/V}$ is in essence the dipole 
moment of the oscillator. The dipole moment can be derived from first principles using classical or quantum theory. 
It would be interesting to 
test a microscopic quantum mechanical prediction for the size of the splitting through experiment, 
the concentration $\sqrt{N/V}$ would need to be accurately known. 
We will come back to this issue and whether it has been considered in the reported SPP strong coupling experiments in 
section \ref{quantum2} after the fully quantum description of strong coupling has been presented. In that section, 
we discuss the details of the quantum mechanical prediction for the size of the splitting. 

\subsection{A note on detecting strong coupling in plasmonics using reflectometry measurements, including a discussion about back-bending}
\label{ReflectometryForSC}

To understand the physics of the SPP+emitter system, we would like to know the dispersion, that is we would like to know the function $\omega (k)$. Measuring the reflectance, as discussed in section \ref{basics}, provides a convenient technique for this purpose. If a mode exists for a certain frequency and wavevector, incident light may be coupled into the system and a corresponding reduction in the reflectance may occur. Most recent work in which splittings are determined from reflectance data are based on examining the reflectance when plotted as a full dispersion curve, i.e. plotting the reflectance as a 2-dimensional data set in $\omega-k$ space, such as that shown in figure \ref{numericalPlotPT}. Note that one should not work in $\omega-\theta$ space,  as was pointed out by Symonds {\it et al.} \cite{symonds2008}. 

\begin{figure}
\includegraphics[width=0.8\textwidth]{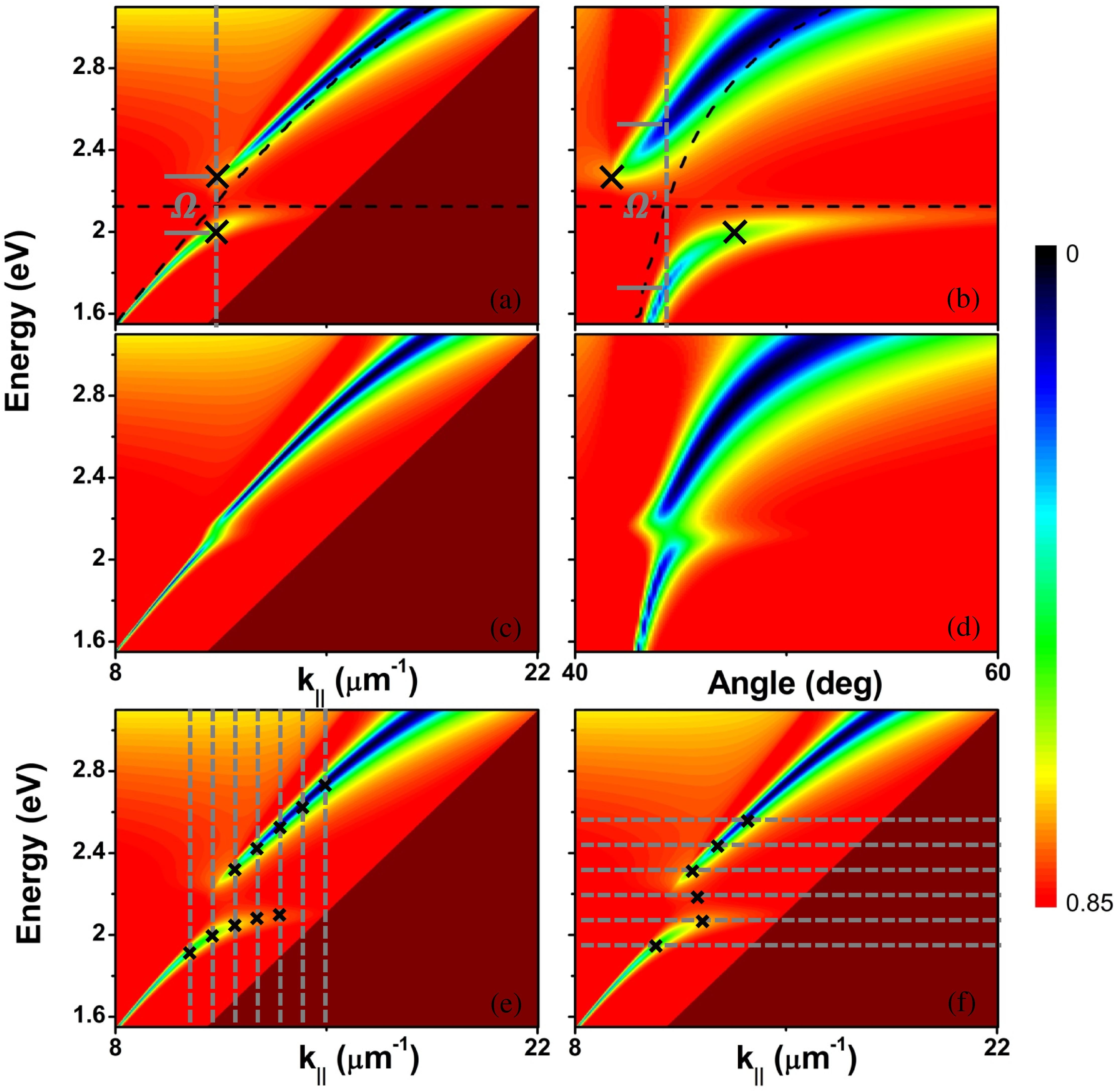}
\caption{a)-b) Reflectance plotted as a function of: left column, frequency (energy) {\it vs.} in-plane wavevector; and right column, frequency (energy) {\it vs.} incident angle. Due to the mapping $k= (2\pi/\lambda) n_p \sin \theta$ in reflectometry experiments, points corresponding to one $k$-vector but different frequencies become shifted with respect to each other in an angle plot. The Rabi splitting/normal mode splitting $\Omega$ is determined as the difference of the two branches at the resonance point, see a). Attempting to determine the splitting $\Omega'$ from the angle-plot will lead to an overestimation of the splitting, see b). e)-f) The vertical/horisontal lines depict wavelength (frequency)/angle scans, respectively: the minima determined from these scans differ as shown by the black crosses in e)-f). Finite linewidth of the normal modes is required for this difference to arise. c)-d) Distortions of the dispersion when approaching strong coupling, but splitting not yet being visible like in c), may lead to slight back-bending c.f.\ d) when the results are shown as an angle-plot with angle-scan used for plotting minima.} 
\label{backbendingPT}
\end{figure}

Historically it was often easier to record a set of angle or wavelength scans, especially before the ready availability of imaging spectrometers. Experimentalists often faced the choice of recording data by sweeping the in-plane wavevector (angle, $\theta$) for a fixed frequency of incident light, a $k$-scan, or sweeping the frequency for a fixed in-plane wavevector (more often fixed incident angle), an $\omega$-scan. As seen figure \ref{backbendingPT}, the presence of an anti-crossing (splitting) is better seen in the latter case.

There is another subtlety related to reflectometry measurements which is relevant in understanding the early experimental results concerning the strong coupling regime, and which highlights the value of plotting the reflectance as a 2-dimensional data set in $\omega-k$ space. If angle-scans for a number of fixed wavelengths are conducted there is the possibility of observing what looks like {\it back-bending}, see figure \ref{backbendingPT}. In the early experiments \cite{Pockrand1978,Pockrand21978} by Pockrand {\it et al.}\ only angle scans are reported, and back-bending of the modal dispersion is evident. However, in \cite{Pockrand1982} both angle and wavelength scans were carried out and it was shown that back-bending in the former is connected with having a splitting in the latter (although the reason for the difference was not discussed). The back-bending was originally also called "anomalous dispersion". However, as the word anomalous hints, the back-bending is not a true modal dispersion; the back-bent curve originates simply from the finite linewidth of the split modes, in the region of the gap they overlap. Now if one does the angle scan for a fixed frequency, see the horizontal lines in figure \ref{backbendingPT}f, these points appear as maxima, even though when making the wavelength scan, see the vertical lines in figure \ref{backbendingPT}e, the same points would appear as minima, or at least below the maximum. Mathematically speaking, a saddle point may look like a minimum or a maximum depending on the direction one crosses the saddle point. Note that if one plots the whole data set, e.g.\ with a colour scale the problem, as noted above, no longer arises; it is when only the minima are plotted that the difference arises. (Similar care is needed in examining the surface plasmon-polariton band gaps discussed in section \ref{propagationSPP} \cite{Barnes_PRB_1996_54_6227}.)

Concerning the earliest papers on the topic, it is of interest to ask whether any observed  back-bending (in angle-scans) actually implies strong coupling. This is not necessarily the case: one can consider a situation where there is no clear splitting between the two branches, but already some broadening and reshaping of the mode structure near the resonance point, see figures \ref{backbendingPT}c-d. If such a system were probed by the angle-scanning technique, back-bending might be visible in the evolution of the minimum. Thus the back-bending does not necessarily imply strong coupling, but it does indicate that the strong coupling regime is near since the dispersion has become distorted.

\section{Review of the experiments} \label{experiments}

Before looking at the coupling between SPPs and different molecular electronic states in detail, we should mention some early work. Agranovich and Malshukov appear to have been the first to predict splitting phenomena of the kind discussed here \cite{Agranovich1974}. The first observation of surface polariton splitting appears to be that due to Yakovlev {\it et al.} \cite{Yakovlev1975}, who looked at how the lattice vibrations (TO phonon mode) of a thin overlayer of LiF modified the surface phonon-polariton mode supported by an underlying rutile ($TiO_2$) substrate. Although the work of Yakovlev {\it et al.} concerned phonon-polaritons rather than plasmon-polaritons, the similarity of the resulting dispersion curves, and of much of the underlying physics, makes this an important historical reference point.

\subsection{J-aggregated systems} \label{J-aggregates}

Strong coupling between the excitonic absorption resonances of dye molecules and surface plasmon polaritons came a few years later, it was first observed by Pockrand {\it et al.} \cite{Pockrand_JChemPhys_1982_77_6289} who looked at a cyanine-based J-aggregated dye molecules deposited on a silver film. J-aggregates of cyanine dyes are self-organised molecular crystals and have partially delocalised excitons that have narrow, red-shifted absorption bands when compared to that of the dye monomer. The higher oscillator strength of J-aggregates results from the way the dipole moments associated with many molecular units act coherently, providing an effective 'super' moment \cite{Kobayashi_J-aggregates}. Something similar has been suggested to explain strong coupling results for quantum dots in cavities \cite{Marsden_NJP_2013_15_025013}. Pockrand {\it et al.} used the attenuated total reflection (ATR) technique (Kretschmann-Raether prism-coupling) to observe strong coupling between the SPP mode supported by the silver film and the exciton mode of the J-aggregates, finding a splitting of ~70 meV (3\%). These authors stressed the importance of the details of the measurement process, as we discussed in section \ref{ReflectometryForSC}. This work came after Pockrand and Swalen had shown that a squarylium dye deposited onto a silver film resulted in back-bending of the SPP dispersion curve \cite{Pockrand_JOSA_1978_68_1147}. Again, as noted in section \ref{ReflectometryForSC}, whether strong coupling or back-bending is seen depends to some extent on the details of the measurement process.

In what follows we look at strong coupling using J-aggregate molecules in the context of different kinds of SPP modes: propagating SPP modes on planar metal surfaces, SPP modes on nano-structured metal films, and localised SPP modes associated with metallic nanoparticles.

\subsubsection{Propagating surface plasmon-polaritons}

The initial work of Pockrand {\it et al.} \cite{Pockrand_JChemPhys_1982_77_6289} was followed up more than 20 years later by Bellessa {\it et al.} \cite{Bellessa2004} who showed evidence of strong coupling, based on reflectivity measurements that had been compiled into a dispersion diagram - an anti-crossing of 180 meV was observed. In addition these authors also looked at how the luminescence from the J-aggregates was modified by the strong coupling. In common with experiments on J-aggregates in microcavities \cite{Lidzey_PRL_1999_82_3316} they found that the luminescence tracked the position of the lower polariton branch, but there was no evidence of the upper polariton branch in the emission. Bellessa {\it et al.} attributed this lack of upper polariton branch emission to uncoupled excitons. A detailed study by Agranovich {\it et al.} suggests that a significant fraction of the J-aggregates are not coupled because they involve incoherent states that do not couple to polariton modes \cite{Agranovich_PRB_2003_67_085311}.

In more recent work Symonds {\it et al.} looked in more detail at SPP -- J-aggregate strong coupling for planar metal surfaces, examining J-aggregated systems \cite{Symonds_NJP_2008_10_065017} and excitons based on a mixed organic-inorganic system \cite{Symonds_NJP_2008_10_065017,Symonds_APL_2007_90_091107}. As noted above, these authors made the important point that in evaluating the extent of the (energy) splitting associated with strong coupling, i.e. the extent of the anti-crossing, it is important to look at data where the in-plane wavevector is held constant and the frequency swept \cite{Symonds_NJP_2008_10_065017}. In doing so they noted that trying to evaluate the splitting from fixed angle scans can lead to an overestimate of the splitting by a factor of up to 2; c.f.\
figure \ref{backbendingPT} (a)-(b) where this issue is illustrated.

As noted in sections \ref{classical}, \ref{quantum} and \ref{quantum2} of this review, the extent of the splitting depends on a number of factors, including the spectral width (damping) of the plasmon mode involved, and the number density of J-aggregated molecules. Balci {\it et al.} \cite{Balci2012} looked at both of these aspects in an arrangement very similar to Pockrand {\it et al.} \cite{Pockrand_JChemPhys_1982_77_6289}. They varied the concentration of J-aggregates in the host PVA layer, finding that, as expected (see sections \ref{classical}, \ref{quantum} and \ref{quantum2}, e.g.\ equation (\ref{OmegaClassicalPT2})) that the extent of the splitting was proportional to the square root of the concentration of the molecules. Balci {\it et al.} \cite{Balci2012} also showed that the width (damping) of the SPP mode influences the extent of the splitting; to do this they varied the thickness of the silver film.

\subsubsection{Surface plasmon-polaritons on nanostructured metal surfaces} \label{Nanostructured}

We have focussed so far on SPPs associated with flat metal films, prism coupling being used to allow incident light to be coupled to the SPPs. Periodically modulated surfaces may also be used when the period is of order the wavelength of light - grating coupling. Symonds {\it et al.} \cite{Symonds_NJP_2008_10_065017} showed that a traditional diffraction grating type surface could be used successfully to explore strong coupling, their (sinusoidal profile) grating being produced by an embossing technique. Vasa {\it et al.} \cite{Vasa2010,Vasa2013} used focussed ion-beam milling to produce rectangular profile gratings and also observed strong coupling between SPP modes of these structures and J-aggregated molecules placed in an adjacent layer. A full discussion of the results of Vasa {\it et al.} will be deferred until later since their focus was on strong coupling dynamics. Dintinger {\it et al.} looked an alternative nano-structured metal surface, a sub-wavelength hole array \cite{Dintinger2005}. Again J-aggregates were introduced by spin coating a layer of J-aggregate doped PVA onto the hole array. Dintinger {\it et al.} explored the strong coupling in two ways. First, they varied the array period, allowing the in-plane wavevector to be varied as a consequence of scattering arising from the presence of the grating (hole array): second, they varied the angle of incidence. In both cases a splitting was observed. They also showed that the extent of the splitting varies with the square root of the absorbance (concentration) of the aggregated molecules, finding a maximum splitting of $\sim$ 250 meV ($\sim$ 14\%).

\subsubsection{Localised surface plasmon-polaritons} \label{Localized}

Strong coupling between localised surface plasmon-polaritons and J-aggregated molecules was investigated in 2006 both theoretically by Ambjornsson {\it et al.} \cite{Ambjornsson_PRB_2006_73_085412}, and experimentally by Sugawara {\it et al.} \cite{Sugawara2006}. Sugawara {\it et al.} exploited the localised modes associated with spherical nanovoids, structures that they produced by electrochemical deposition of gold through a template of self-assembled latex spheres. Once the metal had been deposited the latex spheres were chemically removed to leave metallic voids, the J-aggregated molecules being added by drop-casting. Localised SPP modes are more often explored in plasmonics for metallic nanoparticles rather than voids. Fofang {\it et al.} \cite{Fofang2008} looked at gold nanoshells coated with a layer of J-aggregated molecules. They observed splitting associated with both the dipolar plasmon mode and the quadrupolar plasmon mode of the nanoshells. They also increased the loading of the dye molecules in the surface of the metallic shell to demonstrate the effect of concentration on the extent of the splitting, but the splitting observed saturated very quickly, something they suggested resulted from a limitation of their fabrication approach. 

Wurtz {\it et al.}  \cite{Wurtz2007} coated arrays of densely packed gold nano-rods with J-aggregates. They also observed significant splitting ($\sim$ 14\%) and noted that the use of plasmon modes associated with nano-rods offered a good way to control the coupled system since the plasmon modes are easily controlled through rod size (and rod proximity). These authors attributed their results to the strong coupling of the J-aggregated molecules with the L-mode (dipole-dipole interaction mediated collective mode) of their nanorod assembly.

More recently Bellessa {\it et al.} \cite{Bellessa2009} looked at the case of periodic arrays of metallic nanoparticles by using electron-beam lithography to produce the particle arrays, the J-aggregate film being deposited directly on the array by spin-coating. They found a large splitting of $\sim$ 450 meV ($\sim$ 20\%), which they attributed to a high concentration of molecules in these samples. These authors also conducted a set of measurements and analysis of the linewidths of the two polariton branches in the vicinity of the anti-crossing, showing that the widths of the individual modes (which are quite different) take the same value at the anti-crossing point, providing a further demonstration of the hybrid nature of the modes at this point. We note that these authors also saw significant evidence of uncoupled excitons in their data. This is to be expected since they placed J-aggregated molecules across their samples, but the optical fields associated with the plasmon modes of the nanoparticles only extend a limited distance from the particles \cite{Murray_NL_2006_6_1772} and the nanoparticles are spatially separated by much more than this distance. (Note that aggregates being spatially located outside the field associated with the plasmon mode is not the only reason that excitons may not contribute to strong coupling, there may also be a significant fraction of excitons that do not take part because they are associated with incoherent states \cite{Agranovich_PRB_2003_67_085311,Coles_PRB_2011_84_205214}). Strong coupling of J-aggregates with individual metallic dimers was observed in \cite{Schlather2013}, and with individual nanorods \cite{Zengin_SciRep_2013_3_304}.

Having looked at the time-independent properties of strongly coupled systems involving SPP modes and J-aggregated molecules we now turn our attention to the dynamic behaviour in J-aggregate strong coupling.

\subsubsection{Dynamics in J-aggregate strong coupling}  \label{Dynamics}

So far we have focussed on the spectral response of coupled plasmonic-emitter systems, we turn our attention now to dynamics. The primary method of probing the dynamics of many photo-physical systems is that of pump-probe spectroscopy (see for example \cite{Ulbricht_RMP_2011_83_543}). Vasa {\it et al.} used pump-probe spectroscopy to investigate samples comprising J-aggregate layers deposited onto gold gratings \cite{Vasa2010}. They first showed that strong coupling was present in such systems, an anti-crossing appearing on their dispersion diagram, a diagram produced using experimentally determined reflectivity data, anti-crossing appearing where the grating-scattered SPP mode crossed the exciton absorption. They then looked at transient changes to the reflectivity ($\Delta R/R$) using pump-probe spectroscopy, finding a nonlinearlity, i.e. pump intensity dependent response as measured through $\Delta R/R$ that they attributed to bleaching of the exciton absorption. (Schwartz {\it et al.} point out that care should be exercised in interpreting $\Delta R/R$ type data if the samples under investigation also transmit light, a better measure in such circumstances is to determine the (transient) change in absorption \cite{Schwartz2013}.) These investigations of nonlinear behaviour stop short of demonstrating oscillations in the time-domain that are expected if Rabi oscillations are taking place.

\begin{figure}
\includegraphics[width=0.6\textwidth]{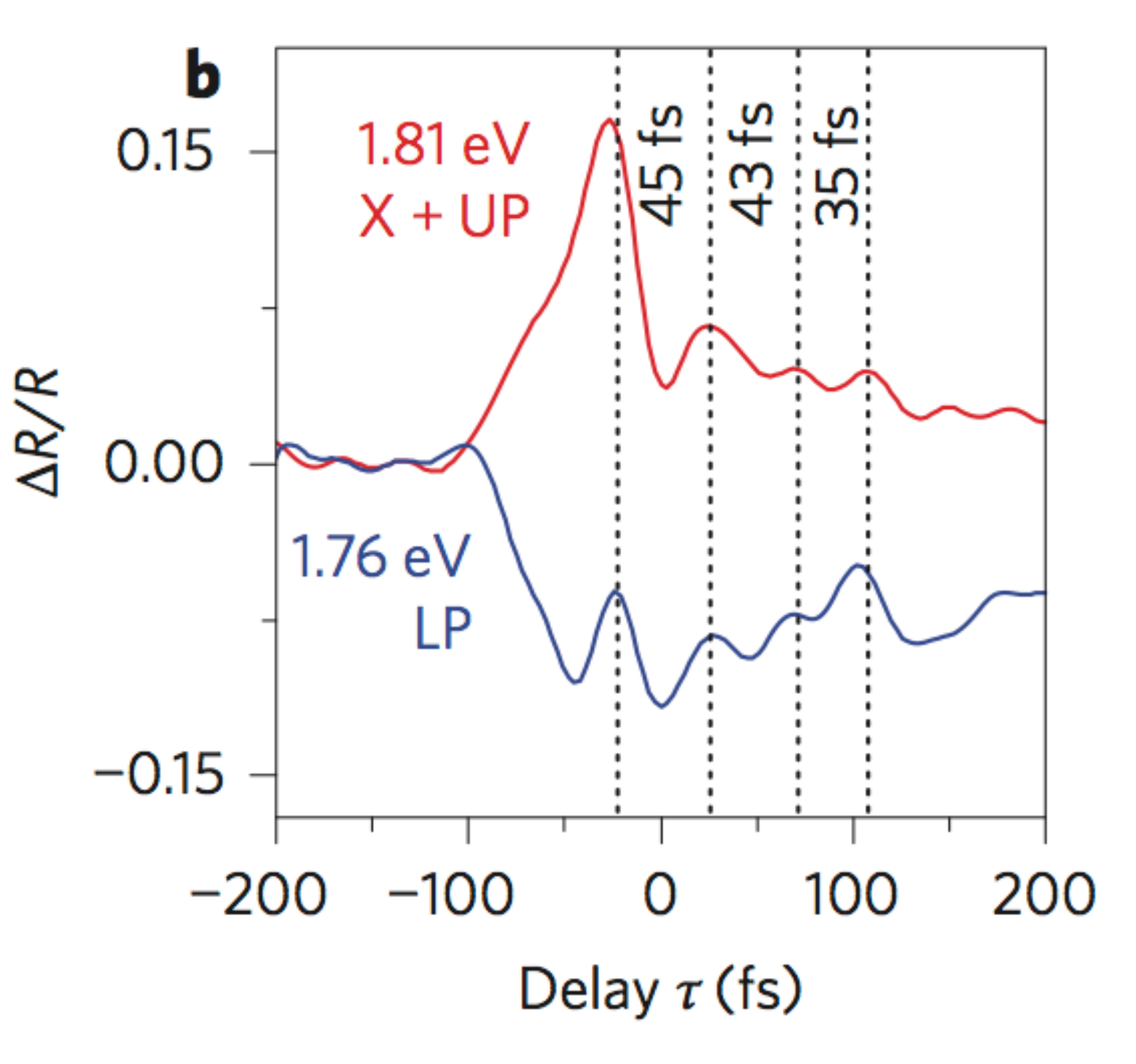}
\caption{The measured differential reflectivity of a J-aggregate-coated gold grating. For the experimental conditions used the splitting was $\sim$ 100 meV. Clear oscillations of period $\sim$ 45 fs are seen in the differential reflectivity associated with the upper branch (UP) and the lower branch (LP). Reproduced with permission from \cite{Vasa2013}.} 
\label{fig_Vasa}
\end{figure}

In what is probably the most extensive investigation so far of the dynamics of SPP-based strong coupling, Vasa {\it et al.} used pump-probe spectroscopy to observe temporal oscillations in their measured $\Delta R/R$ data \cite{Vasa2013}. Through comparison of their transient reflectivity data with simulations of the time evolution of the exciton and SPP population densities, the authors suggest that the oscillations they see are a direct manifestation of Rabi-oscillations; some of their data are reproduced in figure \ref{fig_Vasa}. As these authors point out in their supplementary material, there are many subtleties to consider if their data are to be fully understood. The major challenge in observing these oscillations is that the decay times for plasmon modes are very short $\sim$ 100fs. Rabi oscillation times thus have to be even shorter if they are to be observed. Here the advantage of J-aggregates becomes apparent. The large splitting that may be achieved, e.g. 200 meV, allows oscillation times as short as $\sim$ 20 fs to be produced, short enough to be seen against the plasmon decay. The effect of electron tunneling on single emitter strong coupling in plasmonic nanostructures has been theoretically investigated \cite{Marinica2013}, indicating that such tunneling processes may act to prevent the observation of strong coupling.\\

\subsection{Dye and photochromic molecules}   \label{DyePT}

According to the basic theory of strong coupling, the linewidths of the absorption/emission resonance and the 
optical/plasmonic mode should be smaller than the strength of the coupling in order to clearly
observe strong coupling phenomena, that is, for the avoided crossing in the dispersion to be visible, 
c.f.\ the formula introduced in section \ref{SPP+Lorentzian},

\begin{equation}
\frac{Ne^2}{V \epsilon_0 m}>\frac{\gamma^2}{2}+\frac{\gamma_{SPP}^2}{2}  .  \label{OmegaNgamma}
\end{equation}

\noindent One would thus anticipate that strong coupling is possible only for molecules with a narrow absorption spectrum, 
such as the J-aggregates discussed above. However strong coupling has also been observed between SPPs and molecules with a 
broad absorption spectrum. Why this is possible is still a question requiring further  study, taking into account 
microscopic details such as the vibrational level structure of the molecules. Of course, one may utilize the large
oscillator strengths of certain molecules as well as high concentrations $N/V$ to make the left hand side of the
inequality (\ref{OmegaNgamma}) large. Furthermore, one should keep in mind that splittings even slightly smaller than the
average linewidth can be observed, especially for Lorentzian profiles, as discussed in section \ref{classicwithdampingPT}. One more important issue is that the broad spectra of dye molecules consist of 
a set of underlying vibrational states, often leading to essentially Gaussian absorption/emission spectra. In this context it is interesting to mention the work \cite{Houdre1996} where 
strong coupling for a set of inhomogeneously broadened oscillators was considered both quantum mechanically and classically. Intriguingly, it was shown that the new normal modes formed via strong coupling have the linewidths of the individual oscillators rather than the width related to the inhomogeneous broadening, the effect of the broadening was only to produce a set of states within the energy split caused by strong coupling. One could speculate that the vibrational states in molecules are similarly a set of inhomogeneously broadened emitters; note that the linewidths of individual vibrational states are much smaller than the typical absorption and emission linewidths of the molecules, and usually of Lorentzian line shape.  

In the early studies of Pockrand et al.\ \cite{Pockrand1978,Pockrand21978} materials such as a monolayer of squarylium \cite{Pockrand21978} or cyanine \cite{Pockrand1978} dye combined with Cd-arachidate were used on top of silver films. These dyes have an absorption spectrum width of about 75 meV for the cyanine \cite{Pockrand_JChemPhys_1982_77_6289}. This compares with the J-aggregate widths reported in later literature of 45-70 meV and that of Rhodamine 6 G (main peak) discussed below of about 180 meV. Reflectometry measurements were performed and results of angle scans (c.f.\ section \ref{ReflectometryForSC}) for each frequency were shown. References \cite{Pockrand1978,Pockrand21978} report only angle scans, showing back-bendings of the reflection minima curves obtained from the angle scans. As discussed in subsection \ref{ReflectometryForSC}, back-bending is not firm evidence of strong coupling; instead it means that the system is either in the strong coupling regime or approaching it. Only in \cite{Pockrand1982} (who used J-aggregates) both angle- and wavelength-scans were presented, the latter displaying a clear splitting and thus confirming the presence of strong coupling. Inspired by this, the attention of the research community was focused on J-aggregates for the next 27 years. The first clear observation of prominent strong coupling for SPPs and molecules with a broad absorption spectrum was reported in 2009 \cite{Hakala2009}.

\begin{figure}
\includegraphics[width=1.0\textwidth]{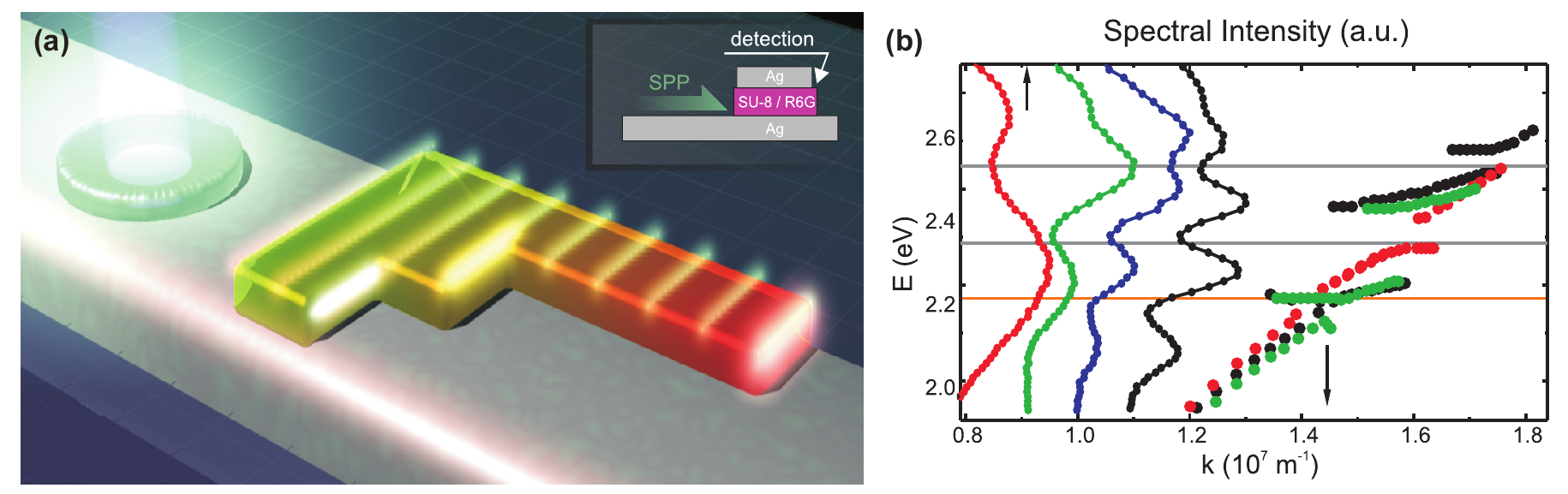}
\caption{(a) Silver waveguides with molecule-containing polymer areas of different lengths were fabricated in \cite{Hakala2009}. SPPs were launched via one molecular area (the disk) and spectra were recorded at the end of the rectangular polymer areas. (b) With increasing length of the area, splittings develop in the spectra (red $1\mu m$, green $2\mu m$, blue $5\mu m$, black $5\mu m$ with silver on top (inset of (a)). These are in good correspondence with the positions where the splittings emerge in the dispersions obtained by reflectometry (red, green, black dots for 4, 25, 50 mM molecular concentrations, horizontal grey lines mark the R6G absorption main and shoulder positions, and the orange the emission main peak). Thus in addition to the concentration, the strong coupling can be controlled by effective interaction time determined by the length of the polymer area. Reproduced with permission from \cite{Hakala2009}.} 
\label{Tommifigure}
\end{figure}

Hakala {\it et al.} \cite{Hakala2009} undertook two types of experiment. First, silver films coated with a PMMA polymer film containing Rhodamine 6G (R6G) dyes were studied using reflectrometry in the Kretschmann-Raether configuration, c.f.\ section \ref{ReflectometryForSC}. The dispersions showed clear avoided crossings, the size of the crossing increasing as the square root of the molecular concentration, as expected according to theory, Equation  \ref{OmegaNgamma}. Splittings up to 230 meV were observed. In the second type of experiment, silver waveguides of transverse and longitudinal sizes of a few microns were fabricated, and polymer areas (containing dye molecules) of different lengths were fabricated on different samples. SPPs were launched at one end of the waveguide, passed through the polymer area, and the spectrum at the end of the polymer covered area was recorded. It was observed that the size and visibility of the splitting was increased for increased length of the polymer area, Fig.\ref{Tommifigure}. For propagating SPPs, the long polymer area corresponded to a longer interaction time with the molecules. One can understand a short interaction time as a large effective 
damping $\gamma$ which according to Equation (\ref{ModesBothGamma}) decreases the Rabi splitting. Thus the variation of length of the polymer area had to be less than the propagation length of the SPP which gives the intrinsic damping of the SPP; only lengths shorter than this are able to increase the damping even further. The results of \cite{Hakala2009} are presented in figure \ref{Tommifigure}; these result have since been reproduced \cite{Moerland2011,Moerland22011}.

In \cite{Cade2009}, strong coupling between dye (Rhodamine 6G) molecules and the plasmon modes supported by an adjacent islandised metal (Ag) film was observed. The novelty of this work was that they also looked at the enhancement of the Raman signal from the dye. They found the Raman signal to be correlated with the strong coupling, observing a maximum Raman signal when a polariton mode matched the Stokes-shifted emission band of the dye. Raman enhancement due to SPP - dye molecule strong coupling has been observed also in \cite{Nagasawa2014}.

An interesting feature of R6G is that it has strongly overlapping monomer and dimer (aggregate) absorption and emission bands \cite{Kuznetsov1973,Levshin1977}. Moreover, the monomer absorption and emission spectra have a double-peaked structure (i.e. a main peak and a shoulder). This double-peaked structure of the monomer may be modified and enhanced by the presence of aggregates at high molecular concentrations. Avoided crossings between both the main absorption peak and the shoulder with SPPs were observed in \cite{Hakala2009}, indicating the formation of hybrid states involving three oscillators: two electronic - the main transition (i.e. the main peak) and shoulder peak of the dye system; and one that is plasmonic, the SPP. 

Strong coupling between SPPs and a number of other molecular systems having broad absorption spectra has recently been demonstrated, for instance with dyes such as Rhodamine 800 \cite{Valmorra2011}, Sulforhodamine 101 \cite{Baieva2012}, Nile Red \cite{Koponen2013} and even bio-molecules such as beta-carotene \cite{Baieva2013}. A systematic study of several different dyes (Rhodamine 640, cresyl violet 670, malachite green, oxazine 720 and 725, methylene blue, DOTCI, HITC) coupled with Au nanorods of various sizes ($\sim 50-100$ nm) and shapes was pursued in \cite{Ni2010}. Due to the inhomogeneous size distribution of the nanorods and other difficulties they were not able to observe clear splittings in the spectra. However, from the shift of the low-energy branch the strength of the coupling could be estimated. It was found that pH and metal ions could be used for reversible control of the shift.       

Very recently, R6G molecules were shown to strongly couple with surface lattice resonances in arrays of metal 
nanoparticles \cite{Rodriguez2013,Vakevainen2013}. Regularly organized metal nanoparticles may display, in addition to 
the localized single particle resonances (LSPR) surface lattice resonances (SLR) corresponding to the diffraction orders 
of the periodic structure \cite{Zou2004,GarciadeAbajo2007,Kravets2008,Auguie2008,Zhou2013}. In previous SPP strong 
coupling studies, on one hand, the effect of periodicity of the nanostructure has been considered 
(c.f.\ section \ref{Nanostructured}), on the other hand strong coupling with localized SPP has been studied 
(section \ref{Localized}). V\"akev\"ainen {\it et al.} \cite{Vakevainen2013} focused on the interplay between the effects 
of periodicity and of the 
localized modes. The mutual couplings of the localized LSPR, the periodicity-dependent SLR, and the molecular exciton resonances were systematically studied by experiments, numerical simulations, and by coupled dipole approximation theory. The observed splittings followed the expected square root of the molecular concentration dependence and were of the order $\sim 150 meV$ for the highest concentrations. One concentration was used in \cite{Rodriguez2013} and similar numbers were obtained. Since the SLR is a collective, delocalized mode, the results mean that molecules located near distant nanoparticles are coherently coupled. The array systems offer long coherence lengths with the possibility of tailoring the field-matter coupling e.g.\ by the shape of the nanoparticle. Furthermore, the SLR can have very narrow lineshapes, the Q-factors can be an order of magnitude more than for propagating or localized SPPs.  

Another interesting class of emitters is that of photochromic molecules. Photochemical effects can be used to induce conformational changes in these molecules which in turn alters the coupling between the molecules and optical fields. This control enables reversible switching from the weak- through to the ultrastrong-coupling regime (for a discussion of ultra-strong coupling see section \ref{ultrastrong} below) using all-optical control \cite{Schwartz2011}. In \cite{Schwartz2011}, spiropyran molecules were used and transformed by UV light into the merocyanin forms for which the dipole moment is large. Strong coupling was observed in two types of systems:\ for low-Q metal microcavities the size of the splitting was about 700 meV (32\% of the molecular transition energy) and for a plasmonic hole-array about 650 meV (30\%). Reversible switching between weak and strong coupling was also demonstrated in \cite{Berrier2011} between porphyrin excitons and surface plasmons, where control of the strength of the dipole moment was achieved by exposing the molecules to NO$_2$ gas; splittings of 130 meV were observed. Another demonstration of switchable strong coupling with spiropyran molecules was given in \cite{Baudrion2013}, where the molecules were strongly coupled to the localized surface plasmon modes of silver nanoparticles. 

Strong coupling results in the system having new normal modes and energies. This modification of the energy landscape may be used to control chemical reactions, as was done in \cite{Hutchison2012}. There, instead if an SPP resonance a low-Q metallic microcavity was used to provide the resonant light mode. We mention this non-SPP strong coupling result here because of the important implications of the concept of modifying chemical landscapes by strong coupling. The reaction considered in \cite{Hutchison2012} was the formation of merocyanin (MC) from the spiropyran (SPI) form of the molecules under UV illumination. The MC ground-excited state excitation was close to the cavity resonance, causing strong coupling. The consequent splitting of the energies of MC modified the chemical energy landscape connecting the two isomers SPI and MC. Modification of the SPI to MC photoisomerization reaction was observed both by optical transmittance measurements and by pump-probe (150 fs pump) spectroscopy.

\subsection{Quantum dots} 

Organic molecules offer a variety of advantages, relative ease of manipulation, and strong dipole moments, especially among the laser dyes. The downside of organic molecules is that they are prone to bleaching and thus will not endure high optical intensities easily. On the other hand, combining the strong coupling regime with high intensity excitation could reveal a lot of interesting physics and applications. Therefore it is of interest to consider emitters that might avoid the bleaching problem. One possibility is quantum dots, also known as semiconductor nanocrystals. A related option are epitaxially grown quantum well materials where shifts ($\sim 7 meV$) of the quantum well exciton energies have been observed \cite{Vasa2008} due to coupling with SPP modes. Describing the experimental data by a coupled-oscillator model predicted couplings of $50 meV$. 

\begin{figure}
\includegraphics[width=0.8\textwidth]{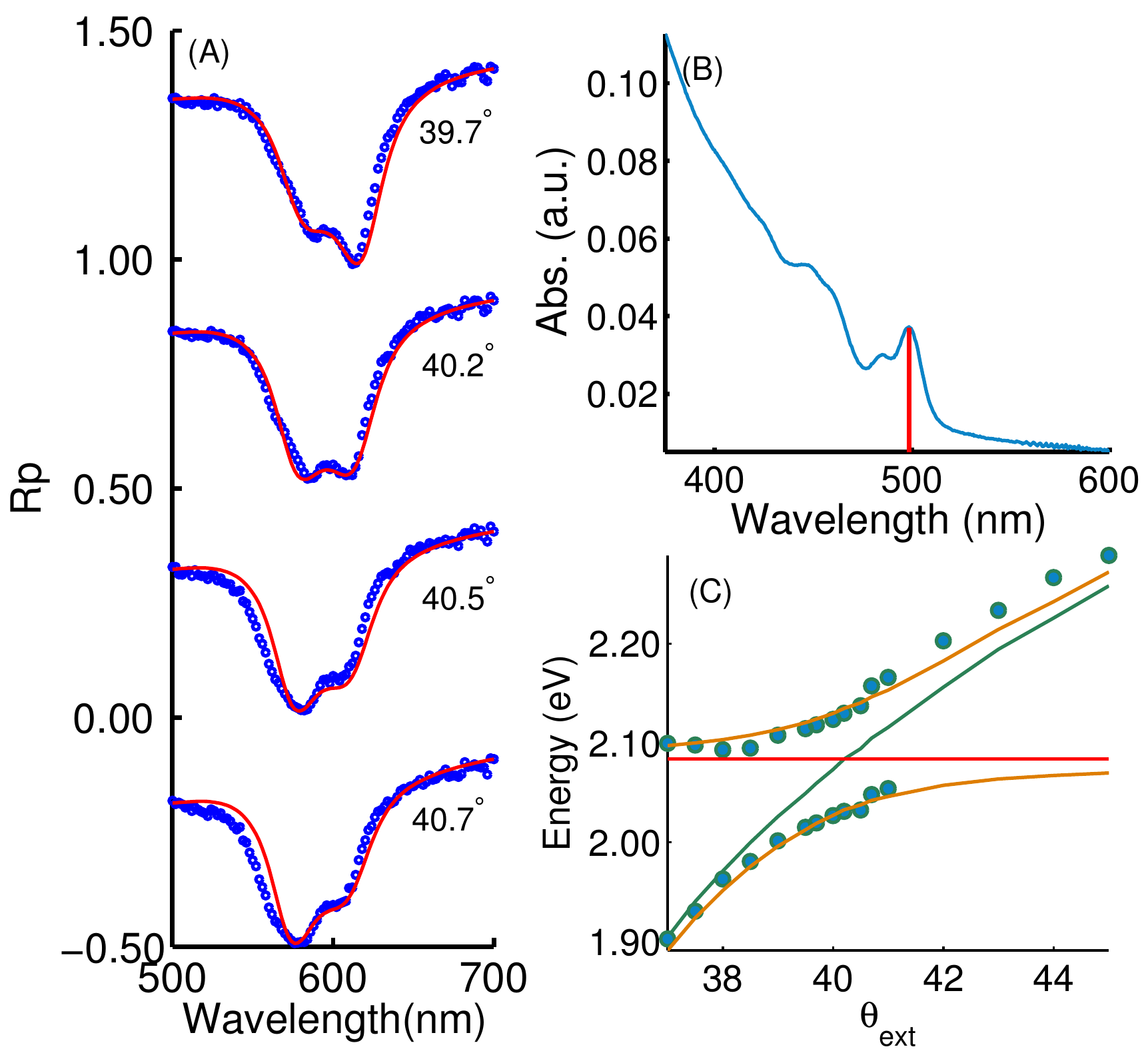}
\caption{(A) Reflectivity spectra of a Ag/CdSe film with
nanocrystals of $\sim$4.3 nm in diameter, for different angles. (B) Normal incidence absorption spectrum of a film of the
CdSe nanocrystals. The vertical line indicates the exciton
transition. (C) Experimental dispersion curve (dots). The green line is
the energy of the uncoupled SPPs, the red line corresponds to the
position of the exciton transition, and the orange lines
are obtained from a fit to a coupled oscillators model, giving a splitting of $\sim$102 meV. Reprinted with permission 
from \cite{Gomez2010a}. Copyright 2010 American Chemical Society.} 
\label{Gomezfigure}
\end{figure}

Strong coupling between SPPs on planar silver films and colloidal CdSe quantum dots was experimentally demonstrated in \cite{Gomez2010a}. Reflectometry measurements were used and the splitting observed was 112 meV. The results are illustrated in figure \ref{Gomezfigure}. In a later work, the same group demonstrated also a double split originating from the SPP mode coupling strongly with two different excitonic modes of the quantum dots (the sample contains quantum dots of two different sizes providing the two excitonic modes) \cite{Gomez2010b}. This corresponds to a hybrid of three modes and is similar to the double split seen in \cite{Hakala2009} where the two excitonic modes were the R6G main peak and shoulder, c.f.\ the dispersions shown in figure \ref{Tommifigure}. Also the dynamics of the quantum dot - SPP system has been studied by steady-state and transient reflectivity measurements in the Kretschmann geometry \cite{Gomez2013} by the group of \cite{Gomez2010a,Gomez2010b}. It was observed that the dynamics is fast whenever the lower hybrid state has predominantly SPP character while when it is predominantly excitonic the dynamics are slower, resembling the typical time scales of 
the CdSe quantum dots. 

It should be noted that, although quantum dots are in many ways emitters with desirable properties, there are fewer reports of strong coupling with quantum dots and SPPs. Moerland {\it et al.} \cite{Moerland2012}, reported only weak coupling regime even under quite high quantum dot densities and large excitation light powers. In the brief review article \cite{Achermann2010} on exciton-plasmon interactions in metal-semiconductor nanostructures the examples of studies of colloidal or epitaxial quantum dots in the weak coupling regime are numerous but in the strong coupling case there are only a few. The quantum dots used in the successful demonstrations \cite{Gomez2010a,Gomez2010b,Gomez2013}
were made by the authors following the approach of \cite{Embden2005}. 

\subsection{Strong coupling and spatial coherence}

\begin{figure}
\includegraphics[width=1.0\textwidth]{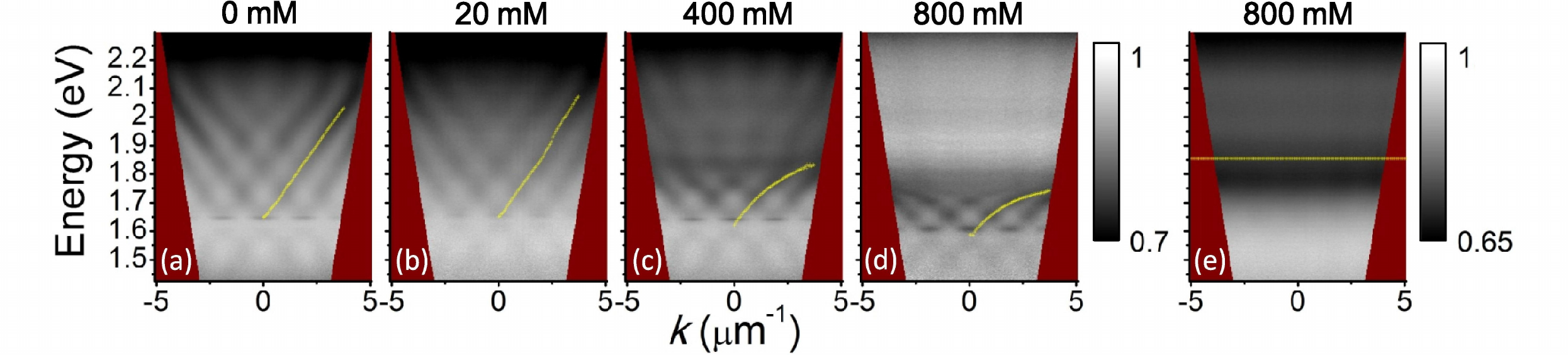}
\caption{(a-d) Spatial coherence images for a silver nanoparticle array covered by a DiD molecule film of different concentrations: 0 mM means no molecules and the increasing concentration means increasing coupling. White colour corresponds to transmission maximum. Interference is observed throughout the weak to strong coupling crossover and the emergence of strong coupling is visible in the bending of the dispersion (yellow line) matching the interference features. (e) A sample having a random distribution of nanoparticles with 800 mM DiD concentration shows no interference fringes. Two transmission minima are seen at 1.85 eV and 2.25 eV, corresponding to DiD absorption and the single particle plasmon resonance, respectively. Reproduced with permission from \cite{Shi2014}.} 
\label{ShiFigure}
\end{figure}

As explained above, the splittings in dispersions, characteristic of strong couping, have been firmly observed for a variety of emitter systems coupling with surface plasmon modes, and even dynamics have been explored. If the new hybrid modes are linear, coherent combinations of the original modes, they should carry the properties of the original modes: in particular the spatial coherence characteristics of an extended light mode. Another way to put this same argument is: in presence of strong coupling, spatially distant emitters should oscillate in phase, creating long-range spatial coherence in the sample. Dispersion measurements alone cannot directly test spatial coherence properties. To be conclusive, one needs to show that the coherence appears in proportion to the weight of the light mode in the hybrid. Spatial coherence of SPP-emitter systems in the strong coupling regime was for demonstrated by Aberra-Guebrou {\it et al.} \cite{AberraGuebrou2012}. This work showed that spatial coherence exists in the strong coupling regime (a different system, namely quantum dots, was given as the weak coupling reference). To prove the connection of the spatial coherence with the weight of the light component of the hybrid mode requires a systematic study of coherence throughout the weak-to-strong coupling crossover: this was done by Shi {\it et al.} \cite{Shi2014}.  In \cite{Shi2014} the spatial coherence properties of a system composed of periodic silver nanoparticle arrays covered with fluorescent organic molecules (DiD) were studied by employing a double slit experiment. The molecule concentration was gradually increased to investigate both the strong and the weak coupling coherence properties within the same system, see Figure \ref{ShiFigure}. Significant spatial coherence lengths in the strongly coupled system are observed even when the mode is very exciton-like. The evolution of spatial coherence was shown to be directly connected to the hybrid mode structure, providing conclusive evidence for the hybrid nature of the normal modes in strongly coupled surface plasmon - emitter systems. 

\section{Quantum description of strong coupling} \label{quantum}

In this section we will first describe strong coupling phenomena in the semiclassical description, section \ref{semiclassical}. There the emitter has a quantum nature, that is, it is described as a two-level system governed by the Schr\"odinger equation, but the field is still a classical electromagnetic field. The ultrastrong coupling regime is briefly discussed in section \ref{ultrastrong}. 
In section \ref{quantum2}, also the field is quantized. Section \ref{semiclassical} is based on elementary one-particle quantum mechanics which we assume the readers of this review are familiar with. The quantization of the electromagnetic field which is the basis of the treatment in section \ref{quantum2} in principle requires knowledge of many-particle quantum physics and the technique of second quantization \cite{Schwabl00}. However, we simply present the results without going through their derivation and describe the main physics so that it is understandable also without a background in quantum optics.  

\subsection{Semiclassical: a quantum emitter interacting with a classical field} \label{semiclassical}

Let us describe the emitter as a quantum two-level system (a spin-half system) with an excited state $|e\rangle$ and a ground state $|g\rangle$, with corresponding energies $E_e$ and $E_g$ (this is just a notation, $|g\rangle$ can be any electronic state not only the actual ground state). We choose $E_e>E_g$. The field, which in our case is the electromagnetic field of the SPP mode, is described by the field amplitude $\mathbf{E} \cos (\omega t)e^{i \mathbf{k} \cdot \mathbf{r}}$. Here $\omega$ and $\mathbf{k}$ are the frequency and wavevector of the SPP mode, respectively, as given by the SPP dispersion, e.g.\ such as in equation (\ref{k_SPP}) and in figures \ref{fig_SPP2} and \ref{SPPdispersionPT} for a planar metal, or in figure \ref{fig_SPP3} for a periodic structure. The vector $\mathbf{E}$ contains the field amplitude and the polarization vector of the field. Note that it is essential to have a well-defined mode which is narrow enough in frequency: damping is not included here, but in the end of this section we will discuss how it can be included. In general, the mode function can be more complicated than the simple plane wave $e^{i \mathbf{k} \cdot \mathbf{r}}$ that we consider here, however, the following derivation can easily be adapted to more complicated forms.   

We apply the standard {\it dipole approximation}, that is, assume that the displacement of the electrons in the emitters (atoms/molecules/quantum dots) due to the field is much smaller than the inverse of $k$ and thus we can approximate $e^{i \mathbf{k} \cdot \mathbf{r}}\simeq e^{i \mathbf{k} \cdot \mathbf{R}}$, where $R$ is the center-of-mass coordinate of the emitter. For an emitter
at a fixed location (i.e.\ it does not feel the light field as a mechanical force) this is a constant irrelevant what comes to the two-level dynamics, and we can choose $\mathbf{R}=0$ for simplicity. 
This approximation should in general be valid, although it is exactly in plasmonics where one might find systems where it is broken down \cite{Andersen2011}, for instance quantum dots with extended electronic states interacting with extremely localized SPP fields.

We first consider a single quantum emitter interacting with the field, a discussion on the many emitters case follows later. We use a vector basis where the excited state is $|e\rangle=\left(\begin{array}{c}
1\\ 0\end{array}\right)$ and the ground state $|g\rangle=\left(\begin{array}{c}
0\\ 1\end{array}\right)$. Transitions from the ground to the
excited state are then provided by the matrix $\sigma_+ = \left(\begin{array}{cc}
0 & 1\\
0 & 0\end{array}\right)$ and the inverse process by
$\sigma_- = \left(\begin{array}{cc}
0 & 0\\
1 & 0\end{array}\right)$. The energies of the states can be expressed using $\sigma_z = \left(\begin{array}{cc}
1 & 0\\
0 & - 1\end{array}\right)$
and the unit matrix $\mathbf{I} = \left(\begin{array}{cc}
1 & 0\\
0 & 1\end{array}\right)$. The Hamiltonian describing the energy of the system is
\begin{equation}
H = \frac{1}{2} E_e (\mathbf{I} + \sigma_z) + \frac{1}{2} E_g (\mathbf{I} - \sigma_z) + \hbar \Omega_0 (\sigma_+ + \sigma_-) \cos
(\omega t) .    
\end{equation} 
Here $\Omega_0$ is the semiclassical Rabi frequency which is proportional to the dipole moment times the field
amplitude, $\Omega_0 = - \mathbf{d}\cdot\mathbf{E}/\hbar$.
If you are interested in the derivation of this Hamiltonian from microscopic principles, the Chapters 3-2.-3-3. of 
\cite{Meystre91} and 2.2 of \cite{Grynberg10} are recommended. To connect with the classical treatment of section \ref{classical},
we denote $(E_e-E_g)/\hbar \equiv \omega_0$. We now perform the so-called rotating 
wave approximation (RWA) (for more information see the above mentioned chapters) in which we first make
a unitary rotation of the basis (in quantum mechanics, this does not change the physics)
by multiplying the wavefunctions by the term $U_1 = \left(\begin{array}{cc}
e^{i\omega_0 t/2} & 0\\
0 & e^{-i\omega_0 t/2} \end{array}\right)$.
Consequently, this transformation has to be applied to the Hamiltonian as well, i.e.\ $U_1 H U^{-1}_1 = U_1 H U^\dagger_1$. 
Expressing furthermore the cosine term  as $\cos (\omega t)
= (e^{i\omega t} + e^{-i\omega t})/2$ one ends up having terms of the form 
$e^{\pm i(\omega - \omega_0)t}$ and $e^{\pm i(\omega + \omega_0)t}$ in the transformed Hamiltonian. The RWA accounts
for neglecting the terms containing $(\omega + \omega_0)t$ since they describe off-resonant processes and 
$\omega + \omega_0$ is a much higher frequency than other frequencies characterizing the dynamics, such as
$\omega - \omega_0$ or $\Omega_0$. The approximation is good reasonably close to resonance $\omega \sim \omega_0$ and
when $\Omega_0$ is not of the same magnitude as $\omega$ and $\omega_0$. It is important to note that this assumption becomes questionable in the ultrastrong coupling
regime, discussed in section \ref{ultrastrong}, where the Rabi frequency $\Omega_0$ is comparable to $\omega$. The signifigance of
the so-called counter-rotating terms $e^{\pm i(\omega + \omega_0)t}$ should be then considered. Here, however, 
they are neglected. Finally, to obtain a convenient form of the Hamiltonian, one more unitary transformation will be
applied namely $U_2 = \left(\begin{array}{cc}
e^{i\delta t/2} & 0\\
0 & e^{-i\delta t/2} \end{array}\right)$ where $\delta = \omega - (E_e-E_g)/\hbar = \omega - \omega_0$. Thus the total transformation
made is $U_T = U_2 U_1$; the actual Hamiltonian in the new basis will contain the transformed Hamiltonian plus an extra term
coming from the Schr\"odinger equation since the transformation $U_T$ is time-dependent (similarly as the Hamiltonian in the interaction
picture is the total transformed Hamiltonian minus the non-interacting Hamiltonian (e.g.\ Chapter 3-1. in \cite{Meystre91})).
After these changes of basis, sometimes referred to as expressing the problem in a rotating frame, and the RWA, the Hamiltonian becomes    
\begin{equation}
H = - \frac{\hbar\delta}{2} \sigma_z + \frac{\hbar \Omega_0}{2} (\sigma_+ + \sigma_-) . \label{semiclassicalHPT}
\end{equation} 
In the literature, the detuning is sometimes defined the other way round than our choice, namely as transition frequency 
minus field frequency, then
of course the minus sign in front of the first term is absent. Furthermore, sometimes $-\Omega_0$ is used instead of 
$\Omega_0$ which is a trivial difference and only exchanges the labeling of the eigenmodes.
The two times two matrix Hamiltonian \ref{semiclassicalHPT} is easy to diagonalize and has the eigenvalues 
\begin{equation}
E_{1}=-\frac{1}{2}\hbar\sqrt{\delta^{2}+\Omega_0^{2}}\label{E1semiclassPT}\end{equation}
\begin{equation}
E_{2}=\frac{1}{2}\hbar\sqrt{\delta^{2}+\Omega_0^{2}},\label{E2semiclassPT}\end{equation}
where the so-called generalized Rabi frequency
\begin{equation}
\Omega=\sqrt{\delta^{2}+\Omega_0^{2}}\label{eq:62}\end{equation}
appears. For writing down the eigenstates, we denote

\begin{equation}
\cos\theta=\frac{\Omega-\delta}{\sqrt{\left(\Omega-\delta\right)^{2}+\Omega_0^2}}\label{eq:63}\end{equation}

\begin{equation}
\sin\theta=\frac{\Omega_0}{\sqrt{\left(\Omega-\delta\right)^{2}+
\Omega_0^2}}.\label{eq:64}\end{equation}
With this definition, the
eigenstates (the so-called {\bf dressed states}) are
\begin{eqnarray}
\left|1\right\rangle &=&-\sin\theta\left|e\right\rangle  +\cos\theta\left|g\right\rangle  \nonumber \\
\left|2\right\rangle &=&\cos\theta\left|e\right\rangle  +\sin\theta\left|g\right\rangle . \label{QeigenvaluesPT}
\end{eqnarray}
Or inversely,
\begin{eqnarray}
\left|g\right\rangle &=&\cos\theta\left|1\right\rangle  +\sin\theta\left|2\right\rangle  \nonumber \\
\left|e\right\rangle &=&-\sin\theta\left|1\right\rangle  +\cos\theta\left|2\right\rangle . \label{QeigenvaluesinversePT}
\end{eqnarray}
If the system is initially in the ground state, we can express its state as a superposition of the new eigenstates:
from equation (\ref{QeigenvaluesinversePT}) we have $\left|g\right\rangle =\cos\theta\left|1\right\rangle  +\sin\theta\left|2\right\rangle$.
Then the time evolution of the state will be easy since the eigenstates evolve with the eigenenergies: the time-dependent wavefunction of the two-level system is
\begin{equation}
| \Psi (t) \rangle = \cos\theta e^{-i E_1 t/\hbar} |1\rangle + e^{-i E_2 t/\hbar} \sin\theta|2\rangle 
\equiv \gamma_g(t) |g\rangle +  \gamma_e(t)  |e\rangle, \label{2leveltimePT}
\end{equation}
where $\gamma_g(t)$, $\gamma_e(t)$ can be calculated using Equation \ref{QeigenvaluesPT} and become 
$\gamma_g(t) = \sin^2 \theta e^{-i E_2 t/\hbar} + \cos^2 \theta e^{-i E_1 t/\hbar}$ and 
$\gamma_e(t) = \sin \theta \cos \theta e^{-i E_2 t/\hbar} - \sin\theta \cos \theta e^{-i E_1 t/\hbar}$. The time-evolution of an initial excited
state and of any superposition of the ground and excited states can be calculated in a similar way.

To understand these results intuitively, let us consider the case on resonance, i.e. when the field and the transition energy are the same, $\delta=0$. Then one has

\begin{equation}
|2\rangle =\frac{1}{\sqrt{2}}\left[|e\rangle  +|g\rangle  \right]\label{eq:68}\end{equation}

\begin{equation}
|1\rangle =\frac{1}{\sqrt{2}}\left[-|e\rangle  +|g\rangle  \right]\label{eq:69}\end{equation}

\begin{equation}
E_{1}= -\frac{\hbar \Omega_0}{2}\label{eq:70}\end{equation}

\begin{equation}
E_{2}=\frac{\hbar \Omega_0}{2}.\label{eq:70}\end{equation}
This means that the eigenstates of the system are actually an equal superposition of the ground and the excited states. Furthermore, the time-evolution 
for an initial ground state is of the form
\begin{equation}
| \Psi (t) \rangle = \cos( \Omega_0 t/2) |g\rangle - i \sin (\Omega_0 t/2) |e\rangle , \label{RabiSemiClPT}
\end{equation}
that is, the system performs {\bf Rabi oscillations} between the ground and the excited states. At the resonance the frequency of these oscillations
becomes the {\bf Rabi frequency} $\Omega_0$ (the frequency is $\Omega_0$ not $\Omega_0/2$ since the probablity of being in the ground state is
$P_g = |\langle g | \Psi (t) \rangle|^2 = \cos^2 ( \Omega_0 t/2) = (1+\cos ( \Omega_0 t))/2$)). Away from resonance, Rabi oscillations take place
at the generalized Rabi frequency $\Omega$ and are smaller in amplitude.

We can now compare the eigenenergies of equations (\ref{E1semiclassPT}) and (\ref{E2semiclassPT}) to the normal modes derived in the classical case of two coupled oscillators, equation (\ref{apprModesPT}) (or (\ref{exactModesPT}) for the less approximate result). The classical and semiclassical results seem to have something in common: in both cases, the two (normal mode/eigenmode) energies are separated by a square root term that contains the detuning squared, here denoted $\delta^2$ and in the classical case $(\kappa - \omega_0)^2$, and another factor, here $\Omega_0^2 = (\mathbf{d}\cdot\mathbf{E})^2$ and $A = (N/V) (e/\sqrt{\epsilon_0 m})^2$ in the classical case (one should consider a single emitter $N=1$ to have a direct comparison here). At resonance, the splitting becomes in the semiclassical case $\Omega_0$ which is the dipole moment times the field amplitude, and in the classical case $e/\sqrt{V\epsilon_0 m}$. Now, there seems to be a puzzling difference: in the classical case the splitting is independent of the field amplitude, whereas in the semiclassical case it is proportional to it, that is, the Rabi frequency and the splitting vanish for a vanishing field. But actually, this is not the right way of making the comparison; we present it only in order to emphasize the point that the splitting derived in the classical case comes from the dispersions where the properties of the oscillator went in to the description only inside the refractive index, that is, in {\it the first order (linear) susceptibility}. The treatment thus basically only describes {\it linear response} to the field: all non-linearities i.e.\ higher order susceptibilities are neglected. In contrast, here in the QM description we presented the {\it exact solution} of the dynamics of a two-level atom interacting with the classical field which naturally takes into account {\it all orders of the field-matter interaction}.

The above issue is further explained by deriving the equivalent of the classical dielectric susceptibility $\chi$, equation (\ref{chiPT}), from the above semiclassical exact solution. For that, one has to calculate the quantum mean value (i.e.\ expectation value) of the induced electric dipole moment

\begin{equation}
\langle \hat{\mathbf{D}} \rangle = \langle \Psi |\hat{\mathbf{D}}| \Psi \rangle , 
\end{equation} 

\noindent where $\Psi$ is the superposition state of the excited and ground states at time $t$, as given by equation (\ref{2leveltimePT}) (we assume here that the two-level system is initially in the ground state). The result becomes (here we follow the derivation and notation in \cite{Grynberg10}, for details see e.g.\ section 2.4.3.\ therein)

\begin{equation}
\langle \hat{D}_i \rangle = d_i \gamma_e^{*} \gamma_g e^{-i \omega_0 t} + c.c. ,
\end{equation} 

\noindent where $d_i$ is the dipole matrix element $d=\langle g | \hat{D}_i | e \rangle$ ($i=x,y,z$). 
Polarization $P_i$ is defined as the dipole moment per unit volume $V$, now $\langle \hat{D}_i \rangle/V$. Let us consider $N$ emitters that interact with the same coherent field. A pumped system is assumed here such that the ground state ("ground" has to be understood as merely a label here) is pumped with the rate $\Lambda_g$ and both states g and e decay with a rate $\gamma$. Then, before turning on the field that causes Rabi oscillations, we have a steady state $N_g = \Lambda_g/\gamma$ and $N_e=0$, and afterwards $N_g+N_e=N=\Lambda_g/\gamma$. The total polarization is obtained by integrating the quantity $\Lambda_g dt_0 \langle \hat{D}_i \rangle/V$ over time, weighted by a factor that describes the decay with rate $\gamma$ (here $\Lambda_g dt_0$ is the number of atoms in state g during the time interval $t_0$, $t_0+dt_0$):

\begin{equation}
P_i = \frac{\Lambda_g}{V} \int_{-\infty}^t dt_0 \langle \hat{D}_i \rangle e^{-\gamma (t - t_0)} .
\end{equation} 

The integral gives (using also $\hbar \Omega_0 = - d E_0$ and assuming isotropic system with polarization parallel to the electric field) 

\begin{equation}
\mathbf{P} = \frac{N}{V} \frac{d^2}{\hbar} \frac{\omega_0-\omega + i \gamma}{\gamma^2 + \Omega_0^2 
+ (\omega - \omega_0)^2} \frac{\mathbf{E}_0}{2} e^{-i \omega t} + c.c.  \label{averagePPT}
\end{equation} 

This is now very close to the polarizability of the classical oscillator case, see equation (\ref{Pwlb}). The difference is the $(\omega - \omega_0)^2$ (Schr\"odinger equation) vs. $(\omega^2 - \omega_0^2)^2 \simeq 2 \omega_0 (\omega - \omega_0)^2$ (classical EOM; Maxwell's equations) as already discussed. Another difference is the term $\Omega_0^2$ in the denominator of equation (\ref{averagePPT}): this comes from the quantum two-level nature of the emitter and describes saturation. However, for weak field intensities it is negligible. Importantly, what is the same is the dependence on the concentration $N/V$. We can now insert this polarizability to the dispersions as we did in equation (\ref{cleanDispersionPT}) of the classical calculation, and obtain the normal mode splitting that is proportional to the square root of concentration. Furthermore, the splitting becomes (just relate the susceptibility from equation (\ref{averagePPT}) to the approximate classical calculation presented earlier) proportional to a term that contains the quantum mechanical dipole moment of the transition as well as $\hbar$, namely $d \sqrt{\omega_0/(\epsilon_0\hbar)}$; this is to be compared to the classical dipole moment $e/\sqrt{\epsilon_0 m}$ of our simple classical Lorentzian model. Thus indeed the semiclassical treatment gives the same result as the classical one what comes to the dependence of the splitting on the concentration, and the same qualitative result on the dependence on a term that contains the dipole moment. This derivation was for the case of both states g and e decaying with $\gamma$; the case where g does not decay could be treated similarly. 

The semiclassical description raises some intriguing questions: Obviously, the individual atoms perform Rabi oscillations with the frequency $\Omega_0 = -d E_0/\hbar$ which goes to zero when the field goes to zero. Nevertheless, the semiclassical description also leads to a splitting 
in the linear absorption spectra that is not proportional to the field but just to $d \sqrt{N/V}$. Will there be any 
dynamics related to the frequency proportional to $d \sqrt{N/V}$? This question was asked already in \cite{Zhu1990}. Their answer, consistent with experimental data, was that in a linear system, time- and frequency domains are connected by a Fourier transform and thus a Fourier transformation of an input pulse $E_{\mathrm{in}}(\omega)$ would be connected to the output pulse $E_{\mathrm{out}}(\omega)$ by the transmission function of the system $t(\omega)$

\begin{equation}
E_{\mathrm{out}}(\omega) = t(\omega) E_{\mathrm{in}}(\omega) . 
\end{equation}
If now $t(\omega)$ contains a double peak structure due to the normal mode splitting (originating from a linear polarizability of the type of equation (\ref{averagePPT})) then
$E_{\mathrm{out}}(\omega)$ will inherit a similar structure, which means that in the time-domain
$E_{\mathrm{out}}(t)$ will display oscillations corresponding to the splitting frequency.  

{\it Thus in the semiclassical description, the normal mode splitting, and related dynamics, for $N$ atoms is not simply $\sqrt{N}$ times the Rabi frequency describing the ground-excited state Rabi oscillations of individual atoms. However, the $N$-atom normal mode splitting and dynamics can be obtained in the linear limit of the individual atom behaviour when the atoms are driven by a coherent field (i.e.\ all atoms that start their oscillation simultaneously remain oscillating in phase).} Especially for the single emitter $N=1$ this is clear: the linear absorption splitting is proportional to $d$ but the frequency of Rabi oscillations is proportional to $E_0 d$. If $N$ atoms perform such oscillations in phase, the linear absorption shows a splitting that is proportional to the square root of the emitter concentration and the
dipole moment. 

To distinguish between the classical and semiclassical cases, one could try to observe saturation effects, but that might be tricky because other non-linearities may step in. The other option would be to probe the difference caused by the classical vs. quantum equations of motion, but as discussed in section \ref{SPP+Lorentzian}, see especially figure \ref{analyticalPlotPT}, it produces a minor effect. One more idea is to have a value for the dipole moment that is either calculated from first principles quantum mechanically, or measured for known quantum emitters, and to show that the size of splitting matches the quantum estimate of the dipole moment rather than the classical estimate.

\subsection{Ultrastrong coupling} \label{ultrastrong}

The ultrastrong regime is characterized by a coupling (here $\Omega_0 = - d E_0/\hbar$) of the same order of magnitude as the frequency of the oscillator and the field (here $\omega$ and $\omega_{0}$). In that case, the rotating wave approximation (RWA) is not justified and one should consider also the counter-rotating terms $e^{\pm i(\omega+\omega_{eg})t}$ neglected above since the time-scales of the dynamics related to the coupling will be now similar to the time-scales of these terms. The ultrastrong coupling regime is extremelly difficult to achieve for traditional quantum optics systems such as atoms in high-finesse cavities.  With superconducting circuits the ultrastrong coupling regime has been realized by enhancing the inductive coupling between a flux qubit and a resonator by using an additional Josephson junction \cite{Niemczyk2010}. It was also noticed that this regime can be simulated with a standard strongly-coupled qubit-resonator system in a rotating frame, under certain conditions for driving (see \cite{Li2012_2} and references therein). For SPPs, there have been recently observations of normal mode splittings that approach the magnitude of
the field frequency \cite{Schwartz2011}.

\subsection{Fully quantum: a quantum emitter interacting with a quantized field} \label{quantum2}

Let us first briefly present the results for the quantized field and the single emitter; the real subtleties 
related to recent SPP experiments will be mentioned with the many-emitter case treated later. In the case of the single emitter, the equivalent of the Hamiltonian \ref{semiclassicalHPT} becomes within the RWA

\begin{equation}
H=\frac{1}{2}\hbar\omega_{0}\sigma_{z}+\hbar\omega\hat{a}^{\dagger}
\hat{a}+\hbar\left(g\hat{a}\sigma_{+}+h.c.\right).\label{eq:53}
\end{equation}

\noindent This is a generic form of the Hamiltonian for a quantized field interacting with a two-level system, the so called Jaynes-Cummings Hamiltonian (see e.g.\ Chapter 13 of \cite{Meystre91}). Here $g$ is proportional to the dipole moment, and $\hat{a}$ is the annihilation operator describes the quantized field, i.e.\ it corresponds to the destruction of a photon, $\hat{a}^\dagger$ corresponds to the creation of a photon. The Hamiltonian $H$ only couples the states $\left|e\right\rangle \left|n\right\rangle $ and $\left|g\right\rangle \left|n+1\right\rangle $ where $n$ refers to the photon number, i.e. one photon is emitted/absorbed when the atom makes a transition between the ground and the excited states. Therefore one can write the Hamiltonian as (remember that the state of the light field may have a distribution of photon numbers)

\begin{equation}
H=\sum_{n}H_{n}.\label{eq:57}\end{equation}
In the basis $\left(\begin{array}{c}
1\\
0\end{array}\right)=\left|e\right\rangle \left|n\right\rangle $, $\left(\begin{array}{c}
0\\
1\end{array}\right)=\left|g\right\rangle \left|n+1\right\rangle $, the Hamiltonian $H_{n}$ is

\begin{equation}
H_{n}=\hbar\left(n+\frac{1}{2}\right)\omega\left(\begin{array}{cc}
1 & 0\\
0 & 1\end{array}\right)+\frac{\hbar}{2}\left(\begin{array}{cc}
-\delta & 2g\sqrt{n+1}\\
2g\sqrt{n+1} & \delta\end{array}\right)\label{eq_L6:58}\end{equation}

\begin{equation}
\delta=\omega-\omega_{0}.\label{eq_L6:59}\end{equation}

\noindent Now diagonalizing equation (\ref{eq_L6:58}) one obtains the following eigenvalues:

\begin{equation}
E_{1n}=\hbar\left(n+\frac{1}{2}\right)\omega-\frac{1}{2}\hbar\sqrt{\delta^{2}+4g^{2}\left(n+1\right)}\label{eq:60}\end{equation}

\begin{equation}
E_{2n}=\hbar\left(n+\frac{1}{2}\right)\omega+\frac{1}{2}\hbar\sqrt{\delta^{2}+4g^{2}\left(n+1\right)},\label{eq:61}\end{equation}
where we can define the generalized Rabi frequency of the quantum case as
\begin{equation}
\mathcal{R}_{n}=\sqrt{\delta^{2}+4g^{2}\left(n+1\right)}.\label{eq:62}\end{equation}
For writing down the eigenstates, we denote,
\begin{equation}
\cos\theta_{n}=\frac{\mathcal{R}_{n}-\delta}{\sqrt{\left(\mathcal{R}_{n}-\delta\right)^{2}+4g^{2}\left(n+1\right)}}\label{eq:63}\end{equation}
\begin{equation}
\sin\theta_{n}=\frac{2g\sqrt{n+1}}{\sqrt{\left(\mathcal{R}_{n}-\delta\right)^{2}+4g^{2}\left(n+1\right)}}.\label{eq:64}\end{equation}
With this definition, the eigenstates (the dressed states) are,
\begin{equation}
\left|1n\right\rangle =-\sin\theta_{n}\left|e\right\rangle \left|n\right\rangle +\cos\theta_{n}\left|g\right\rangle \left|n+1\right\rangle \label{eq_L6:65}\end{equation}
\begin{equation}
\left|2n\right\rangle =\cos\theta_{n}\left|e\right\rangle \left|n\right\rangle +\sin\theta_{n}\left|g\right\rangle \left|n+1\right\rangle .\label{eq_L6:66}\end{equation}

\noindent On resonance, i.e. when the field and the transition energy are the same, $\delta=0$, we have,

\begin{equation}
\left|1n\right\rangle =\frac{1}{\sqrt{2}}\left[-\left|e\right\rangle \left|n\right\rangle +\left|g\right\rangle \left|n+1\right\rangle \right]\label{eq:68}\end{equation}

\begin{equation}
\left|2n\right\rangle =\frac{1}{\sqrt{2}}\left[\left|e\right\rangle \left|n\right\rangle +\left|g\right\rangle \left|n+1\right\rangle \right]\label{eq:69}\end{equation}

\begin{equation}
E_{1n}=\hbar\left(n+\frac{1}{2}\right)\omega - \hbar g\sqrt{n+1}\label{eq:70}\end{equation}

\begin{equation}
E_{2n}=\hbar\left(n+\frac{1}{2}\right)\omega+\hbar g\sqrt{n+1}.\label{eq:70}\end{equation}
This means that the eigenstates of the system are an equal superposition of the ground state + one extra photon and the excited + no extra photon.

Now we see a clear contrast to the semiclassical case: there is a splitting in the spectrum even for zero photon number $n=0$. This is called the {\bf vacuum Rabi splitting} and its existence is attributed to the electromagnetic vacuum fluctuations. Moreover, the system has a discrete set of states and splittings that are given by consecutive photon numbers $n = 0, 1, 2, ...$: this is called {\it the Jaynes-Cummings ladder}. The system now shows Rabi oscillations just like the semiclassical single 2-level system, but within a certain $n,n+1$ manifold and with a frequency that has a non-zero value even for zero photon number. For a distribution of photon numbers, one would expect to see averaged dynamics.

Let us now go to the many-emitter case which is the one relevant for all existing SPP strong coupling experiments. Consider $N$ two-level systems. The Hamiltonian becomes

\begin{equation}
H=\frac{1}{2}\hbar\omega_{eg}\hat{S}_{z}+\hbar\omega\hat{a}^{\dagger} \hat{a}+\hbar\left(g\hat{a}\hat{S}_{+}+h.c.\right)\label{DickePT}
\end{equation}

\noindent where collective two-level operators $\hat{S}_{z}= \sum_{i=1}^N \sigma^{(i)}_{z}$ and $\hat{S}_{+}=\sum_{i=1}^N \sigma^{(i)}_{+}$ have been introduced. This is the so-called Dicke Hamiltonian. It is also known as the Tavis-Cummings Hamiltonian. It can be analytically solved and it has an interesting, rather complicated energy level structure, see e.g.\ \cite{Garraway2011}, and the model displays phase transitions.

Instead of the full solution of the Dicke model, it is common in various light-matter interaction contexts to take the limit where the total number of emitters $N$ is large but the number of excited emitters is small (in other words low photon numbers/intensities exciting the system). The practicality of this limit can be seen by doing the Holstein-Primakoff transformation to the collective spins. The Holstein-Primakoff transformation in general maps spin-systems (such as two-level atoms) to bosonic systems. In the present case, it accounts for 
\begin{equation}
\hat{S}_{+} = \hat{b}^\dagger (N - \hat{b}^\dagger \hat{b})^{1/2},\hat{S}_{-} = (N - \hat{b}^\dagger \hat{b})^{1/2} \hat{b} , 
\hat{S}_{z} = \hat{b}^\dagger \hat{b} - \frac{N}{2},  
\end{equation}
where now $\hat{b}$ and $\hat{b}^\dagger$ are bosonic operators. In the limit of large $N$, one can then approximate the spins by $\sqrt{N} b$ and the Hamiltonian becomes,

\begin{equation}
\hat{H}\simeq \hbar\omega_{0}\left(- \frac{N}{2}+\hat{b}^\dagger \hat{b}\right) + \hbar\omega\hat{a}^{\dagger}
\hat{a}+\hbar g \sqrt{N} \left(\hat{a}^\dagger\hat{b}+h.c.\right)  .
\end{equation}

This is the quantum equivalent of two coupled Harmonic oscillators, with the bosonic modes $\hat{a}$ and $\hat{b}$. The collection of two-level systems now acts like a {\it giant quantum oscillator}, corresponding to the mode $\hat{b}$. Solving the hybrid eigenmodes of this Hamiltonian leads to a splitting of the size $\Omega = 2 g \sqrt{N}$. The coupling coefficient 
$g = F_{geom} d\sqrt{\omega_0/(V\hbar \epsilon_0)}$. The factor $F_{geom}$ is simply a number that depends on details such as how $V$ is defined and the orientation of the dipoles; 
it may have values such as 
$1/2$, $1/\sqrt{2}$, $\sqrt{\pi/2}$, etc. For different cases, see for instance \cite{Meystre91,Walls95,Scully97,Cohen-Tannoudji92,Agranovich2003}.
In other words, $g \propto d\sqrt{\omega_0/(V\hbar \epsilon_0)}$ indicates how the coupling depends on the system parameters and gives a good order of magnitude estimate, but for a precise number the
microscopic details of the specific system should be considered. 
Note also that this formula is for emitters in vacuum; for molecular films typical in SPP strong coupling studies one should multiply $\epsilon_0$ by the permittivity of the film $\epsilon_{film}$. We see that the familiar $\sqrt{N/V}$ term appears again. Moreover, the quantities multiplying it are the same as in the semi-classical case. As with the coupled oscillators discussed earlier in this review, one will then obtain eigenmodes, with a splitting proportional, again, to the square root of concentration and the dipole moment. The eigenmodes are basically superpositions of having a photon in the mode or having an excitation in the giant oscillator mode. The set of oscillators can be understood as a "superatom" with a large dipole moment. This approach, where the set of two-level systems is essentially approximated by one bosonic giant oscillator mode was applied e.g.\ in \cite{Agranovich2003} in the context of semiconductors in microcavities, rather similar to the SPP + organic emitters systems considered in this review. The same basic approach was used in \cite{Gonzalez-Tudela2013} where strong coupling between SPPs and molecules was considered. Both works, naturally, ended up with the 
$d \sqrt{N/V}$ dependence of the splitting in the energy spectrum. In \cite{Gonzalez-Tudela2013}, the specific features of plasmonic systems such as the 2D and near-field (exponentially decaying) nature of the SPP modes on planar metal surfaces were incorporated in the microscopic model. An interesting cavity QED treatment of interactions between a metal nanoparticle and a dipole emitter was presented in \cite{Waks2010}. Quantized treatment of an SPP mode interacting with two spatially separated quantum dots was given in \cite{Chen2011}.

Although the size of the eigenmode splitting in the fully quantum treatment is the same as that in the semiclassical treatment (and the classical treatment, provided we accept a difference in the value of the classical and quantum dipole moments), there is a fundamental difference in the (semi)classical and in the fully quantum (boson-approximated Dicke) results. From the giant quantum oscillator viewpoint one assumes there are two distinct modes that interact with a coupling of strength proportional to $d\sqrt{N/V}$: the photon field, and the giant quantum oscillator made up of $N$ atoms. A splitting of size $d\sqrt{N/V}$ in the energy spectrum corresponds to oscillations between the modes at this frequency, i.e.\ Rabi oscillations. Thus in the fully quantum case, one expects (giant) Rabi oscillations with the frequency $d\sqrt{N/V}$. In contrast, in the semiclassical model one assumes individual two-level systems to oscillate with the single particle Rabi frequency $-d E_0/\hbar$ (not the giant one 
$\propto g\sqrt{N}$), yet the linear susceptibility and thus the splitting visible in the absorption spectrum follows a $d\sqrt{N/V}$ dependence, and one might hope to see the corresponding dynamics in transient spectra.   

In summary, those observations of normal mode splittings in SPP systems which show the $\sqrt{N/V}$ dependence of the splitting basically demonstrate that the emitters are acting coherently. The observations so far are consistent with classical, semiclassical and fully quantum descriptions, since the $\sqrt{N/V}$ dependence of the normal mode splittings seen in linear absorption experiments is the same in all three cases; it is indeed the $\sqrt{N/V}$ dependence that is quantitatively tested in experiments. A further possible quantitative comparison is that of the sizes of the splittings to values given by microscopic quantum theory, namely the $\propto d\sqrt{\omega_0/(\hbar \epsilon_0 \epsilon_{film})}$ dependence. However, this has only been done in \cite{Shi2014} where a quantitative agreement with the microscopic prediction was found, giving evidence for the quantum nature of the emitters. Other differences also exist such as saturation effects in the semiclassical and quantum cases and the $\omega$ vs. $\omega^2$ differences from the quantum and classical equations of motion. One could also try to see the discrete dependence on the photon number $n$, i.e.\ the Jaynes-Cummings ladder. Furthermore, one could also study nonclassical effects by considering the second order correlation function $g^{(2)}$, although the regime of bosonic approximation of the Dicke model does not produce interesting results, instead one has to include non-linearities inherent in the Dicke model \cite{Gonzalez-Tudela2013}.

In some other systems, for instance Rydberg atoms (for reviews see e.g.\ \cite{Saffman2010,Comparat2010,Low2012}), phenomena related to the giant oscillator (superatom) behaviour have been observed. Phase transitions related to the Dicke model have been observed for ultracold atoms in optical cavities \cite{Baumann2010}, for a review see \cite{Ritsch2013}. 

\subsection{Damping in the quantum case}

In the quantum case the effects of damping are usually described by a system-reservoir approach, where the system of interest is coupled to a bath (reservoir) and a so-called master equation is derived (see e.g.\ Chapter 14 in \cite{Meystre91} and Chapter 10 in \cite{Walls95}). The result is that the Rabi oscillations will be damped. Naturally, if the damping is too large i.e.\ of order or faster than the Rabi oscillations, a Fourier transform of the dynamics will not produce a clear split in the spectrum. Thus the condition to observe strong coupling effects is the same as in the classical case: the damping rate has to be smaller than the Rabi frequency, i.e.\ the linewidth of the modes smaller than the splitting in energies, as a general rule. However, careful studies of particular systems can reveal more complicated effects, such as the case of inhomogeneously broadened oscillators  
\cite{Houdre1996} discussed in section \ref{DyePT}. 

\subsection{Thresholdless lasing}

The coherent interaction of an electromagnetic field with quantum emitters is the basis of an accurate description for phenomena such as stimulated and spontaneous emission, gain, and lasing (or the spaser \cite{Bergman2003} in plasmonics). These phenomena can usually occur both in the strong and weak coupling regimes, although for instance lasing is primarily considered in the weak coupling limit. To restrict the scope of this review to something manageable, we will not discuss these issues here; they each deserve their own review in the context of plasmonics. However, it is of relevance to mention that {\it thresholdless lasing} is expected to occur in the strong coupling regime \cite{Rice1994}. Rice and Carmichael \cite{Rice1994} showed that the concept of laser threshold is well defined only in the thermodynamic limit (as is the case in any phase transition phenomenon), thermodynamic limit here meaning small fluctuations, that is, small amount of spontaneous emission to the lasing mode. In the strong coupling limit, spontaneous emission to the lasing mode becomes large and the lasing threshold not just goes to zero but actually ceases to exist by definition. There is thus a fundamental conceptual difference between lasing in the weak and strong coupling limits. It would be of interest to explore the possibility of thresholdless lasing in the case of strongly coupled SPP-emitter systems.

\section{Conclusions, open questions and future directions} \label{conclusions}

When we began work on this review more than two years ago we did so in the knowledge that this was a topical area, and that a review might provide colleagues with a useful summary of the field. We had not anticipated the extent to which interest in this area would pick up over the intervening two years. What drives this blossoming of interest? Maybe one reason is that strong coupling spans several scientific realms that are deeply connected by common underlying physics. Thus whilst strong coupling originated in the areas of ultracold atomic physics and cavity quantum electrodynamics, other areas of science are now involved. For example, Hutchison {\it et al.}  \cite{Hutchison2012} have shown that the combination of strong coupling between quantum emitters and surface plasmon polaritons offers the fascinating prospect of 'engineering' electronic energy levels relevant to chemical processes. An indication of the wide scope the strong coupling phenomenon can be seen from reports on SPP - quantum emitter strong coupling in the context of: vibrational transitions \cite{Nagasawa_JPCL_2014_5_14}, ionization potentials and work functions \cite{Hutchison_AdvMat_2013_25_2481}, and thermodynamic processes \cite{Canaguier-Durand_AngChemIntEd_2013_52_10533}; these are in addition to the more 'obvious' areas of quantum information processing and thresholdless lasing mentioned above. One of the major advantages that the SPP - quantum emitter provides is that of an 'open cavity', i.e. enabling easy access to the mode volume in which the strong coupling takes place.

\begin{figure}
\includegraphics[width=1.0\textwidth]{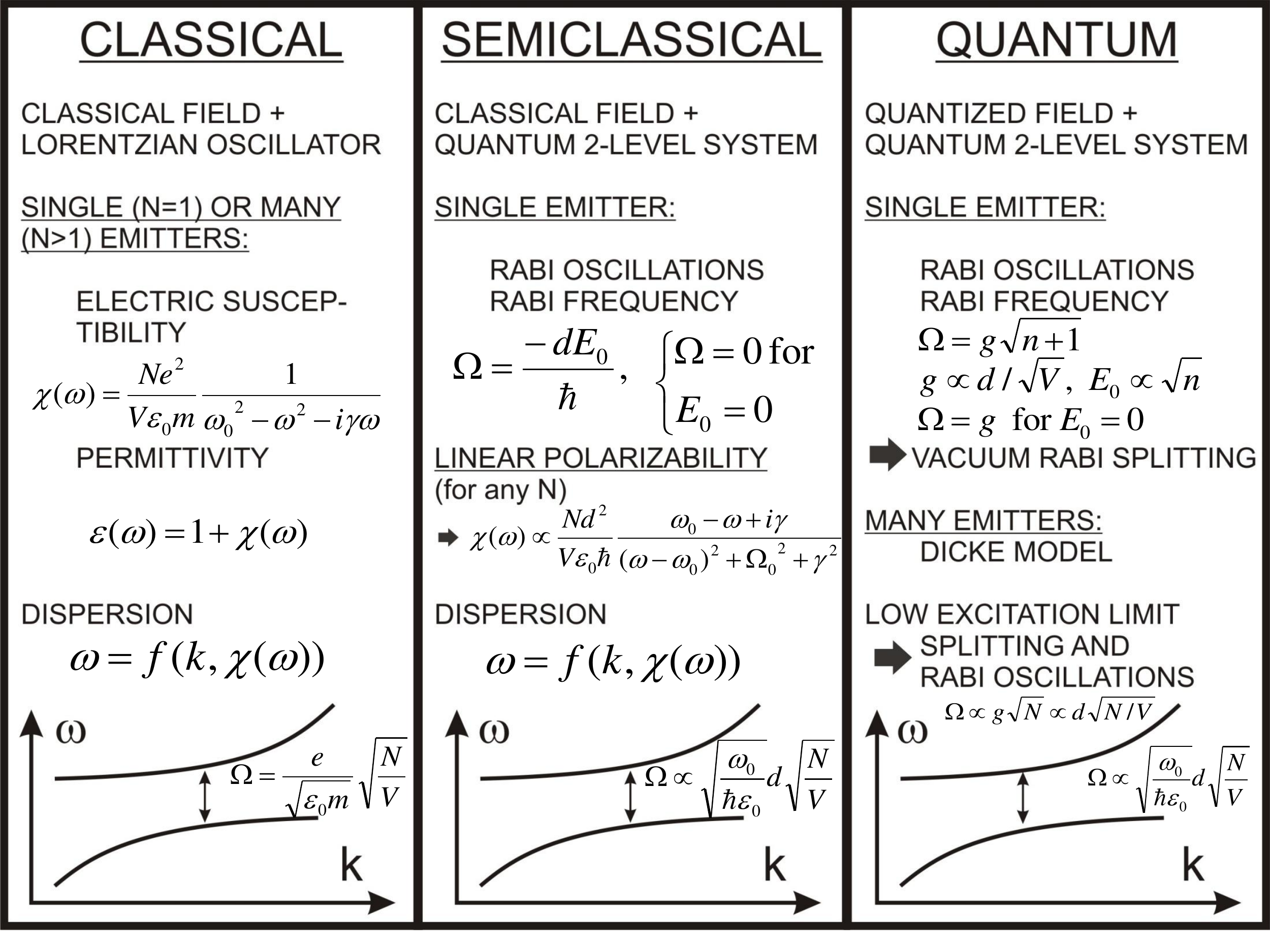}
\caption{Comparison of classical, semiclassical and quantum descriptions of strong coupling
for single and many emitters. Low excitation (linear) regime has to be assumed to obtain the same 
emitter concentration $\sqrt{N/V}$ dependence in all three cases; higher light (SPP) intensities
would lead to clear differences. Other differences, and similarities, are visible in the formulas.}
\label{ClassSemiQuantum}
\end{figure}

One of our aims in writing this review was to provide a single source for the background physics that is needed to better appreciate recent developments in this field. In this context we have been keen to discuss different ways of looking at the strong coupling phenomenon: classical (section \ref{classical}), semi-classical (section \ref{semiclassical}) and quantum (section \ref{quantum2}). Figure \ref{ClassSemiQuantum} summarizes these three approaches. In the classical description the energy level splitting associated with the strong coupling between a quantum emitter and an optical mode (here a surface plasmon polariton) arises when the susceptibility of the emitter, based on a classical Lorentzian oscillator is substituted into the dispersion relation for a SPP. In the semi-classical case the susceptibility is derived by considering a two-level system rather than a Lorentzian oscillator, and then substituting this susceptibility into the (classical) SPP dispersion relation. In the fully quantum case both emitter (two level system) and mode (quantum field) are considered quantum mechanically and the splitting comes naturally via solutions to the appropriate Hamiltonian.

We have sought to show that all three approaches lead to the same prediction for the extent of the splitting. Where the theories differ is in the non-linear behaviour and in the exact relationship between the dipole moment of the emitter and the extent of the coupling. This last point is hard to check since it requires an accurate measurement for the dipole moment of the emitter of interest. Testing for differences between the classical and quantum descriptions through comparison with experiment is an area we expect to see develop in the future. We also expect to see further tests of the quantum/classical comparison as further experiments showing Rabi oscillations in the time-domain emerge.

A key open question is whether strong coupling can be achieved between a plasmon mode and a single emitter at room temperature. Quantum dots are obvious candidates for this since they are nanometer scale in size, similar to molecular sizes, but it is easier to envisage ways of positioning single quantum dots at desired locations near the metallic structure that hosts the SPP mode. It has been theoretically predicted that the vacuum Rabi splitting might be observable for a single quantum dot in the centre of a dimer nanoantenna \cite{Savasta2010}. In this work, no specific type of quantum dot was considered; dipole moments $\mu/e$ of 0.3-0.5 nm were assumed in the calculations. Tr\"ugler and Hohenester \cite{Trugler_PRB_2008_77_115403} examined this question theoretically using a master equation approach. Their conclusion was that strong coupling should be both possible and measurable between the localised surface plasmon-polariton mode of a suitably shaped metallic nanoparticle and a single molecule. They did not consider quantum dots specifically but rather single dipoles in general, having dipole moments of 10 a.u., which is 0.68 nm normalized with respect to the elementary charge. This work also predicted Rabi splittings of order 50 meV for singe emitters placed a few nanometers from the tip of a cigar-shaped nanorod. Similar nano-antennas were considered in a later work \cite{Slowik2013} where strong coupling was also also predicted. In \cite{Hummer2013} various plasmonic waveguide structures were considered for single emitter strong coupling, it was concluded that cryogenic temperatures would be needed to observe strong coupling. The results of this work \cite{Hummer2013} also suggest the need for nanostructures with very small dimensions and sharp features to provide small enough mode volumes for room temperature single emitter strong coupling. Sharp nano-tips were predicted to be advantageous for coupling light from an emitter into a plasmonic channel \cite{Chang2006} whilst in \cite{Agostino2013} conically shaped nanoparticles were predicted to be extremely useful for achieving strong coupling with single emitters. Experiments to confirm such single-emitter strong-coupling have yet to be reported at the time of writing. If/when such an experiment is reported it will be a major landmark in the field.

What other developments might we anticipate? Rich new physics is also expected if various types of interactions between the emitters are considered. For instance in \cite{Salomon2012} a theoretical study of a SPP-emitter system, beyond the usual approach namely including dipole-dipole interaction, predicted that a collective mode emerges in the middle of the split. It would be also interesting to see if Dicke super-radiance can be realised. Yet another intriguing possibility is to use the high field enhancement associated with plasmon modes to see if non-linear Rabi splitting can be seen. The extent of the splitting depends on the number of (plasmon) quanta. For low intensity excitation the vacuum state is the most probable, but for higher excitation the splitting should scale as $\sqrt{n+1}$, as discussed in section \ref{quantum2}. Such effects have been seen, for instance, at microwave frequencies in circuit QED \cite{Fink_Nature_2008_454_315,Bishop_NatPhys_2009_5_105}, and more recently at near-IR frequencies for quantum dots in a micro-pillar cavity \cite{Kasprzak_NJP_2013_15_045013}. Plasmonic components are key elements in building metamaterials, and strong coupling is also of relevance in the metamaterials  context \cite{Benz_NatCom_2013_4_2882}.

Harnessing plasmonics in the field of strong coupling seems a natural way to exploit the key advantages that plasmonics offers in pushing our control of light down to the nanoscale, namely optical field enhancement and optical field confinement. Given the rapid recent advances in plasmonics it is perhaps not surprising that strong coupling of quantum emitters and SPPs is such a keenly pursued topic. Given the level of current interest, the wide range of specialist fields involved and the rapidly developing capabilities afforded by a range of nanotechnologies we can expect to see many new and unexpected developments emerge in the years ahead.

\ack
We thank Tommi Hakala, Jani-Petri Martikainen, Heikki Rekola and Lei Shi for careful reading of the manuscript, useful comments, and help with the figures. We are also indebted to many colleagues for valuable discussions on many of the topics discussed here. This review was conceived in Benasque, Spain at Nanolight12. It was (almost) completed two years later back in Benasque, at Nanolight14. The stimulating conditions provided by both colleagues and the organisers at both of these meetings have been invaluable in helping us complete this task. WLB would like to acknowledge the financial support of The Leverhumle Trust. PT would like to acknowledge the financial support by the Academy of Finland through its Centres of Excellence Programme (projects No.\ 251748, No. 135000, No.\ 141039, No.\ 263347 and No.\ 272490) and by the European Research Council (ERC-2013-AdG-340748-CODE). As a final note we would like to apologise to all those whose work we have not managed to include, recently, each week has brought a new clutch of papers, and we have had to stop somewhere.


\section*{References}

\bibliographystyle{iopart-num}
\bibliography{bib_SCreview,bib_SCreview_wlb}

\end{document}